\documentclass[final,3p,times]{elsarticle}

\usepackage{amssymb}

\usepackage{amsmath}
\usepackage{cancel}
\usepackage{hyperref}
\usepackage{epstopdf}
\usepackage[mathscr]{eucal}

\usepackage{packagesAndStyles/tex/general}
\usepackage{packagesAndStyles/tex/tensorCommon}
\usepackage{packagesAndStyles/tex/tensorEquation}
\usepackage{packagesAndStyles/tex/tensorOperator}
\usepackage{packagesAndStyles/tex/finiteVolume}

\usepackage{subcaption}

\let\originaleqref\eqref
\renewcommand{\eqref}{Eqn.~\originaleqref}

\usepackage{tikz}
\usetikzlibrary{shapes,arrows}

\tikzstyle{decision} = [diamond, draw, fill=blue!20,
    text width=5em, text badly centered, node distance=2cm, inner sep=0pt]
\tikzstyle{myBlock} = [rectangle, draw, fill=blue!20,
    text width=8em, text centered, minimum height=2em,
    node distance = 2cm]
\tikzstyle{line} = [draw, -latex']
\tikzstyle{cloud} = [draw, ellipse,fill=red!20, node distance=2cm,
    minimum height=2em, text centered]
\tikzstyle{none} = [minimum height=0pt, text width = 0em, node distance=0cm]

\usepackage{tikz}
\usetikzlibrary{shapes,arrows}

\tikzstyle{block} = [rectangle, draw, fill=white!5,
text width=21em, text centered, rounded corners, minimum height=3em, node
distance = 7cm]
\tikzstyle{line} = [draw, -latex']

\usepackage{color}

\usepackage{setspace}
\usepackage{packagesAndStyles/tex/specialsymbolsfonts}

\hypersetup{
   breaklinks,
   pdfcreator    = \LaTeX{},
   pdfproducer   = \LaTeX{},
   bookmarksopen = true,
   colorlinks = true,
   urlcolor   = cyan,
   citecolor  = red,
   linkcolor  = green,
   filecolor  = blue,
}

\journal{Journal of Computational Physics}

\begin{document}

\newcommand{\divIndex}[2] {\ensuremath{\frac{\partial #1}{\partial #2}}}

\begin{frontmatter}


\title{A Sharp Free Surface Finite Volume Method Applied to Gravity Wave Flows}

\author[address1]{Vuko Vuk\v{c}evi\'{c}}
\ead{vuko.vukcevic@fsb.hr}

\author[address2]{Johan Roenby}
\ead{johan.roenby@gmail.com}

\author[address1]{Inno Gatin}
\ead{inno.gatin@fsb.hr}

\author[address1,address3]{Hrvoje Jasak\corref{cor1}}
\ead{hrvoje.jasak@fsb.hr}
\ead{h.jasak@wikki.co.uk}

\address[address1]{University of Zagreb, Faculty of Mechanical Engineering and
    Naval Architecture, Ivana Lu\v{c}i\'{c}a 5, Zagreb, Croatia}
\address[address2]{Stromning, Luftmarinegade 62, DK-1432, Copenhagen K, Denmark}
\address[address3]{Wikki Ltd, 459 Southbank House, SE1 7SJ, London, United
    Kingdom}

\cortext[cor1]{Corresponding author.}

\begin{abstract}

This paper presents a sharp free surface method for fully nonlinear flow of
two immiscible phases for wave propagation problems in the Finite Volume
framework. The method resolves a sharp interface between two phases by combining
the geometric reconstruction Volume--of--Fluid scheme \texttt{isoAdvector} for
accurate advection of the free surface with the Ghost Fluid Method for
the consistent treatment of density and pressure gradient discontinuities at the
free surface. The method uses a compact computational stencil irrespective of
cell shape and is formally second--order accurate in time and space. The primary
focus of this work is to present the combined method and verify and validate it
for wave--related problems in ocean sciences, marine and coastal engineering, by
considering the following test cases: i) wave propagation of a
two--dimensional wave with moderate steepness, ii) green water
(water--on--deck) simulations for a ship model with violent free surface
flow patterns. The method is implemented in OpenFOAM, an open source
software for Computational Fluid Dynamics.

\end{abstract}

\begin{keyword}

Free surface flow \sep
Geometric reconstruction Volume--of--Fluid method \texttt{isoAdvector} \sep
Ghost Fluid Method \sep
OpenFOAM \sep
Wave propagation and loads \sep
Verification and validation

\end{keyword}

\end{frontmatter}


\section{Introduction}
\label{sec:intro}

Computational Fluid Dynamics (CFD) tools for free surface flows are becoming
increasingly popular due to increased availability of computational resources
required for this type of simulations in both scientific and industrial setting.
Two--phase CFD simulations represent an important addition to Experimental Fluid
Dynamics (EFD), where highly resolved Direct Numerical Simulations
(DNS)~\cite{yuEtAl2016} provide detailed information of flow features, which may
be used to investigate underlying physical mechanisms in free surface flows with
complex topological changes (\eg plunging breaking wave~\cite{deikeEtAl2016}).
The CFD methods for free surface flows have a wide range of applications for
industrially relevant problems, such as marine
hydrodynamics~\cite{larssonEtAl2013, sternEtAl2012}, wave propagation and load
assessment~\cite{higueraEtAl2015, paulsenEtAl2014b}, seakeeping of offshore
platforms and ships~\cite{vukcevicEtAl31SNH2016, vukcevicJasakObliqueTokyo2015},
\etc\\
\indent The presence of a free surface implies a
discontinuity in pressure and velocity gradient tangential to the
interface~\cite{batchelor1967}. In recent years, substantial research effort has
been undertaken in order to numerically handle the discontinuities, where a
number of different methods has emerged, most of them relying on the Eulerian
approach. One of the first methods adopted by many authors is based on diffusive
interface modelling and conditional averaging~\cite{dopazo1977}, where two sets
of governing equations for two fluids are combined in a single set of equations
with density and viscosity varying across the interface based on the volume
fraction~\cite{ubbinkIssa1999}. Similar methods are still popular for
simulations of gravity waves~\cite{jacobsenEtAl2012, higueraEtAl2013a,
paulsenEtAl2014a, lupieriEtAl2014}, where the numerical smearing of the
interface dictates the resolution of pressure discontinuities. In sharp
interface methods, two similar approaches are used to
treat the discontinuities at the free surface: the Ghost Fluid Method
(GFM)~\cite{fedkiwEtAl1999b, fedkiwEtAl1999} and the Embedded Free Surface (EFS)
method~\cite{wangEtAl2013}. Both methods assume a single--fluid formulation with
special interpolation schemes handling the discontinuities,
while their difference lies in the numerical discretisation of interface jump
conditions~\cite{wangEtAl2013}. The GFM has been first introduced by
Fedkiw~\etal~\cite{fedkiwEtAl1999b} in a finite difference framework. The
GFM has been used alongside the Level Set (LS) interface capturing method
to ensure sharp interface reconstruction, where layers of ghost cells
(nodes) have been used to extrapolate the fields from one--side of the
interface. The GFM has been extended to free surface incompressible flows
with large density variations~\cite{olssonKreiss2005}, where it is also used
alongside a variant of LS method that ensures mass
conservation~\cite{desjardinsEtAl2008}. More recent works include the
extension of the method to compressible two--phase flows by Bo and
Grove~\cite{boGrove2014} using the Volume--of--Fluid (VOF) method and an
extensive review of different treatments of tangential stress balance at the
interface in two--phase incompressible flow by
Lalanne~\etal~\cite{lalanneEtAl2015}. Most of the publications use either finite
differences or finite volumes for spatial discretisation on structured grids.
Queutey and Visonneau~\cite{queuteyVisonneau2007} present a method for handling
free surface discontinuities in arbitrary polyhedral Finite Volume (FV)
framework, where the interface is assumed to be aligned with the grid faces.
Such a procedure imposes restrictions in terms of grid refinement
near the free surface. The method resembles the GFM as used by other authors in
the same field (\eg Huang~\etal~\cite{huangEtAl2007}), although no comments have
been made on their similarities and differences. The GFM has been recently
extended to arbitrary polyhedral FV framework by
Vuk\v{c}evi\'{c}~\etal~\cite{vukcevicEtAl2017}, where no a--priori assumptions
have been made on the location of the interface. Although the method ensures
consistent treatment of discontinuities across the interface even with the
algebraic VOF method for interface capturing~\cite{ubbinkIssa1999,
ruschePhD2002}, the numerical diffusion of the volume fraction field cannot be
easily controlled.\\
\indent In order to avoid the problem of controlling the numerical diffusion of
the volume fraction field in case of complex flow patterns, a better alternative
to interface advection scheme needs to be considered. The two most established
methods for interface advection in Eulerian framework found in the
literature~\cite{tryggvasonEtAl2011} are: geometrically reconstructed
Volume--of--Fluid (VOF) method~\cite{aulisaEtAl2013, roenbyEtAl2016} and the
Level Set (LS) method~\cite{sethian1996, osherFedkiw2003, olssonEtAl2007}. In
LS, a colour function based on signed distance profile~\cite{osherFedkiw2003} is
used, where the zero level set represents the sharp interface. Without a
special treatment, the LS method is not conservative~\cite{olssonKreiss2005,
olssonEtAl2007} and the colour function often loses the signed distance property
during the numerical advection step due to discretisation errors. The latter
problem is by various redistancing algorithms. The most direct
approach is the actual calculation of the smallest distance to the zero level
set surface~\cite{gomezEtAl2005}. Costly direct calculation can be avoided if
one solves an Eikonal equation which forces the gradient of the colour function
to be equal to unity~\cite{hartmannEtAl2008}. Both strategies are often
performed only in a narrow band near the interface in order to minimise the
computational effort.  Sun and Beckermann~\cite{sunBeckermann2007,
sunBeckermann2008} presented an alternative LS advection equation derived from
diffusive interface Phase Field equation. The resulting transport equation has
additional terms that serve to keep the signed distance profile of a colour
function during the advection step. The method has been recently extended to
unstructured polyhedral FV grids~\cite{vukcevicEtAl2016a}, where the authors
have shown that all the additional terms can be treated in an implicit manner.
Such a procedure works well for wave propagation problems since the diffusion of
the interface is a user--defined parameter and the signed distance field is well
preserved~\cite{vukcevicEtAl2016b}, although there is no guarantee that phase
mass conservation will be preserved.\\
\indent Geometric VOF methods rely on the volume fraction field in order to
reconstruct and advect the interface. These methods have proven to be very
accurate and efficient~\cite{aulisaEtAl2013, Lopez2008, Hernandez2008} on
structured, hexahedral grids. Substantial research has also been devoted to
interface reconstruction on unstructured grids~\cite{Ahn2007, Xie2014}. Although
the methods have proved to be accurate even on unstructured grids, their
performance is significantly limited by complicated geometric operations they
require. Recently, Roenby~\etal~\cite{roenbyEtAl2016} presented the
\texttt{isoAdvector} algorithm developed in OpenFOAM~\cite{Roenby2017} which is
computationally efficient since it keeps costly geometric operations to a
minimum. The geometric reconstruction step uses the volume fraction field
interpolated from cell--centres to all points of arbitrary polyhedral cells,
while the advection step is based on analytically evaluating the flux of one
fluid through a polygonal face within a time--step.\\
\indent The aim of this work is to combine the \texttt{isoAdvector}
method~\cite{roenbyEtAl2016} and the Ghost Fluid Method~\cite{vukcevicEtAl2017}
and apply it to accurate simulations of gravity wave related two--phase flows.
In recent years, CFD tools have been extensively used for wave
propagation
problems~\cite{jacobsenEtAl2012, higueraEtAl2013a, higueraEtAl2013b,
higueraEtAl2015, paulsenEtAl2014a, paulsenEtAl2014b, lupieriEtAl2014}, with a
focus on establishing accuracy and maturity of the applied toolboxes. Although
some of the
authors indeed perform grid sensitivity studies, they do not report achieved
orders of accuracy and the corresponding numerical uncertainty following the
guidelines in the established literature~\cite{roache1997, sternEtAl2001}. In
this work, we follow the least squares procedure by E\c{c}a and
Hoekstra~\cite{ecaHoekstra2014} to assess the numerical uncertainty and achieved
order of accuracy of the proposed, formally second order accurate approach.\\
\indent The paper is organised as follows. First, the mathematical
model of incompressible, free surface flow in a single--fluid formulation is
presented. Second, the numerical details of the Ghost Fluid Method are
presented, followed by the formulation of the \texttt{isoAdvector} method for
interface advection. Third, a set of relevant verification and validation test
cases is presented.  Verification is first performed on a wave propagation case,
where the achieved order of accuracy and numerical uncertainty are assessed in
detail. The estimates on dissipation and dispersion errors during wave
propagation in a long numerical wave tank are also provided. Additional test
case presents verification and validation study for a violent free surface
flow occurring on a deck of a simplified ship model in regular waves, comparing
the results to experimental data by Lee~\etal~\cite{leeEtAl2012}.


\section{Mathematical Model}
\label{sec:mathematicalModel}

\indent We consider two incompressible, Newtonian fluids separeted by a sharp
interface in a gravitational field. The motion of each fluid is governed by
Navier--Stokes equations in primitive
form together with the incompressibility constraint~\cite{desjardinsEtAl2008}:
\begin{equation}
\label{eq:2}
    \ddt{\U}
  + \div{(\U \U)}
  - \div \left( \nu \grad \U \right)
  =
  - \frac{1}{\rho} \grad p_d
    \mcomma
\end{equation}
\begin{equation}
\label{eq:1}
    \div \U = 0.
\end{equation}
\noindent Here $\U$ is the velocity field, and $\nu$ is the kinematic viscosity
field assumed to take different constant values, $\nu^+$ and $\nu^-$, in each of
the two fluids. Similarly, $\rho$ is the density field taking different constant
values, $\rho^+$ and $\rho^-$, in each fluid. The quantity $p_d$ in \eqref{eq:2}
is the dynamic pressure defined as the pressure field, $p$, with the hydrostatic
potential subtracted:
\begin{equation}
\label{eq:3}
    p_d = p - \rho \vec{g} \dprod \vec{x} \mcomma
\end{equation}
where $\vec{g}$ is the gravitational acceleration and $\vec{x}$ is
the position vector. For two--fluid problems we must also account for the
position and motion of the fluid interface, on which appropriate boundary
conditions must be imposed. In what follows, we will work with a slightly
simplified form of jump conditions at the interface, neglecting surface tension
effects and the effect of tangential stress
balance compared to normal stress balance~\cite{vukcevicPhD2016}. This
assumption is justified for flows with large Weber and Reynolds numbers that are
of interest in this work, as discussed by Huang~\etal~\cite{huangEtAl2007} in
detail. The jump conditions are briefly outlined here, while the reader is referred
to~\cite{vukcevicPhD2016} for a detailed analysis.
\begin{itemize}
    \item Density discontinuity:
    \begin{equation}
    \label{eq:4}
        \left[ \rho \right]
      =
        \rho^-
      - \rho^+
        \mcomma
    \end{equation}
    where $[ \dprod ]$ notation is taken from the GFM
    literature~\cite{desjardinsEtAl2008, huangEtAl2007} and denotes the jump in
    variables across the free surface. Superscripts $^+$ and $^-$ denote
    the values infinitesimally close to the free surface in heavier
    and in lighter fluid, respectively.

    \item Kinematic boundary condition:
    \begin{equation}
    \label{eq:5}
        \left[ \U \right]
      =
        \U^- - \U^+
      =
        \vec{0}
        \mfstop
    \end{equation}
    Kinematic boundary condition ensures the continuity of the velocity field at
    the free surface.

    \item Simplified tangential stress balance:
    \begin{equation}
    \label{eq:6}
        \left[ \grad_n \U_t \right] = \vec{0} \mcomma
    \end{equation}
    stating that the normal gradient of the tangential velocity field does not
    have a jump. This simplified form is obtained by neglecting surface
    divergence of surface tension force and surface gradient of the normal
    velocity component~\cite{vukcevicPhD2016}.

    \item Dynamic boundary condition:
    \begin{equation}
    \label{eq:7}
        [ p_d ]
      =
      - \left[ \rho \right] \vec{g} \dprod \vec{x}
        \mfstop
    \end{equation}
    The dynamic boundary condition is obtained by neglecting surface tension
    effects and using the pressure decomposition given by~\eqref{eq:3}.

    \item Additional dynamic boundary condition:
    \begin{equation}
    \label{eq:8}
        \left[ \frac{\grad p_d}{\rho} \right]
      =
        \vec{0}
        \mcomma
    \end{equation}
    follows from the inspection of Navier--Stokes equations (\eqref{eq:2})
    when one assumes the simplified form of the tangential stress balance given
    by~\eqref{eq:6}.
\end{itemize}
\indent It is important to clearly state that the jump conditions given by
Eqns.~(\ref{eq:4})--(\ref{eq:8}) have been derived with the following
assumptions:
\begin{itemize}
    \item Surface tension effects are neglected ($\sigma = 0$), \ie we only
    consider flows with high Weber numbers;
    \item High Reynolds number flows investigated in this work allow us to
    assume that the tangential stress balance is of minor importance compared to
    normal stress balance~\cite{huangEtAl2007}. The kinematic viscosity is
    therefore defined in terms of volume fraction
    function~\cite{vukcevicPhD2016}:
    \begin{equation}
    \label{eq:9}
        \nu
      =
        \alpha \nu^+
      + (1 - \alpha) \nu^-
        \mcomma
    \end{equation}
    where $\nu^+$ is the kinematic viscosity of heaviour fluid and $\nu^-$ is
    the kinematic viscosity of lighter fluid.
\end{itemize}
\indent In the VOF method, the volume fraction $\alpha$ is defined as:
\begin{equation}
\label{eq:10}
    \alpha
  =
    \frac{V^+}{V} \mcomma
\end{equation}
\noindent where $V^+$ is the volume occupied by water inside a control volume
$V$. The mass conservation equation for one phase (fluid $+$) reduces to the
well--known VOF advection equation:
\begin{equation}
\label{eq:11}
    \ddt{\alpha}
  + \div{(\U \alpha)}
  =
    0 \mcomma
\end{equation}
\noindent where solenoidal velocity field has been assumed, as given
by~\eqref{eq:1}.


\section{Numerical model}
\label{sec:numericalModel}

\begin{figure}[h]
\begin{center}
\includegraphics[width=4.5cm]{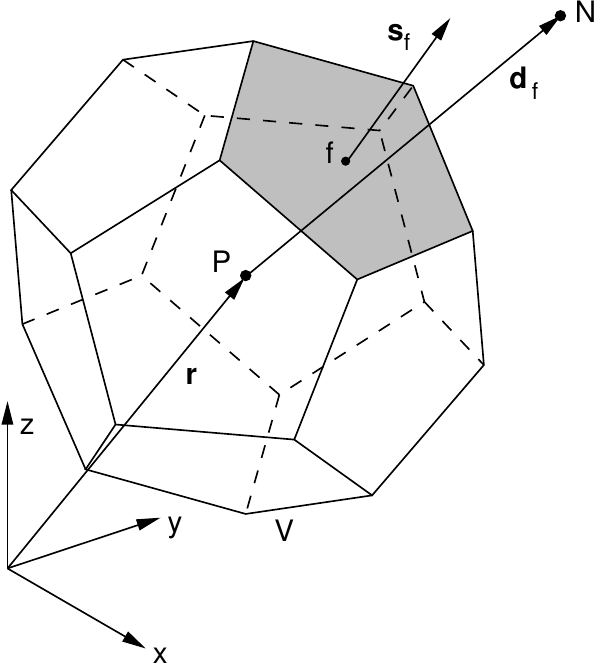}
\caption{Polyhedral control volume. Control volume $P$ shares a common face with
its immediate neighbour $N$.}
\label{fig:polyhedralCV}
\end{center}
\end{figure}

\begin{figure}[h]
\begin{center}

\begin{tikzpicture}[node distance = 2cm, auto,%
    scale=0.7, every node/.style={transform shape}]
    \node [cloud, text width=3.6cm] (startCoupling)
        {Start $\alpha - \U - p_d$ coupling (SIMPLE)};
    \node [myBlock, below of=startCoupling, node distance=1.5cm] (alphaEqn)
        {Solve $\alpha$, \eqref{eq:11}};

    \node [cloud, right of=alphaEqn, text width=3cm, node distance = 5cm]
        (startpUCoupling) {start $\U - p_d$ coupling (PISO)};
    \node [myBlock, below of=startpUCoupling, node distance = 2cm] (UEqn)
        {Solve $\U$, \eqref{eq:2}};
    \node [myBlock, below of=UEqn, node distance=1.5cm] (pdEqn)
        {Solve $p_d$, \eqref{eq:1}};

    \node [decision, below of=pdEqn, node distance=2.5cm] (endPISO)
        {PISO converged?};
    \node [decision, left of=endPISO, node distance=5cm] (endSIMPLE)
        {SIMPLE converged?};
    \node [none, right of=pdEqn, node distance=2.5cm, text width=0cm]
        (dummy1) {};

    \node [cloud, below of=endSIMPLE, text width=2cm, node distance = 3cm]
        (nextTime) {Advance time--step};
    \node [none, left of=UEqn, node distance=8cm, text width=2cm]
        (dummy2) {};

    \path [line] (startCoupling) -- (alphaEqn);
    \path [line] (alphaEqn) -- (startpUCoupling);
    \path [line] (startpUCoupling) -- (UEqn);
    \path [line] (UEqn) -- (pdEqn);
    \path [line] (pdEqn) -- (endPISO);
    \path [line, dashed] (endPISO) -- node [near start] {yes} (endSIMPLE);
    \path [line, dashed] (endPISO) -| node [near start] {no} (dummy1);
    \path [line, dashed] (dummy1) -- (pdEqn);
    \path [line, dashed] (endSIMPLE) -- node [near start] {yes} (nextTime);
    \path [line, dashed] (endSIMPLE) -| node [near start] {no} (dummy2);
    \path [line, dashed] (dummy2) |- (alphaEqn);

\end{tikzpicture}

\caption{Flow chart of the segregated solution algorithm.}
\label{fig:flowChart}
\end{center}
\end{figure}
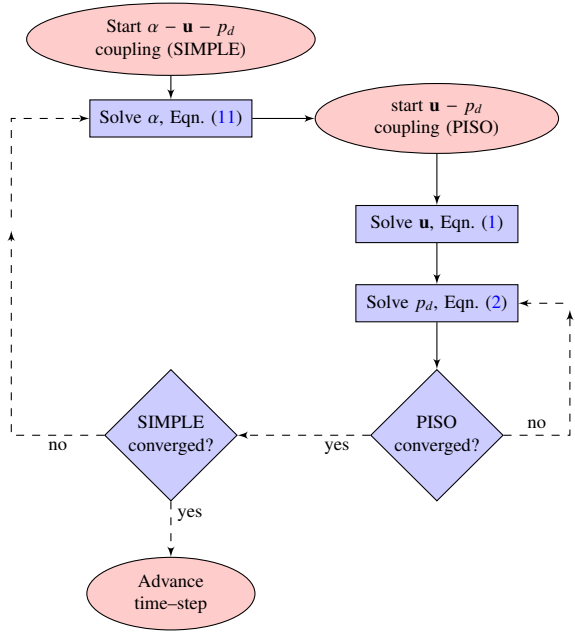

\indent Continuity and Navier--Stokes equations (\eqref{eq:1} and~\eqref{eq:2})
are discretised in space using a second--order accurate, collocated FV method
for general unstructured/structured grids~\cite{jasakPhD1996}. An arbitrary
polyhedral control volume (CV) has a number of neighbours, each defined with
surface area vector $\Sf$ and distance vector $\df$ from cell centre $P$ to
neighbouring cell centre $N$, as shown in~\autoref{fig:polyhedralCV}. The
governing equations for the flow field (\eqref{eq:1} and~\eqref{eq:2}) and the
free surface advection equation (\eqref{eq:11}) consitute a nonlinear system of
coupled partial differential equations. The pressure--velocity--free surface
coupling is achieved using a combination of SIMPLE~\cite{patankarSpalding1972}
and PISO~\cite{issa1986} algorithms, where a number of PISO correctors can be
used within each SIMPLE (nonlinear, or outer) correction step to ensure faster
convergence without relaxation factors, as discussed by
Vuk\v{c}evi\'{c}~\etal~\cite{vukcevicEtAl2017}. The continuity equation is used
to derive a dynamic pressure equation using the Rhie--Chow
interpolation~\cite{rhieChow1983} as a filter for spurious pressure
oscillations. Time derivative term in the Navier--Stokes
equations is discretised using a second--order accurate, two--time levels
backward Euler scheme (see \eg Queutey and Visonneau~\cite{queuteyVisonneau2007}
or Tukovi\'{c} and Jasak~\cite{tukovicJasak2012}). The convection term is
discretised using the Gauss theorem, where the face values are obtained with
second--order accurate linear upwind scheme~\cite{jasakPhD1996}. The diffusion
term is discretised using the Gauss theorem and central differencing, where the
non--orthogonal correction is treated using the over--relaxed approach in an
explicit manner~\cite{jasakPhD1996, demirdzic2015}. The pressure gradient term
in the Navier--Stokes equations and the pressure Laplacian resulting from the
continuity equation are discretised using the Gauss theorem, where
interface--corrected interpolation schemes take into account the jump conditions
at the free surface with the Ghost Fluid Method. The reader is referred to
Jasak~\cite{jasakPhD1996} and Ferziger and Peri\'{c}~\cite{ferzigerPeric1996}
for the details regarding polyhedral FV discretisation and the solution
algorithms. The combined solution algorithm is presented
in~\autoref{fig:flowChart}. The next section is devoted to
interface--corrected interpolation for dynamic pressure which has
discontinuities across the free surface given by~\eqref{eq:7}
and~\eqref{eq:8}.\\
\indent The introduction of waves in the computational domain and prevention of
their reflection on outgoing boundaries is achieved with relaxation zones as
introduced by Jacobsen~\etal~\cite{jacobsenEtAl2012}. In Jacobsen's approach,
a part of the computational domain is dedicated to relaxation zones where
the CFD solution is gradually forced towards a target solution with a smoothly
varying exponential blending function. In the present study, the target solution is
the potential flow solution for an incident wave. Using relaxation zones that
are long enough (usually one and a half of the dominant wave length,
~\cite{vukcevicEtAl2016b}) ensures negligible wave reflection. In the present work,
the volume fraction is blended explicitly with the target solution after the
transport due to geometrically reconstructed advection, while the momentum
equation is blended implicitly as described by
Jasak~\etal~\cite{jasakEtAlCFDWTOT2015}. Specific settings for each of the test
cases will be given, while the reader is referred to
Jacobsen~\etal~\cite{jacobsenEtAl2012}, Jasak~\etal~\cite{jasakEtAlCFDWTOT2015}
and Vuk\v{c}evi\'{c}~\etal~\cite{vukcevicPhD2016} for additional details.\\

\subsection{Interface--corrected dynamic pressure interpolation with the Ghost
Fluid Method}
\label{sec:gfmInterpolation}

\begin{figure}[!b]
\begin{center}
  \begin{subfigure}[b]{0.45\textwidth}
  \includegraphics[width=\textwidth]{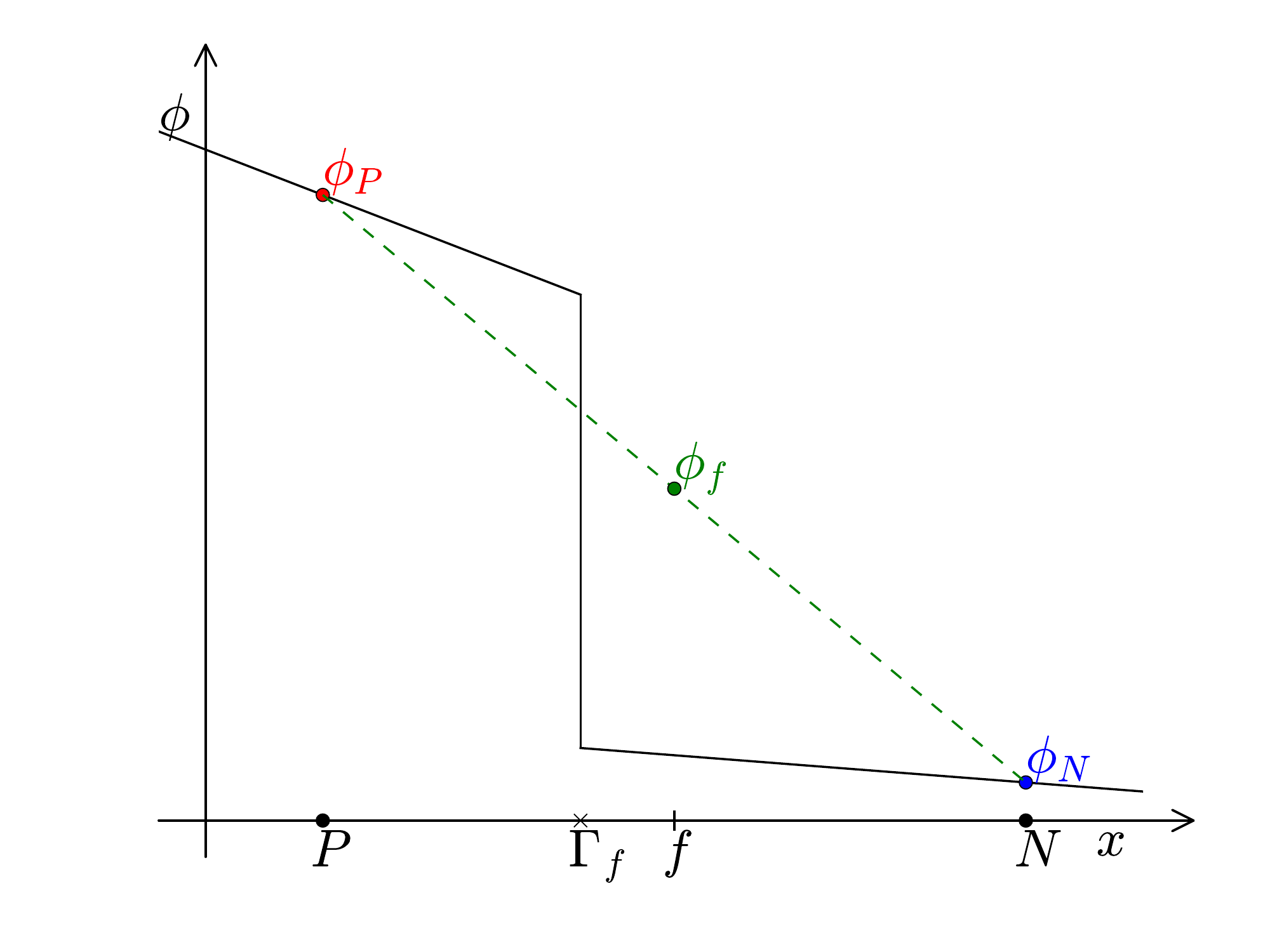}
  \caption{Ill--defined linear interpolation.}
  \label{fig:discontinuityTreatmentOrdinary}
  \end{subfigure}%
  \begin{subfigure}[b]{0.45\textwidth}
  \includegraphics[width=7cm]{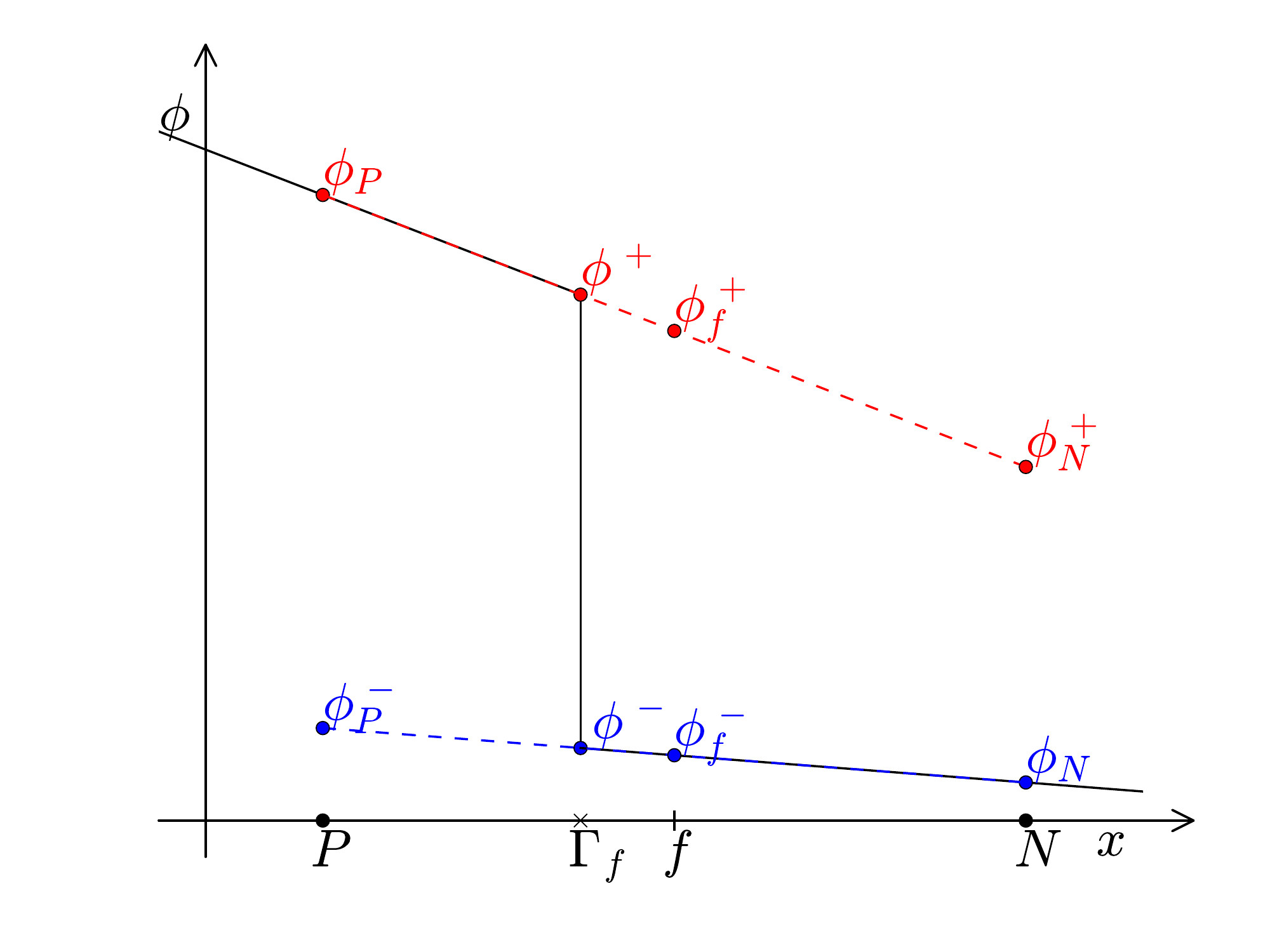}
  \caption{Ghost Fluid Method interpolation.}
  \label{fig:discontinuityTreatmentGFM}
  \end{subfigure}%
\end{center}
\caption{Cell centre to face centre interpolation schemes in the presence of a
discontinuity at the interface. $f$ is the face between cells $P$ and $N$ and
$\Gamma_f$ is the location of the interface (discontinuity).}
\label{fig:discontinuityTreatments}
\end{figure}

\indent One of the fundamental steps in the collocated FV method is the
interpolation of fields from cell centres to face centres. Linear
interpolation (or central differencing), based on Taylor series expansion
becomes erroneous in presence of a discontinuity, since the expansion assumes
sufficiently smooth spatial variation ($C^1$ continuity). This can be easily
demonstrated in a simplified, one--dimensional case presented
in~\autoref{fig:discontinuityTreatmentOrdinary}, where $P$ and $N$ denote cell
centres, $f$ is the face between them and $\Gamma_f$ represents the interface
where the discontinuity in $\phi$ and $\grad \phi$ is present. Simple linear
interpolation of cell centred values $\phi_P$ and $\phi_N$ to face centred value
$\phi_f$ yields an incorrect value since the discontinuity is not taken into
account. The idea behind the GFM is to use one--sided extrapolates to define
"ghost" values at the other side of the interface by second--order accurate
discretisation of interface jump conditions. Since two equations (\eqref{eq:7}
and~\eqref{eq:8}) for dynamic pressure discontinuities exist, one can introduce
two additional unknowns: $p_d^+$ and $p_d^-$, infinitesimally close to the
free surface from both sides. These values can be solved for and expressed in
terms of cell--centred values (\ie $p_d^+ = p_d^+(p_{dN}, p_{dP})$), providing
correct gradients. Using the correct gradients, the second--order extrapolation
from the interface towards the neighbouring cell centre is carried out. This
procedure is presented in~\autoref{fig:discontinuityTreatmentGFM} for a general
discontinuous variable $\phi$, defining one--sided extrapolates respecting the
jump conditions at the interface. It is important to stress that no assumption
has been made so far on the location of the interface.

\begin{figure}[!t]
\begin{center}
\includegraphics[clip, trim=2.5cm 2.5cm 1.5cm 2.5cm, width=8cm]%
    {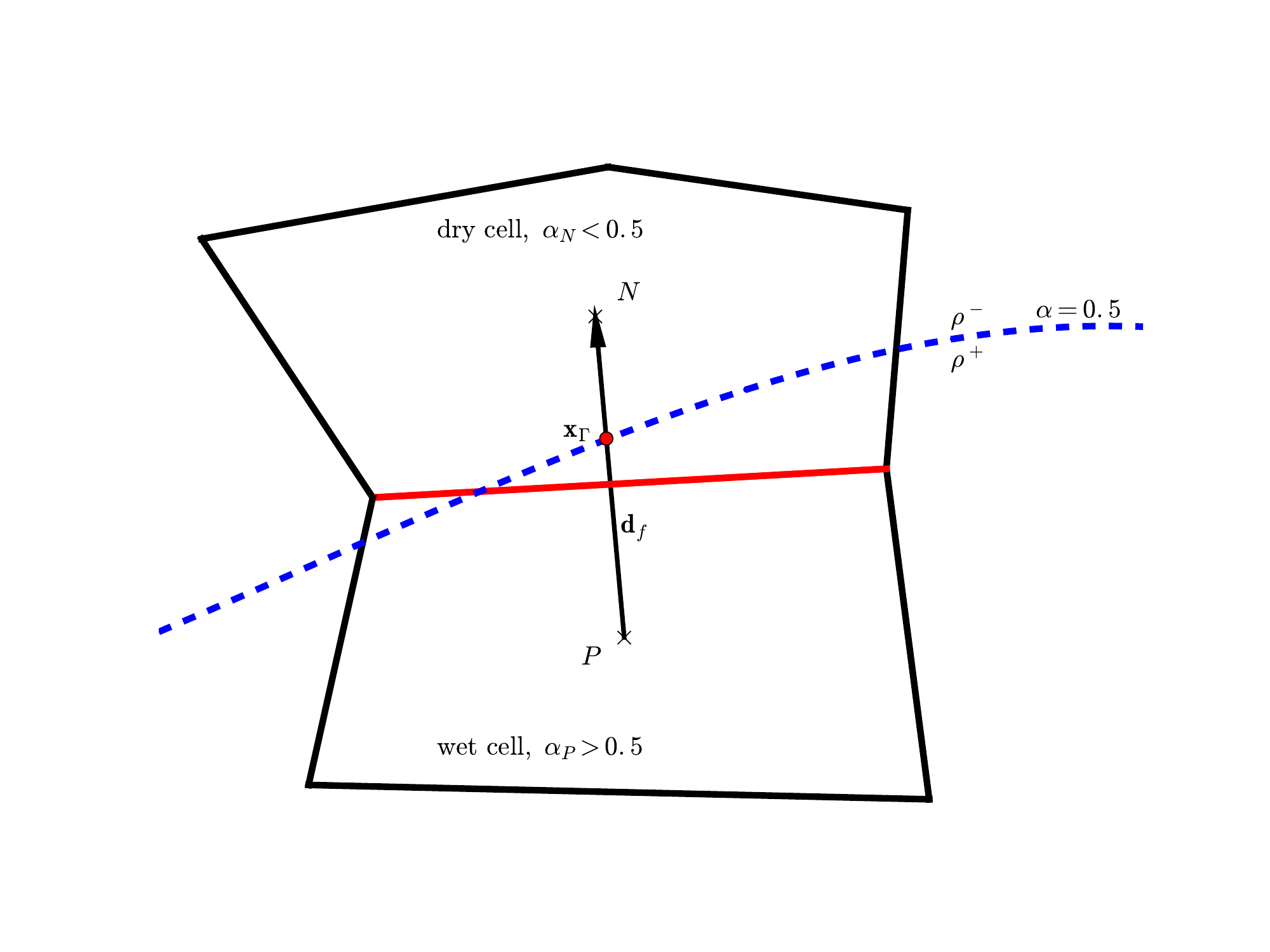}
\caption{Unstructured interface stencil in
two--dimensions~\cite{vukcevicEtAl2017}.}
\label{fig:interfaceStencil}
\end{center}
\end{figure}

\indent The GFM interpolation is only required in the presence of
discontinuities, while far from the free surface, ordinary interpolation is
sufficiently accurate. In the following analysis, we assume that the free
surface location can be readily estimated from the volume fraction field,
provided that it remains sharp (see~\autoref{sec:isoAdvector}).\\
\indent Consider a computational stencil on polygonal two--dimensional mesh for
clarity,~\autoref{fig:interfaceStencil}. The free surface, denoted by 
blue dashed line is defined with volume fraction contour $\alpha = 0.5$. Cells
with $\alpha
> 0.5$ are marked as "wet cells", while the cells with $\alpha < 0.5$ are marked
as "dry cells". $\vec{x_{\Gamma}}$ represents the location of the interface
somewhere along the distance vector $\df$ between adjacent cell centres. The
exact location of the interface between $P$ and $N$ can be defined as:
\begin{equation}
\label{eq:12}
    \vec{x}_{\Gamma} = \vec{x}_P + \lambda \df
    \mcomma
\end{equation}
\noindent where the parametrised distance to the interface $\lambda$ can be
readily estimated using the volume fraction~\cite{vukcevicEtAl2017}:
\begin{equation}
\label{eq:13}
    \lambda
  =
    \frac{\alpha_P - 0.5}{\alpha_P - \alpha_N}
    \mfstop
\end{equation}
\noindent Note that such procedure of defining the location of the interface
does not require reconstruction using the Level Set signed distance
field~\cite{huangEtAl2007}. Formally, the location estimate given
by~\eqref{eq:13} is of the same order of accuracy as the advection step.\\
\indent Before discretising the jump conditions at the interface, we introduce a
substitution for the inverse density following Huang~\etal~\cite{huangEtAl2007}:
\begin{equation}
\label{eq:14}
    \beta = \frac{1}{\rho} \mfstop
\end{equation}
\noindent The dynamic pressure jump conditions, written in terms of inverse
density $\beta$ reads:
\begin{equation}
\label{eq:15}
\begin{split}
    \left[ p_d \right]
  = 
    p_d^- - p_d^+
& =
    \left( \frac{1}{\beta^-} - \frac{1}{\beta^+} \right)
        \vec{g} \dprod \vec{x}_{\Gamma} \\
&  =
    \mathscr{H}
    \mcomma
\end{split}
\end{equation}
\noindent After the advection step, the location of the interface is
calculated for each pair of interface cells using~\eqref{eq:12}. The jump in
dynamic pressure can then be evaluated explicitly using~\eqref{eq:15}.\\
\indent Following Huang~\etal~\cite{huangEtAl2007} and previous work by
Vuk\v{c}evi\'{c}~\etal~\cite{vukcevicEtAl2017}, the dynamic pressure gradient
jump condition is discretised in a second--order accurate manner using
one--sided gradient evaluations based on parametrised distance to the
interface~\eqref{eq:13}:
\begin{equation}
\label{eq:16}
    \begin{split}
    [\beta \grad p_d]
& =
    \beta^- (\grad p_d)^-
  - \beta^+ (\grad p_d)^+ \\
& =
    \beta^- \frac{p_{dN} - p_d^-}{1 - \lambda}
  - \beta^+ \frac{p_d^+ - p_{dP}}{\lambda}
  =
    0
    \mfstop
    \end{split}
\end{equation}
\noindent The system of equations given by discretised jump
conditions (\eqref{eq:15} and~\eqref{eq:16}) can be easily solved for $p_d^+$
and $p_d^-$. Here, the complete procedure for extrapolation is presented for
cell $P$ using $p_d^+$, while the procedure for cell $N$ using $p_d^-$ is
analogous and can be easily inferred.
\begin{equation}
\label{eq:17}
    p_d^+
  =
    \frac{\lambda \beta^-}{\overline{\beta_w}} p_{dN}
  + \frac{(1 - \lambda) \beta^+}{\overline{\beta_w}} p_{dP}
  - \frac{\lambda \beta^-}{\overline{\beta_w}} \mathscr{H}
    \mcomma
\end{equation}
\noindent where $\overline{\beta_w}$ is the weighted inverse density:
\begin{equation}
\label{eq:18}
    \overline{\beta_w}
  =
    \lambda \beta^-
  + (1 - \lambda) \beta^+
    \mfstop
\end{equation}
\noindent It is interesting to note that $\overline{\beta_w}$ actually
represents the harmonic interpolation of density based on the actual distance to
the free surface.
\noindent Once the $p_d^+$ is known, the dynamic pressure field is extrapolated
from the heavier fluid (fluid "$^+$") at the location infinitesimally close to
the free surface, towards the neighbouring cell centre:
\begin{equation}
\label{eq:19}
    p_{dN}^+
  = 
    p_d^+
  + \frac{1 - \lambda}{\lambda} \left( p_d^+ - p_{dP} \right)
    \mfstop
\end{equation}
\noindent Substituting~\eqref{eq:17} into~\eqref{eq:19} yields the extrapolated
value at the neighbouring cell centre given in terms of two cell centred values,
inverse density and the explicit jump term $\mathscr{H}$:
\begin{equation}
\label{eq:20}
    p_{dN}^+
  =
    \frac{\beta^-}{\overline{\beta_w}} p_{dN}
  + \left( 1 - \frac{\beta^-}{\overline{\beta_w}} \right) p_{dP}
  - \frac{\beta^-}{\overline{\beta_w}} \mathscr{H}
    \mcomma
\end{equation}
\indent \eqref{eq:20} (and analogous expression for $p_{dP}^-$) are used
whenever the discretisation requires cell--centred values from the other side of
the interface. The jump conditions at the free surface are therefore taken into
account for the pressure gradient in the Navier--Stokes equations (\eqref{eq:2})
and the pressure Laplacian in the continuity equation (\eqref{eq:1}).  The
procedure has been derived using a compact computational stencil, respecting the
collocated FV framework with face--based connectivity~\cite{jasakPhD1996}.
This procedure results in a symmetric discretisation of the Laplacian operator,
thus preserving the symmetry of the underlying differential operator. This has
been discussed in detail by Vuk\v{c}evi\'{c}~\etal~\cite{vukcevicEtAl2017},
while in--depth derivation of the procedure is presented by
Vuk\v{c}evi\'{c}~\cite{vukcevicPhD2016}, and shall not be repeated here.\\
\indent The proposed method belongs to a family of balanced force
methods (see \eg~\cite{Denner2014}), where the coupling between density field
and dynamic pressure is resolved within the pressure equation instead of the
momentum equation. The procedure assumes a--priori known location of the
interface defined by the volume fraction field $\alpha$, making it suitable for
segregated (or partitioned) solution algorithms as the one used in this work
(see \autoref{fig:flowChart}).  The method presented so far is at most
second--order accurate, although the accuracy will directly depend on the
accuracy of the advection step,~\eqref{eq:13}. The next section is devoted to
second--order accurate advection of the free surface.


\subsection{Interface advection with the isoAdvector scheme}
\label{sec:isoAdvector}

\noindent The implicit representation of a fluid interface via volume fractions
is the natural one in the FV framework. The task of advancing the interface in
time becomes a matter of modelling the composition (heavy and light fluid) of
the total volume of fluid passing from one cell into its neighbour during a
time step. Typically, the available information
consists of volume fractions in cells at the beginning of the time step,
$\alpha_P$, and the velocity field represented in two ways, namely by the cell
averaged velocity, $\U_P$, and by the volumetric face flux, $\phi_f$. These
velocity field representations are available at the beginning of the time step
(see \autoref{fig:flowChart}), and since the nonlinear (outer) iterations are
performed, we may also have estimates for them at the end of the time
step. The challenge of advecting the fluid interface becomes a question of using
$\alpha_P(t)$, $\U_P(t)$ and $\phi_f(t)$, and possibly available estimates of
$\U_P(t+\Delta t)$ and $\phi_f(t+\Delta t)$, to predict $\alpha_P(t + \Delta
t)$. In the following, we will describe how this task is performed using the
\texttt{isoAdvector} algorithm by Roenby~\etal~\cite{roenbyEtAl2016}.

The starting point of \texttt{isoAdvector} is the continuity equation for the
density field integrated over the volume of an interface cell:
\begin{equation}
    \label{eq:dV/dt}
	\frac{\d \ }{\d t}\int_{V}\rho(\vec x,t)\d V + \sum_f \int_{S_f} \rho(\vec x,t)\vec u(\vec x,t) \cdot \d \vec S = 0.
\end{equation}
Here $V$ is the cell volume, $S_f$ is the surface of one of the faces comprising
the cell boundary and the sum $\sum_f$ is over all the cell's faces.
Without loss of generality, define a normalised and shifted density field, or
indicator function as:
\begin{equation}\label{eq:Hdef}
	H(\vec x,t) = \frac{\rho(\vec x,t)-\rho^-}{\rho^+ - \rho^-},
\end{equation}
where $\rho^-$ and $\rho^+$ are considered constant.
Isolating $\rho(\vec{x}, t)$ in \eqref{eq:Hdef} and inserting it into
\eqref{eq:dV/dt}, after some rearrangement it follows:
\begin{equation}\label{eq:dintH/dt}
	\frac{\d \ }{\d t}\int_{V}H(\vec x,t)\d V + \sum_f \int_{S_f} H(\vec x,t)\vec u(\vec x,t) \cdot \d \vec S = - \frac{\rho^-}{\rho^+ - \rho^-}\sum_f \int_{S_f} \vec u(\vec x,t) \cdot \d \vec S.
\end{equation}
So far no assumption of incompressibility has been made. Assuming two
constants $\rho^+$ and $\rho^-$ are indeed the densities
of the heavy and light fluid, respectively, then both fluids are incompressible,
causing the right hand side in \eqref{eq:dintH/dt} to vanish. The indicator
function, $H(\vec{x},t)$, becomes a 3--dimensional Heaviside function taking the
values 0 and 1 in the region of space occupied by the light and the heavy fluid,
respectively. With the definitions of the volume fraction of cell $P$:
\begin{equation}
\alpha_P = \frac{1}{V_P}\int_{V_P}H(\vec x,t)\d V,
\end{equation}
\eqref{eq:dintH/dt} can be written as:
\begin{equation}\label{eq:dadt}
\frac{d\alpha_P}{dt} + \frac1{V_P}\sum_f \int_{S_f} H(\vec x,t)\vec u(\vec x,t) \cdot \d \vec S = 0.
\end{equation}
This equation is exact for incompressible fluids. The key to accurate interface
advection is to realise that the discontinuous nature of the problem demands
geometric modelling involving considerations of the shape and orientation of the
face, as well as of the local position, orientation and motion of the
interface. We formally integrate \eqref{eq:dadt} over time from time $t$
to time $t + \Delta t$:
\begin{equation}
\alpha_P(t+\Delta t) = \alpha_P(t) - \frac1{V_P}\sum_f \Delta V_f(t,\Delta t)
\end{equation}
where $\Delta V_f(t,\Delta t)$ denotes the volume of heavy fluid transported
through the face $f$ during the time step $[t,t+\Delta t]$:
\begin{equation}\label{eq:dVdef}
\Delta V_f(t,\Delta t) = \int_t^{t + \Delta t}\int_f H(\vec x,\tau)\vec u(\vec x,\tau) \cdot \d \vec S d\tau.
\end{equation}
If the flow was steady and face $f$ completely immersed in the heavy fluid
during the entire time step, this will just be $\Delta V_f(t,\Delta t) = \phi_f
\Delta t$. Likewise, if the face was in the light fluid throughout the time
step, $\Delta V_f(t,\Delta t)$ would be zero. But even for steady flows, some
faces will in general be fully or partially swept by the interface in a
non-trivial manner during a time step. In the \texttt{isoAdvector} advection
step we model the face-interface intersection line sweeping the face during the
time step. This approach is geometric in nature, but novel compared to existing
geometric advection methods that focus on calculation of flux polyhedra and
their intersection with the mesh cells~\cite{Xie2014, Lopez2008, Hernandez2008,
Ahn2007}.

The first step in our modelling process is to realise that the rapid changes in
$\Delta V_f$ during a time step is typically not due to an abruptly varying
velocity field but due to the passage of the interface through the cell face.
Hence, we will assume that $\vec u(\vec x,\tau)\cdot \d \vec S$ in \eqref{eq:dVdef}
can be written in terms of an averaged flux over the face and over the time
step:
\begin{equation}\label{eq:constUAssump}
	\vec u(\vec x,\tau)\cdot d\vec S \approx \overline{\vec u}_f \cdot \vec n_f \d A = \frac{\overline{\phi}_f}{A_f}\d A, \text{ for } \vec x \in S_f \text{ and } t\in [t,t+\Delta t].
\end{equation}
Here $\overline{\vec u}_f$ and $\overline{\phi}_f$ can be thought of as
averages over both time step and face area. At the beginning of the
algorithm, stepping forward from time $t$, we may use the available $\phi_f(t)$
as the estimate of the average flux over the time step, $\overline{\phi}_f$.
However, during nonlinear iterations in a single time step, the averaged flux is
readily available due to the availability of $\phi_f(t + \Delta t)$. In any case,
inserting \eqref{eq:constUAssump} into \eqref{eq:dVdef} we can write:
\begin{equation}\label{eq:dV}
\Delta V_f(t,\Delta t) \approx \overline{\phi}_f \int_t^{t + \Delta t} \alpha_f^+(\tau) \d\tau,
\end{equation}
where we have defined the quantity:
\begin{equation}\label{eq:alphaf}
\alpha_f^+(t) = \frac{1}{A_f}\int_f H(\vec x,t) \d A,
\end{equation}
which is the instantaneous ``Area-Of-Fluid'' of face $f$, i.e. the fraction of
the face area submerged in heavy fluid. If the velocity field is constant in
space and time and the face is planar, the approximation in \eqref{eq:dV}
becomes exact.

\begin{figure}[h]
\begin{center}
  \begin{subfigure}[b]{0.4\textwidth}
  \includegraphics[width=0.9\textwidth]{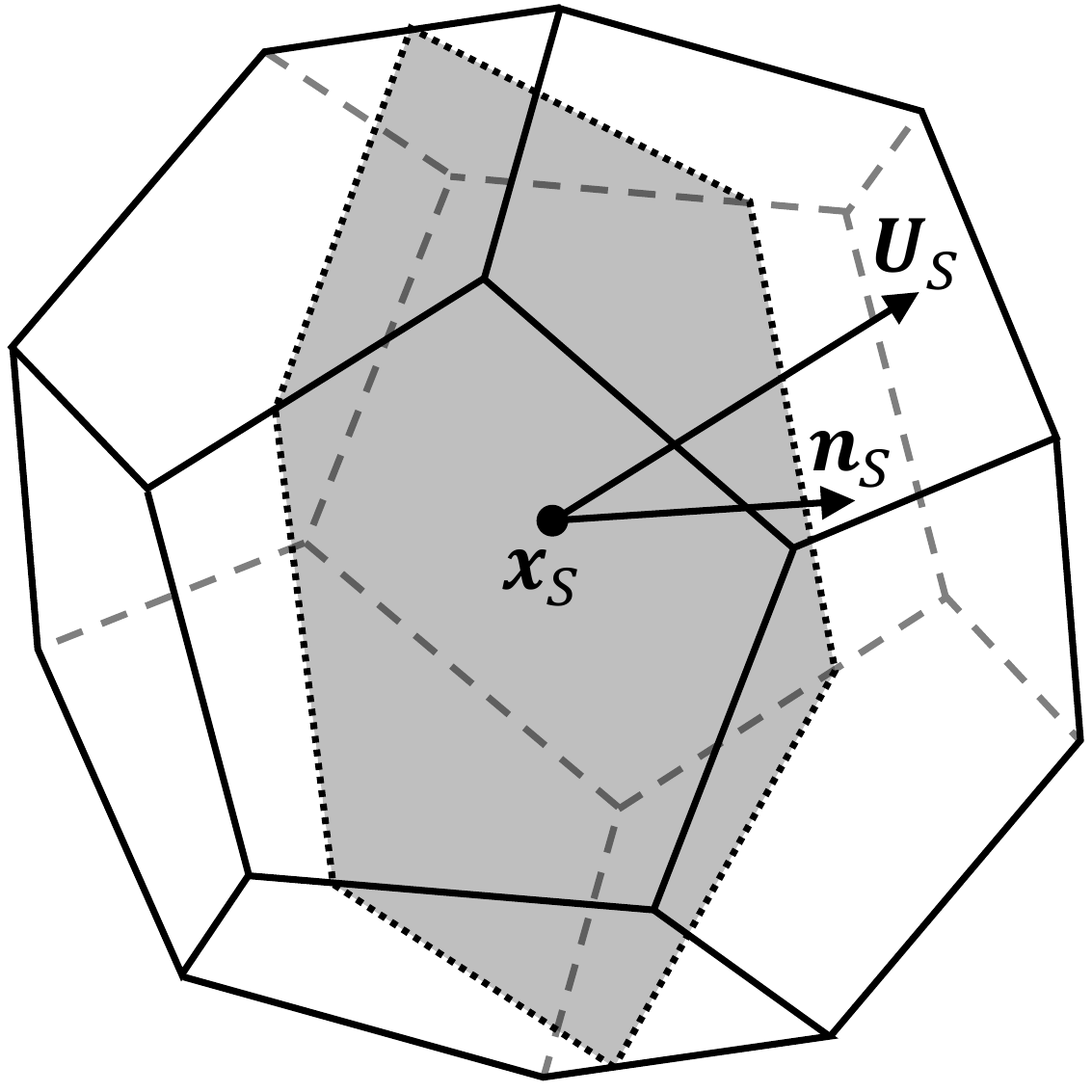}
  \caption{Reconstructed "isoface" in interface cell.}
  \label{fig:isoface}
  \end{subfigure}%
  \begin{subfigure}[b]{0.4\textwidth}
  \includegraphics[width=0.9\textwidth]{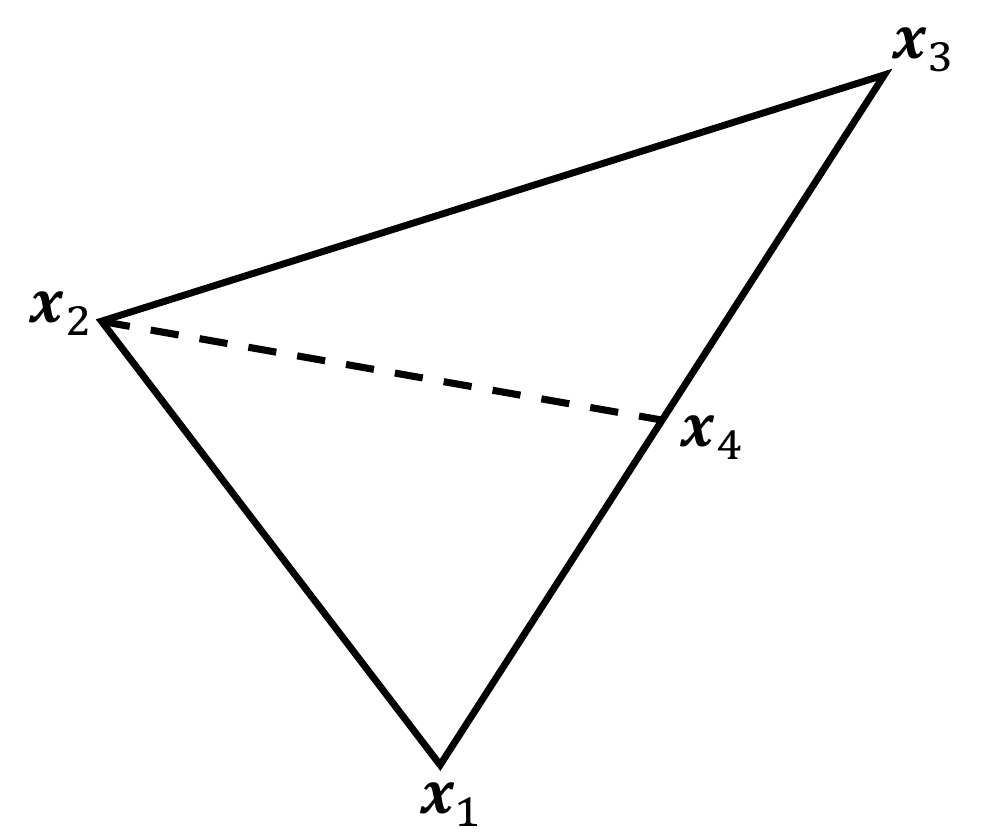}
  \caption{Triangular subface cut by planar isoface at face--interface
  intersection line.}
  \label{fig:triangle}
  \end{subfigure}%
\end{center}
\caption{Reconstructed isoface within a polyhedral cell.}
\label{fig:isofaceIsoAdvector}
\end{figure}

To progress, we now assume that the interface has been reconstructed within the
interface cell from which face $f$ receives fluid (upwind cell). The
reconstructed interface is represented by an internal polygonal face.
We will call such a cell cutting face an \emph{isoface}, for reasons to become
clear below. The isoface cuts the cell into two disjoint subc--ells occupied by
the heavy and light fluid, respectively, as illustrated in \autoref{fig:isoface}.

The isoface will intersect some cell faces, cutting them into two subfaces
immersed in heavy and light fluid, respectively, while others will be fully
immersed in one of the two fluids. This is the state at time $t$. However,
\eqref{eq:dV} requires $\alpha_f^+$ for the whole interval $[t,t+\Delta t]$. To
obtain an estimate of this, we first note that the isoface will have a
well-defined face centre, $\vec x_S$ and a well defined unit normal, $\vec n_S$,
the latter by convention pointing away from the heavy fluid. We may then
interpolate the cell averaged velocity field, $\vec u_P$ to the isoface
centre, $\vec x_S$, to obtain the isoface velocity $\U_S$. If the fluid
interface is a plane with unit normal $\vec n_S$ starting at $\vec x_S$ at time
$t$ and moving with constant velocity $\U_S$, then the interface will arrive
at a given point $\vec x_v$ at time:
\begin{equation}\label{eq:tv}
	t_v = t + \frac{(\vec x_v - \vec x_S)\cdot \vec n_S}{\vec \U_S\cdot \vec n_S}.
\end{equation}
In particular, this holds true for all points on the general polygonal
(N--sided) face $f$, including its vertices $\vec x_1,...,\vec x_N$, and
therefore defines the face-interface intersection line at any
$\tau\in[t,t+\Delta t]$ as required in \eqref{eq:dV}. We will now use this to
explicitly calculate the time integral in \eqref{eq:dV}. 

\begin{figure}[h]
\label{fig:bla}
\begin{center}
  \begin{subfigure}[b]{0.45\textwidth}
  \includegraphics[width=\textwidth]{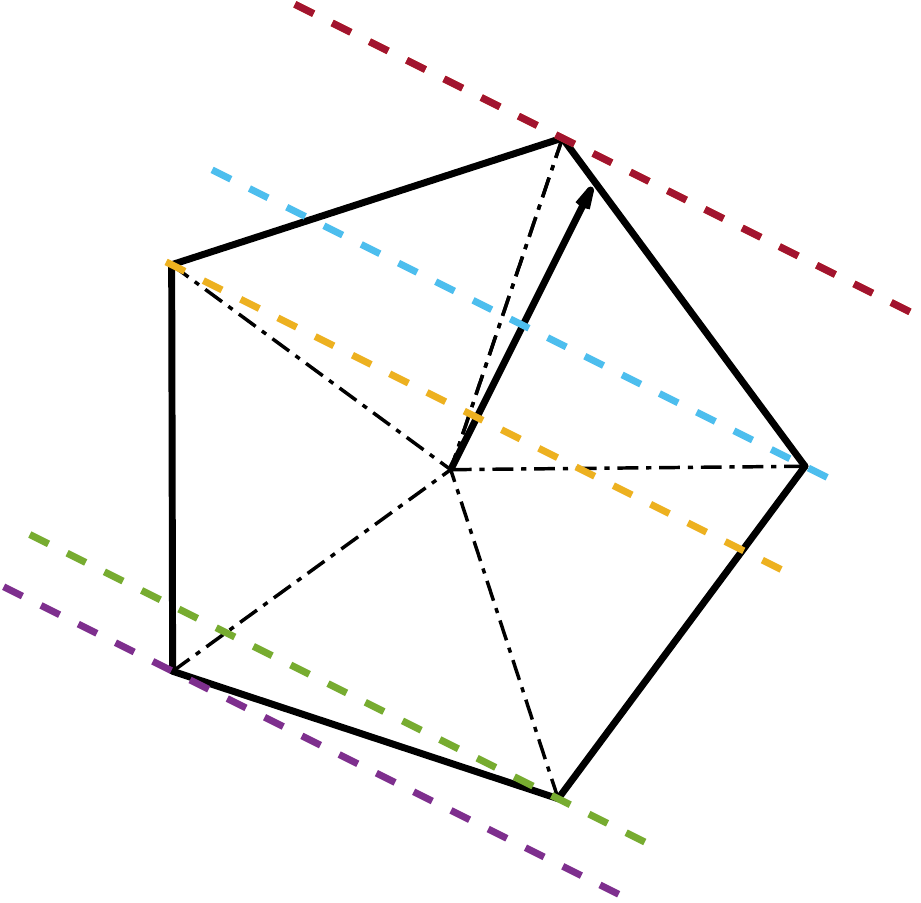}
  \caption{Face--interface intersection line sweeping a polygonal face and passing by its vertices.}
  \label{fig:FIIL}
  \end{subfigure}%
  \begin{subfigure}[b]{0.45\textwidth}
  \includegraphics[width=\textwidth]{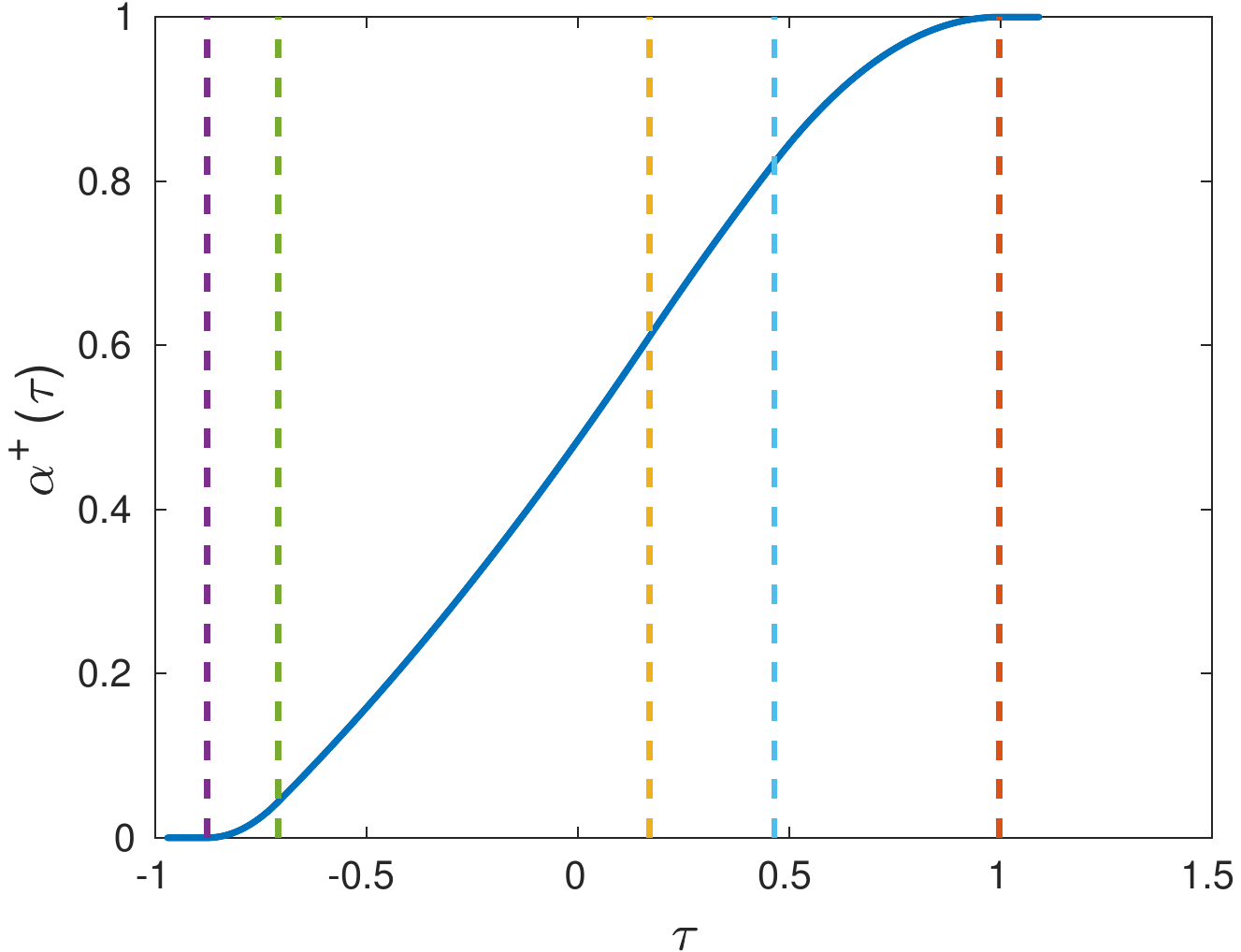}
  \caption{The evolution of the area--of--fluid as the face is swept. Quadratic dependency on $\tau$ with different coefficients on each subinterval.}
  \label{fig:alphaplus}
  \end{subfigure}%
\end{center}
\caption{\texttt{isoAdvector} algorithm sweeping a polygonal face.}
\end{figure}

First note that a planar polygonal face may be triangulated in a number of ways,
with the triangles lying exactly on the surface of the face. For a non-planar
polygonal face we must define its surface, which we do by estimating a face
centre and using that as the apex for N triangles with the N face edges as base
lines. The face surface is then defined by the union of these N triangles.
In other words, any polygonal face may be represented as a union of triangles.
Our analysis can therefore be confined to a triangular subface since the
contribution from these can subsequently be accumulated to obtain the time
integral in \eqref{eq:dV} for the whole face. Therefore, we consider a triangle
with vertices $\vec x_1, \vec x_2$ and $\vec x_3$. The interface arrival times
from \eqref{eq:tv} can be calculated and we may assume without loss of
generality that the points are ordered such that $t_1 \leq t_2 \leq t_3$. The
interface enters the triangle at time $t_1$ at the point $\vec x_1$, and then
sweeps the triangle reaching $\vec x_2$ at time $t_2$, where it also intersects
the edge $\vec x_1 - \vec x_3$ at a point we shall call $\vec
x_4$,illustrated in~\autoref{fig:triangle}. In what follows, we denote an edge
between $\vec x_i$ and $\vec x_j$ as $\vec x_{ij} = \vec x_i - \vec x_j$. Then
for $\vec x_4$ we have:
\begin{equation}\label{eq:x_41}
	\vec x_{41} = \frac{\vec x_{21}\cdot\vec n_S}{\vec x_{31}\cdot\vec n_S}\vec x_{31}.
\end{equation}
Finally, at time $t_3$, the interface leaves the face through $\vec x_3$. We
note that in general the three times $t_1$, $t_2$ and $t_3$ and the two times
$t$ and $t + \Delta t$ can be distributed in various ways. For instance if $t <
t_1 < t_2 < t + \Delta t < t_3$, then the triangle is completely immersed in the
light fluid from time $t$ to time $t_1$ at which point the isoface will enter
the triangle sweeping it and ending up at time $t + \Delta t$ on the triangle.
The correct ordering must be taken into account, when doing the time
integration in \eqref{eq:dV}. Let us for the sake of simplicity consider the
case where the triangle is entirely swept during the time step, i.e. where $t <
t_1$ and $t_3 < t + \Delta t$. We will derive an expression for $\alpha^+(\tau)$
under the assumption that $\vec U_s\cdot \vec n_S > 0$, meaning that the
interface is moving towards the light fluid region within the cell. If this is
not the case, what we have derived is instead an expression for $1-\alpha^+$,
which is equally useful. At a time $\tau$ between $t_1$ and $t_2$, the immersed
part of the triangle will have area:
\begin{equation}
	A^+(\tau) = \frac{1}{2}|\vec x_{41}\tilde t \times \vec x_{21}\tilde t| \quad \textrm{ where } \quad \tilde t = \frac{\tau-t_1}{t_2 - t_1}.
\end{equation}
With a total area of the triangle of $A = \frac12 |\vec x_{31}\times\vec
x_{21}|$, we can then write:
\begin{equation}\label{eq:aplus1}
	\alpha^+(\tau) = \frac{|\vec x_{41}\times \vec x_{21}|}{2A}\left(\frac{\tau -t_1}{t_2 - t_1}\right)^2 \quad \textrm{ for } \quad t_1 < \tau < t_2.
\end{equation}
In a similar manner we find:
\begin{equation}\label{eq:aplus2}
	\alpha^+(\tau) = \alpha^+(t_2) + \frac{|\vec x_{43}\times \vec x_{23}|}{2A}\left[1-\left(1 - \frac{\tau-t_3}{t_2 - t_3}\right)^2\right] \quad \textrm{ for } \quad t_2 < \tau < t_3.
\end{equation}
From \eqref{eq:aplus1} and \eqref{eq:aplus2} it is evident that $\alpha^+$ for
the sub-triangles of a polygonal face are quadratic polynomials in $\tau$ with
coefficients changing at the intermediate time $t_2$. The coefficients are
uniquely determined by the face vertex positions, $\vec x_1, \vec x_2$ and $\vec
x_3$, the isoface velocity, $\U_s$, the unit normal, $\vec n_S$, and the
isoface centre at the beginning of the time step, $\vec x_S$. In
\autoref{fig:FIIL} and \autoref{fig:alphaplus}, we show an example of the time evolution of $\alpha_f^+(t)$ for a polygonal face as it is swept by a planar interface. 

If we call the polynomial coefficients for the first sub time interval of an
polygon's $i$'th triangle $A_{i,1}, B_{i,1}$ and $C_{i,1}$ (see
\eqref{eq:aplus1}), and the coefficients for its second sub interval $A_{i,2},
B_{i,2}$ and $C_{i,2}$ (see \eqref{eq:aplus2}), then the time integral in
\eqref{eq:dV} takes the form:
\begin{equation}\label{eq:intaplus}
\int_t^{t + \Delta t} \alpha_f^+(\tau) \d\tau \approx \sum_{i = 1}^N \sum_{j = 1}^2 \frac13 A_{i,j}(t_{i,j+1}^3-t_{i,j}^3) + \frac12 B_{i,j}(t_{i,j+1}^2 - t_{i,j}^2) + C_{i,j}(t_{i,j+1} - t_{i,j}).
\end{equation}
Here $t_{i,1}, t_{i,2}$ and $t_{i,3}$ are the arrival times for the $i$'th
triangle of our polygonal face (see \eqref{eq:tv}). The approximation in
\eqref{eq:intaplus} is exact if the interface is in fact a plane with normal
$\vec n_S$ starting at position $\vec x_S$ at time $t$ and travelling with
constant velocity $\U_S\cdot \vec n_S$ normal to itself.

This concludes our description of the \texttt{isoAdvector} advection step. We
will now briefly describe the \texttt{isoAdvector} reconstruction step giving
rise to the first syllable, the "iso", in the method name. The reconstruction
step is used to obtain the isoface at the beginning of a time step including
its centre $\vec x_S$ and unit normal, $\vec n_S$. As suggested by the name,
this is done by representing the isoface as the intersection between the cell
and a numerically calculated isosurface of the volume fraction field,
$\alpha_P(t)$.  To calculate such isosurface, the volume fraction field is first
interpolated from the cell centres to the vertices of the cell. In the current
implementation the inverse centre-to-vertex distances are used as interpolation
weights. With a volume fraction value associated with each cell vertex, we can
now for a given iso--value, $\alpha_0$, determine for each cell edge, if
$\alpha_0$ lies between the two vertex values of that edge. If this is the case,
we mark a cut point on the edge by linear interpolation. Doing this for all the
cell's edges and connecting the cut points across the cell faces, we obtain the
isoface. Its centre and normal can be calculated by triangulation as for any
other polygonal face.\\
\indent It is important to choose for
each interface cell a distinct iso--value giving rise to an isoface cutting
the cell into sub--cells of volumetric proportions in accordance with the
volume fraction of the cell. The search algorithm for finding the iso--value to
within a user specified tolerance has been optimized by exploiting the known
functional form of a subcell volume as a function of the iso--value. For more
details, the reader is referred to Roenby~\etal~\cite{roenbyEtAl2016}.

The final element in the \texttt{isoAdvector} algorithm is a heuristic bounding
step. It is introduced to correct volume fractions ending up outside the
meaningful interval, $[0,1]$, if the \texttt{isoAdvector} algorithm is stressed
beyond its formal region of validity by taking time steps so large that the
underlying geometric assumptions break down. The bounding step is optional and
contains both a volume--preserving step and an optional non-conservative brute
force chopping of the volume fractions. For more details, the reader is referred
to Roenby~\etal~\cite{roenbyEtAl2016}.

The advection step presented in this section is formally second order accurate
since the time--averaged velocity and volumetric flux fields, $\overline{\U}_f$
and $\overline{\phi}_f$ field, respectively, are used. This corresponds to a
well--known Crank--Nicolson scheme.


\section{Test cases}
\label{sec:testCases}

\noindent This section presents verification and validation of the proposed
numerical approach. The analysis starts by considering a
progressive wave using a set of uniformly refined grids to calculate numerical
uncertainty and achieved order of convergence. In addition, the results are
compared with stream function wave theory based on potential flow
solution~\cite{rieneckerFenton1981}. Finally, we consider wave impact loads on a
deck of a simplified ship in regular waves, comparing the results with recent
experimental data, while also assessing numerical uncertainty for the pressure
impulses. This wave breaking phenomena is often denoted as the "green--water
effect".


\subsection{Wave propagation}
\label{sec:wavePropagation}

\noindent A progressive wave with parameters given
in~\autoref{tab:waveParameters} is considered for verification purposes. The
wave is moderately nonlinear~\cite{deanDalrymple2010} with steepness $kH/2 =
0.174$. The computational domain is thirteen wave lengths long ($L_x = 13\lambda$)
and two water depths high ($L_y = 2d$). Relaxation zones that are $1.5\lambda$
long are used at the left and right boundaries to introduce waves and prevent
wave reflection, while the wave propagates from left to right boundary. Such a
long domain allows us to have ten wave lengths of full CFD solution unaffected
by relaxation zones, which gives us the possibility to investigate the
dissipation (loss of wave amplitude) and dispersion errors (phase drift) along
the numerical wave tank.\\

\begin{table}[!b]
\caption{Progressive wave parameters.}
\begin{center}
\begin{tabular}{*{3}{r}}
\hline
    Wave height& $H$, m & 0.3 \\
    \hline
    Wave period & $T$, s & 2 \\
    \hline
    Wave radian frequency & $\omega$, rad/s & $\pi$ \\
    \hline
    Wave length & $\lambda$, m & 5.409 \\
    \hline
    Water depth & $d$, m & 1 \\
    \hline
    Wave steepness & $kH/2$ & 0.174 \\
\hline
\end{tabular}
\end{center}
\label{tab:waveParameters}
\end{table}

\indent Six structured, uniform grids outlined in~\autoref{tab:wavePropagationGrids}
are used to calculate the achieved order of accuracy and numerical uncertainty
using the least squares method according to E\c{c}a and
Hoekstra~\cite{ecaHoekstra2014, uncertaintyReFrescoWebsite}. Grids 1 and 2 are
obtained by coarsening Grid 3, while Grids 4, 5 and 6 are obtained by
refining Grid 3. A constant refinement factor of $r = 1.4$ has been used for
both spatial dimensions and for time. Therefore, the smallest computation is
performed on Grid 1 with 11\,865 cells (approximately 26 cells per wave length
and 5 cells per wave height) using 102 time steps per encounter period, while
the largest computation is performed on Grid 6 with 338\,520 cells
(approximately 140 cells per wave length and 28 cells per wave height) using
549 time steps per encounter period.\\

\begin{table}[t!]
\caption{Grids for the wave propagation case.}
\begin{center}
\begin{tabular}{*{7}{r}}
\hline
    Grid index & 1 & 2 & 3 & 4 & 5 & 6  \\
    \hline
    Number of cells & 11\,865 & 23\,226 & 45\,084 & 88\,160 & 172\,900 &
    338\,520 \\
    \hline
    Number of time steps per wave period & 102 & 143 & 200 & 280 & 392 & 549 \\
    \hline
    Number of cells per wave length & 26 & 37 & 51 & 71 & 100 & 140 \\
    \hline
    Number of cells per wave height & 5.2 & 7.3 & 10.2 & 14.3 & 20.0 & 28.0 \\
\hline
\end{tabular}
\end{center}
\label{tab:wavePropagationGrids}
\end{table}

\begin{figure}[b!]
\begin{center}
  \begin{subfigure}[b]{0.45\textwidth}
  \includegraphics[width=\textwidth]{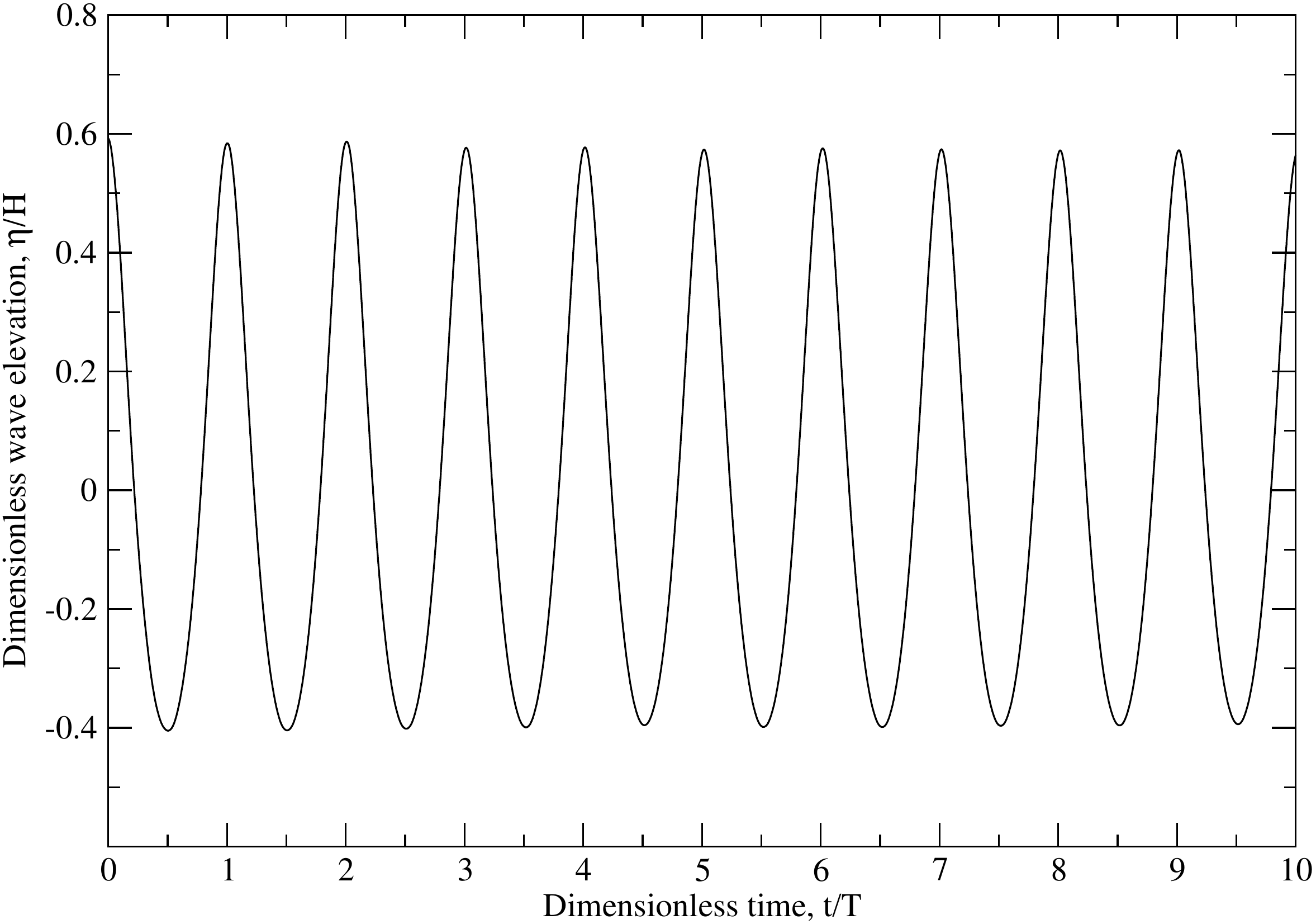}
  \caption{Wave elevation signal in time.}
  \label{fig:waveGauge6Signal}
  \end{subfigure}%
  \begin{subfigure}[b]{0.45\textwidth}
  \includegraphics[width=7cm]{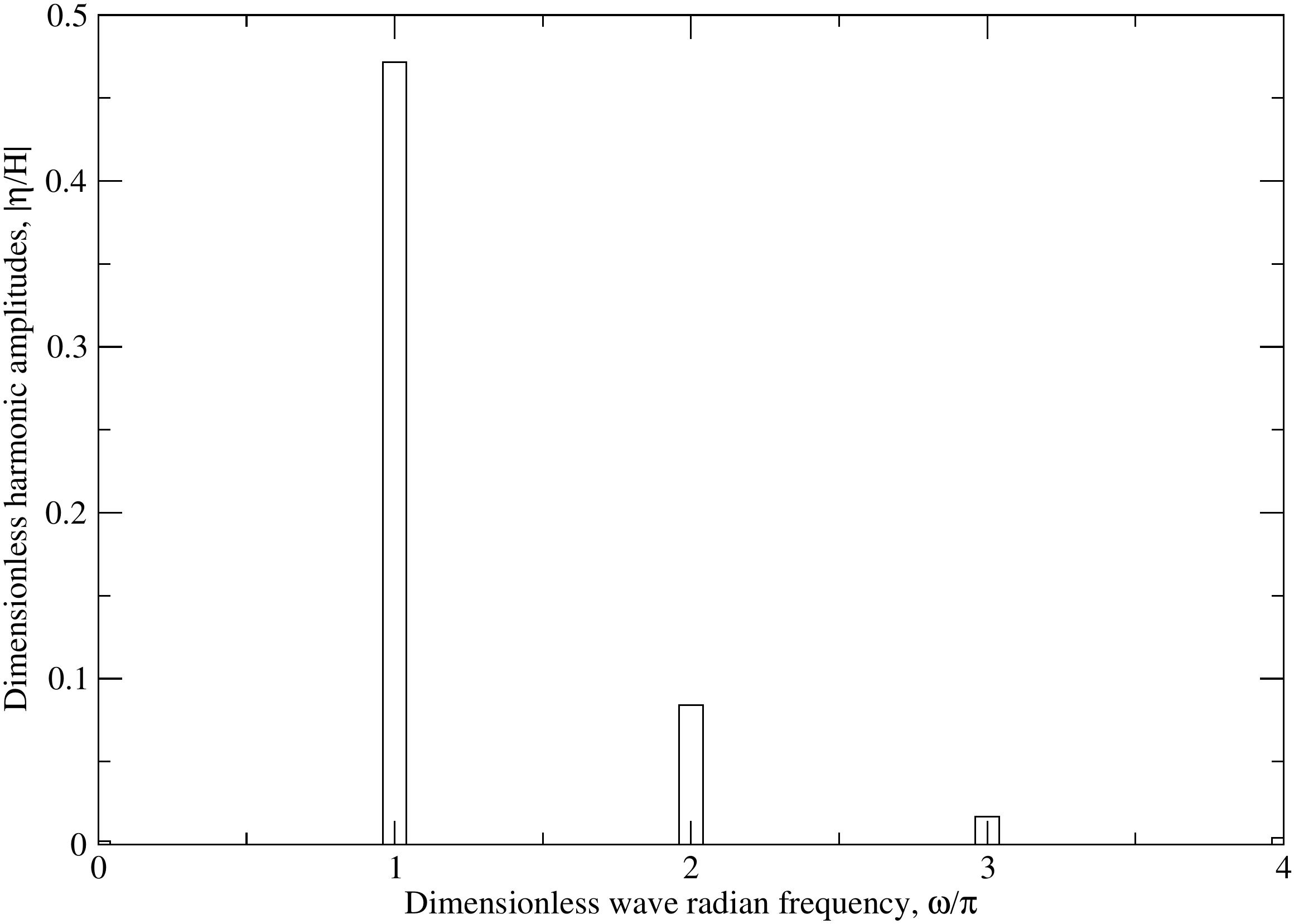}
  \caption{Wave elevation spectrum in frequency domain.}
  \label{fig:waveGauge6Spectrum}
  \end{subfigure}%
\end{center}
\caption{Wave elevation at wave gauge $6\lambda$ obtained with Grid 6.}
\label{fig:waveGauge6}
\end{figure}

\begin{figure}[!t]
\begin{center}
    \includegraphics[width=\textwidth]{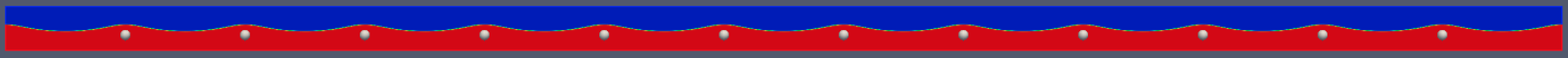}
    \caption{Location of the velocity probes and wave gauges.}
    \label{fig:probesAndGauges}
\end{center}
\end{figure}

\indent Wave elevation is measured at twelve locations, $1\lambda, 2\lambda$
\ldots $12\lambda$ from the inlet. A representative signal obtained using Grid 6
at wave gauge $6\lambda$ is shown in~\autoref{fig:waveGauge6Signal}, with its
frequency content presented in~\autoref{fig:waveGauge6Spectrum}. The
frequency content indicates that most of the energy, $(\eta/H)^2$ is contained
within the first harmonic, while approximately 10\% is contained within higher
harmonics. Velocity is also measured at twelve locations corresponding to
$1\lambda, 2\lambda \ldots 12\lambda$ below the calm free surface at $z = -0.3$
metres. The locations of the velocity probes are presented
in~\autoref{fig:probesAndGauges}. The initial condition for the simulation
corresponds to the solution from the stream function wave
theory~\cite{rieneckerFenton1981}.\\

\subsubsection{Verification analysis for the progressive wave}
\label{sec:verification}

\begin{figure}[!b]
\begin{center}
  \begin{subfigure}[b]{0.45\textwidth}
  \includegraphics[width=\textwidth]{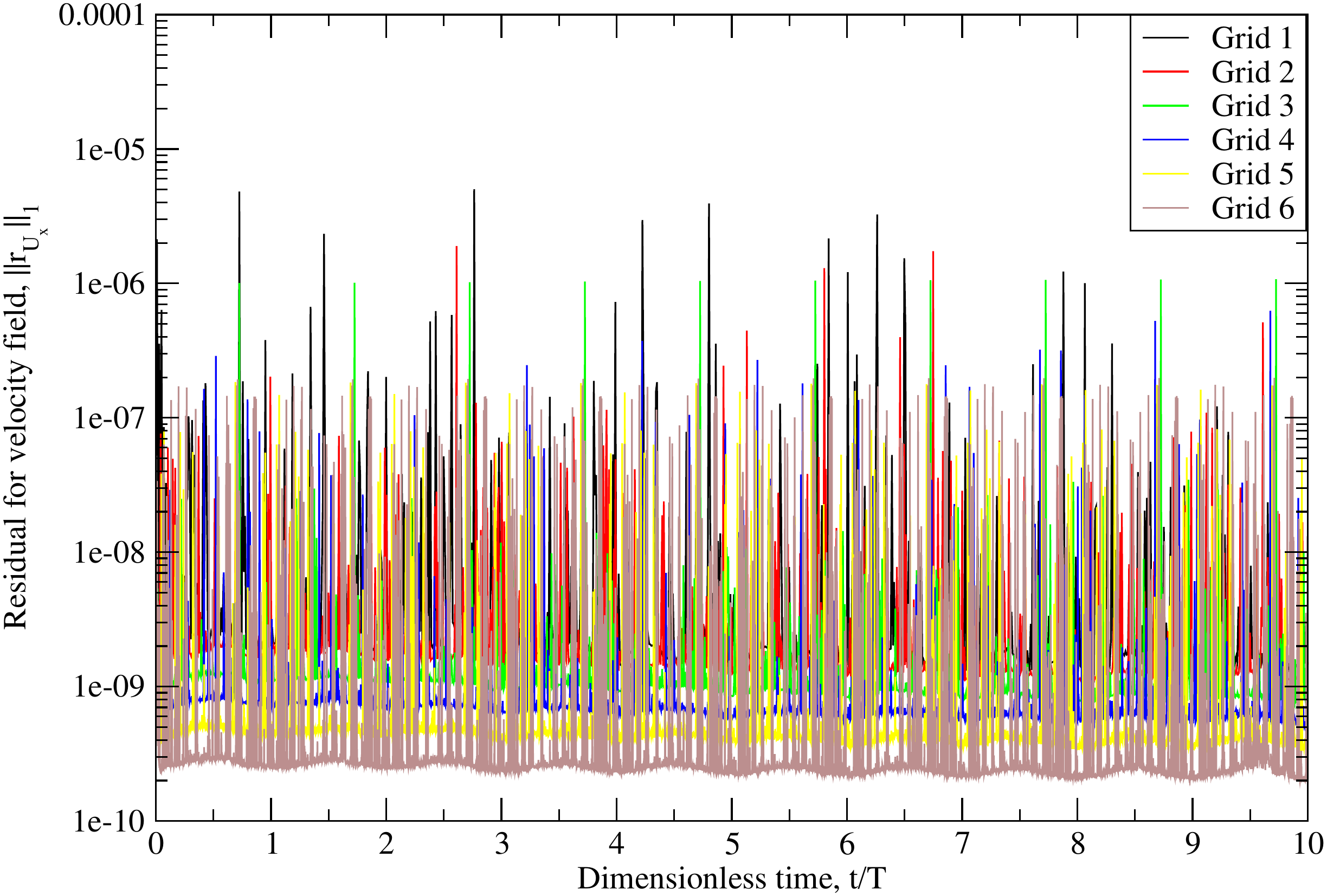}
  \caption{Residual for the velocity field.}
  \label{fig:iterativeConvergenceVelocity}
  \end{subfigure}%
  \begin{subfigure}[b]{0.45\textwidth}
  \includegraphics[width=7cm]{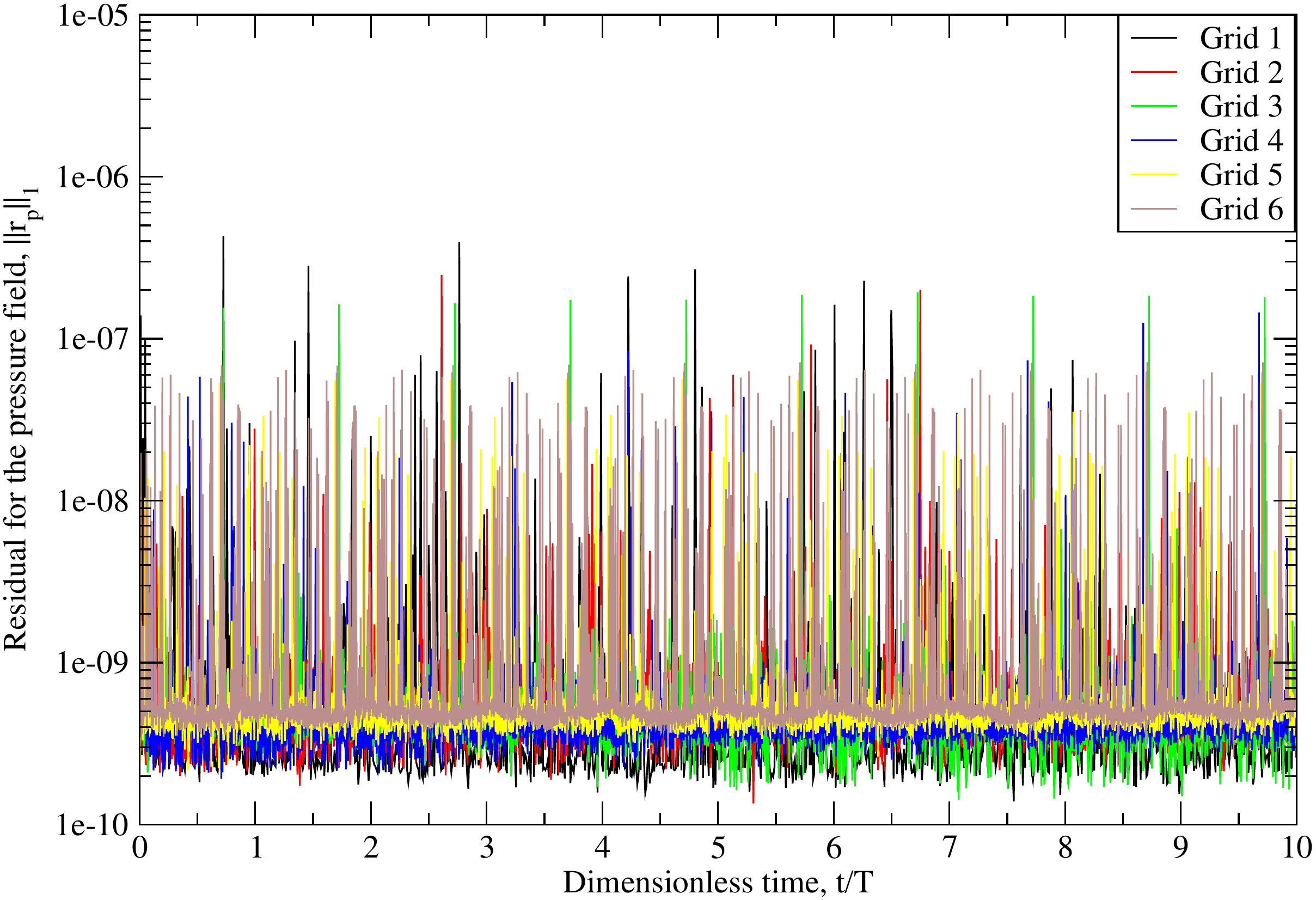}
  \caption{Residual for the pressure field.}
  \label{fig:iterativeConvergencePressure}
  \end{subfigure}%
\end{center}
\caption{$L_1$ residuals during the simulations, indicating the level of
iterative convergence used in present simulations.}
\label{fig:iterativeConvergence}
\end{figure}

\noindent E\c{c}a~\etal~\cite{ecaEtAlNUTTS2017} have recently shown that the
iterative errors in unsteady simulations should be kept to a minimum, otherwise
they can significantly influence the final solution. Therefore, all simulations
are performed with six nonlinear iterations per time step and two additional
pressure correction steps per each nonlinear iteration. Relaxation is not used.
The settings ensure that residuals for all variables measured by the $L_1$
norm in the last nonlinear iteration are always smaller than $5 \times 10^{-6}$,
making iterative uncertainty small. The residuals during the simulations
have oscillatory behaviour,~\autoref{fig:iterativeConvergence}, while on average
they are approximately $\mathcal{O}(10^{-9})$. 

\begin{figure}[b!]
\begin{center}
  \begin{subfigure}[b]{0.45\textwidth}
    \begin{tabular}{l c c c c}
      \hline
      Item & $\eta_{A}$, m & $U_{\eta A}$ & $p$ & $q$ \\
      \hline
      \ensuremath{\eta_{AG1}} & $1.44 \times 10^{-1}$ & $0.4\%$ & $2.00$ & $1.98$ \\
      \ensuremath{\eta_{AG2}} & $1.43 \times 10^{-1}$ & $0.5\%$ & $2.00$ & $2.00$ \\
      \ensuremath{\eta_{AG3}} & $1.42 \times 10^{-1}$ & $0.4\%$ & $1.27$ & $1.26$ \\
      \ensuremath{\eta_{AG4}} & $1.41 \times 10^{-1}$ & $0.6\%$ & $1.53$ & $1.52$ \\
      \ensuremath{\eta_{AG5}} & $1.41 \times 10^{-1}$ & $1.1\%$ & $1.40$ & $1.40$ \\
      \ensuremath{\eta_{AG6}} & $1.43 \times 10^{-1}$ & $3.5\%$ & $1.09$ & $1.07$ \\
      \ensuremath{\eta_{AG7}} & $1.42 \times 10^{-1}$ & $2.9\%$ & $0.91$ & $0.90$ \\
      \ensuremath{\eta_{AG8}} & $1.39 \times 10^{-1}$ & $1.2\%$ & $1.21$ & $1.20$ \\
      \ensuremath{\eta_{AG9}} & $1.45 \times 10^{-1}$ & $5.8\%$ & $0.54$ & $0.54$ \\
      \ensuremath{\eta_{AG10}} & $1.39 \times 10^{-1}$ & $1.2\%$ & $1.45$ & $1.43$ \\
      \ensuremath{\eta_{AG11}} & $1.41 \times 10^{-1}$ & $2.9\%$ & $1.07$ & $1.06$ \\
      \ensuremath{\eta_{AG12}} & $1.48 \times 10^{-1}$ & $6.6\%$ & $1.22$ & $1.20$ \\
      \hline
      Average & $1.42 \times 10^{-1}$ & $2.25\%$ & $1.31$ & $1.30$ \\
    \end{tabular}
    \label{tab:uncertaintyAmplitudeTable}
    \caption{Results for all wave gauges.}
  \end{subfigure}%
  \begin{subfigure}[b]{0.35\textwidth}
  \includegraphics[width=7cm]{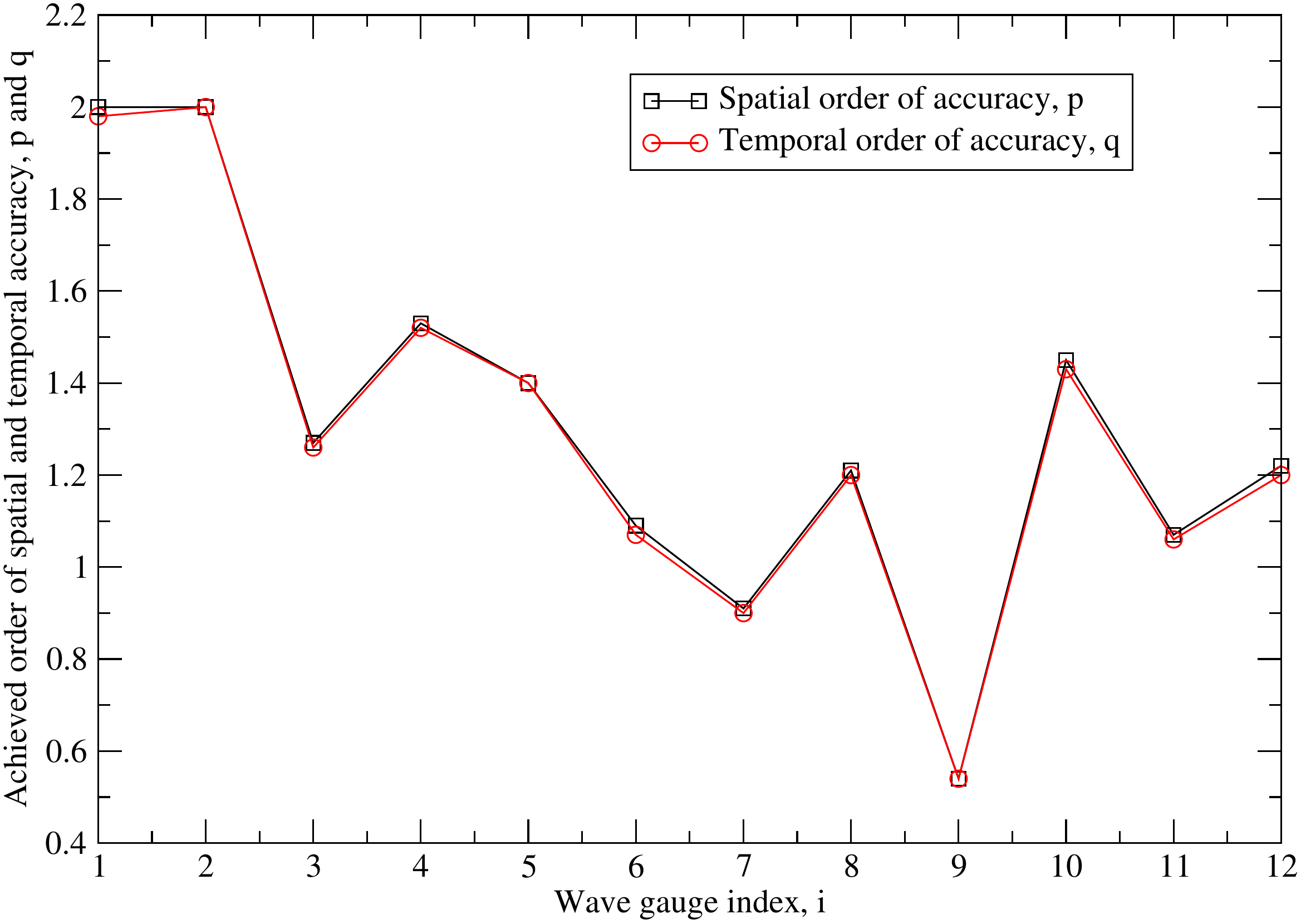}
  \caption{Achieved order of convergence for all wave gauges.}
  \label{fig:amplitudesAccuracy}
  \end{subfigure}%
\end{center}
\caption{Uncertainty analysis for wave elevation: first order amplitudes.}
\label{tab:uncertaintyAmplitudes}
\end{figure}
\begin{figure}[t!]
\begin{center}
  \begin{subfigure}[b]{0.45\textwidth}
    \begin{tabular}{l c c c c}
      \hline
      Item & $\eta_{\theta}$, $^{\circ}$ & $U_{\eta_{\theta}}$, $^{\circ}$ & $p$ & $q$ \\
      \hline
      \ensuremath{\eta_{\theta G1}} & $182$ & $0.2$ & $1.70$ & $1.68$ \\
      \ensuremath{\eta_{\theta G2}} & $180$ & $0.4$ & $2.00$ & $1.98$ \\
      \ensuremath{\eta_{\theta G3}} & $179$ & $1.1$ & $2.00$ & $1.99$ \\
      \ensuremath{\eta_{\theta G4}} & $178$ & $1.6$ & $2.00$ & $2.00$ \\
      \ensuremath{\eta_{\theta G5}} & $179$ & $4.0$ & $1.23$ & $2.00$ \\
      \ensuremath{\eta_{\theta G6}} & $178$ & $5.0$ & $1.98$ & $1.96$ \\
      \ensuremath{\eta_{\theta G7}} & $178$ & $5.4$ & $1.92$ & $1.90$ \\
      \ensuremath{\eta_{\theta G8}} & $177$ & $5.8$ & $1.88$ & $1.87$ \\
      \ensuremath{\eta_{\theta G9}} & $177$ & $6.1$ & $1.90$ & $1.88$ \\
      \ensuremath{\eta_{\theta G10}} & $177$ & $6.3$ & $1.86$ & $1.84$ \\
      \ensuremath{\eta_{\theta G11}} & $177$ & $6.3$ & $1.91$ & $1.89$ \\
      \ensuremath{\eta_{\theta G12}} & $183$ & $4.9$ & $1.93$ & $1.91$ \\
      \hline
      Average & $179$ & $3.92$ & $1.86$ & $1.91$ \\
    \end{tabular}
    \label{tab:uncertaintyPhaseTable}
    \caption{Results for all wave gauges.}
  \end{subfigure}%
  \begin{subfigure}[b]{0.35\textwidth}
  \includegraphics[width=7cm]{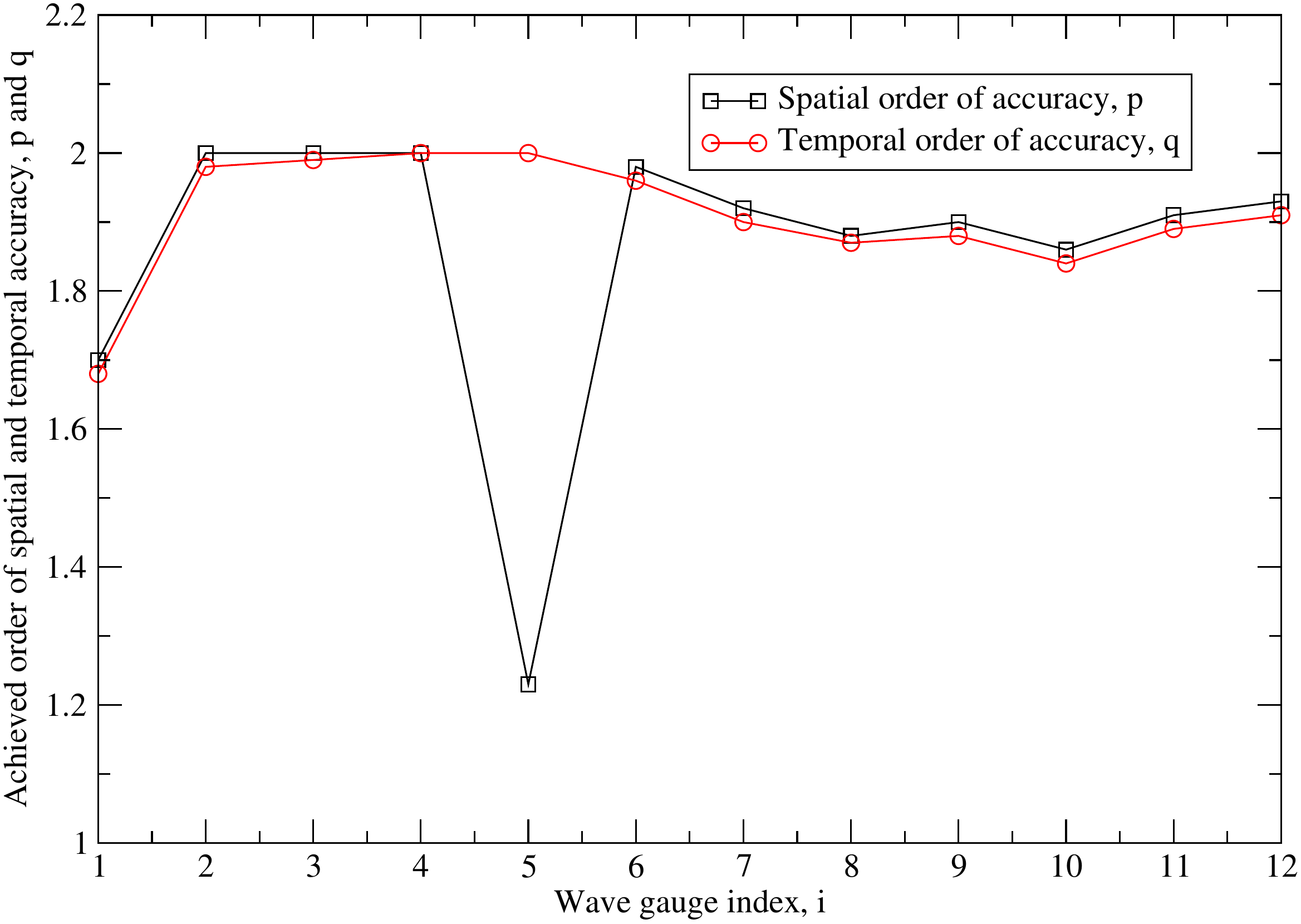}
  \caption{Achieved order of convergence for all wave gauges.}
  \label{fig:phasesAccuracy}
  \end{subfigure}%
\end{center}
\caption{Uncertainty analysis for wave elevation: first order phases.}
\label{tab:uncertaintyPhases}
\end{figure}
\begin{figure}[b!]
\begin{center}
  \begin{subfigure}[b]{0.45\textwidth}
    \begin{tabular}{l c c c c}
      \hline
      Item & $u_{xA}$, m/s & $U_{U_{xA}}\%$ & $p$ & $q$ \\
      \hline
      \ensuremath{u_{xAP1}} & $4.10 \times 10^{-1}$ & $3.4\%$ & $1.93$ & $0.80$ \\
      \ensuremath{u_{xAP2}} & $3.98 \times 10^{-1}$ & $1.1\%$ & $2.00$ & $1.98$ \\
      \ensuremath{u_{xAP3}} & $3.60 \times 10^{-1}$ & $12.0\%$ & $1.00$ & $0.98$ \\
      \ensuremath{u_{xAP4}} & $3.46 \times 10^{-1}$ & $15.5\%$ & $1.00$ & $0.98$ \\
      \ensuremath{u_{xAP5}} & $3.89 \times 10^{-1}$ & $1.3\%$ & $2.00$ & $1.98$ \\
      \ensuremath{u_{xAP6}} & $3.94 \times 10^{-1}$ & $1.9\%$ & $2.00$ & $2.00$ \\
      \ensuremath{u_{xAP7}} & $3.93 \times 10^{-1}$ & $2.1\%$ & $2.00$ & $2.00$ \\
      \ensuremath{u_{xAP8}} & $3.92 \times 10^{-1}$ & $2.0\%$ & $2.00$ & $2.00$ \\
      \ensuremath{u_{xAP9}} & $3.88 \times 10^{-1}$ & $1.9\%$ & $1.75$ & $1.74$ \\
      \ensuremath{u_{xAP10}} & $3.88 \times 10^{-1}$ & $0.7\%$ & $^{*~1, 2}$ & $1.00$ \\
      \ensuremath{u_{xAP11}} & $3.92 \times 10^{-1}$ & $1.8\%$ & $2.00$ & $2.00$ \\
      \ensuremath{u_{xAP12}} & $4.09 \times 10^{-1}$ & $2.4\%$ & $1.56$ & $2.00$ \\
      \hline
      Average & $3.88 \times 10^{-1}$ & $3.84$ & $1.69$ & $1.62$ \\
      \hline
      \multicolumn{5}{l}{\scriptsize $^{*~1, 2}$ Fit was made using first and
      second order exponents}\\
      \hline
    \end{tabular}
    \label{tab:uncertaintyVelocityXAmplitudesTable}
    \caption{Results for all probes.}
  \end{subfigure}%
  \begin{subfigure}[b]{0.35\textwidth}
  \includegraphics[width=7cm]{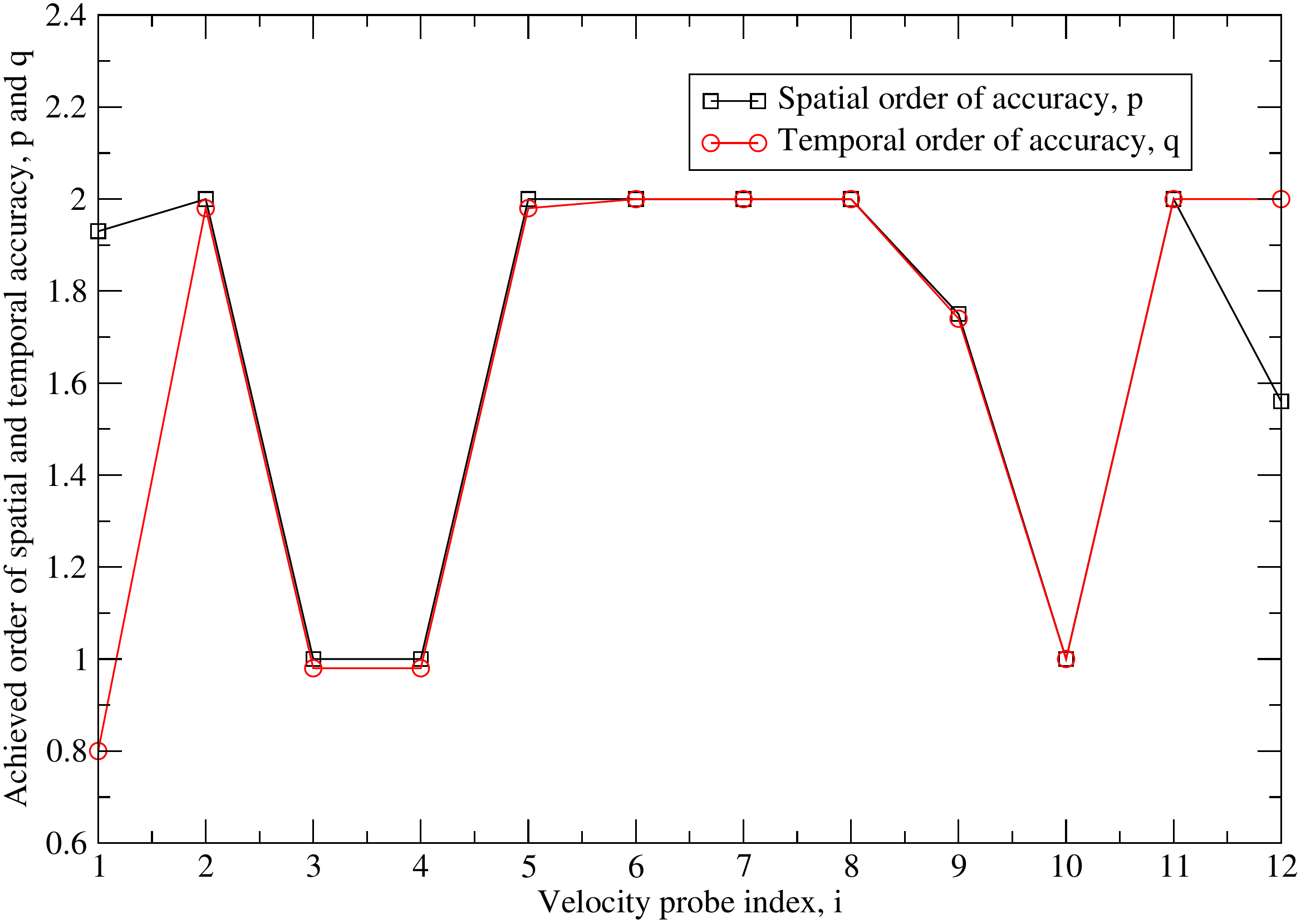}
  \caption{Achieved order of convergence for all probes.}
  \label{fig:velocityAmplitudesXAccuracy}
  \end{subfigure}%
\end{center}
\caption{Uncertainty analysis for $x$ (horizontal) component of the velocity
field: first order amplitudes.}
\label{tab:uncertaintyVelocityXAmplitudes}
\end{figure}
\begin{figure}[t!]
\begin{center}
  \begin{subfigure}[b]{0.45\textwidth}
    \begin{tabular}{l c c c c}
      \hline
      Item & $u_{x\theta}$, $^{\circ}$ & $U_{U_{x\theta}}$, $^{\circ}$ & $p$ & $q$ \\
      \hline
      \ensuremath{u_{x\theta P1}} & $166$ & $18.2$ & $^{*~1, 2}$ & $2.00$ \\
      \ensuremath{u_{x\theta P2}} & $179$ & $2.2$ & $1.00$ & $^{*~1, 2}$ \\
      \ensuremath{u_{x\theta P3}} & $174$ & $3.8$ & $^{*~1, 2}$ & $1.00$ \\
      \ensuremath{u_{x\theta P4}} & $180$ & $4.1$ & $2.00$ & $2.00$ \\
      \ensuremath{u_{x\theta P5}} & $198$ & $28.8$ & $1.09$ & $1.08$ \\
      \ensuremath{u_{x\theta P6}} & $180$ & $12.3$ & $1.78$ & $1.76$ \\
      \ensuremath{u_{x\theta P7}} & $178$ & $10.1$ & $1.81$ & $1.80$ \\
      \ensuremath{u_{x\theta P8}} & $173$ & $7.4$ & $1.00$ & $^{*~1, 2}$ \\
      \ensuremath{u_{x\theta P9}} & $177$ & $7.2$ & $2.00$ & $1.98$ \\
      \ensuremath{u_{x\theta P10}} & $178$ & $18.2$ & $2.00$ & $1.98$ \\
      \ensuremath{u_{x\theta P11}} & $188$ & $20.9$ & $1.68$ & $1.66$ \\
      \ensuremath{u_{x\theta P12}} & $182$ & $6.1$ & $2.00$ & $2.00$ \\
      \hline
      Average & $179$ & $11.6$ & $1.53$ & $1.61$ \\
      \hline
      \multicolumn{5}{l}{\scriptsize $^{*~1, 2}$ Fit was made using first and
      second order exponents}\\
      \hline
    \end{tabular}
    \label{tab:uncertaintyVelocityXPhaseTable}
    \caption{Results for all probes.}
  \end{subfigure}%
  \begin{subfigure}[b]{0.35\textwidth}
  \includegraphics[width=7cm]{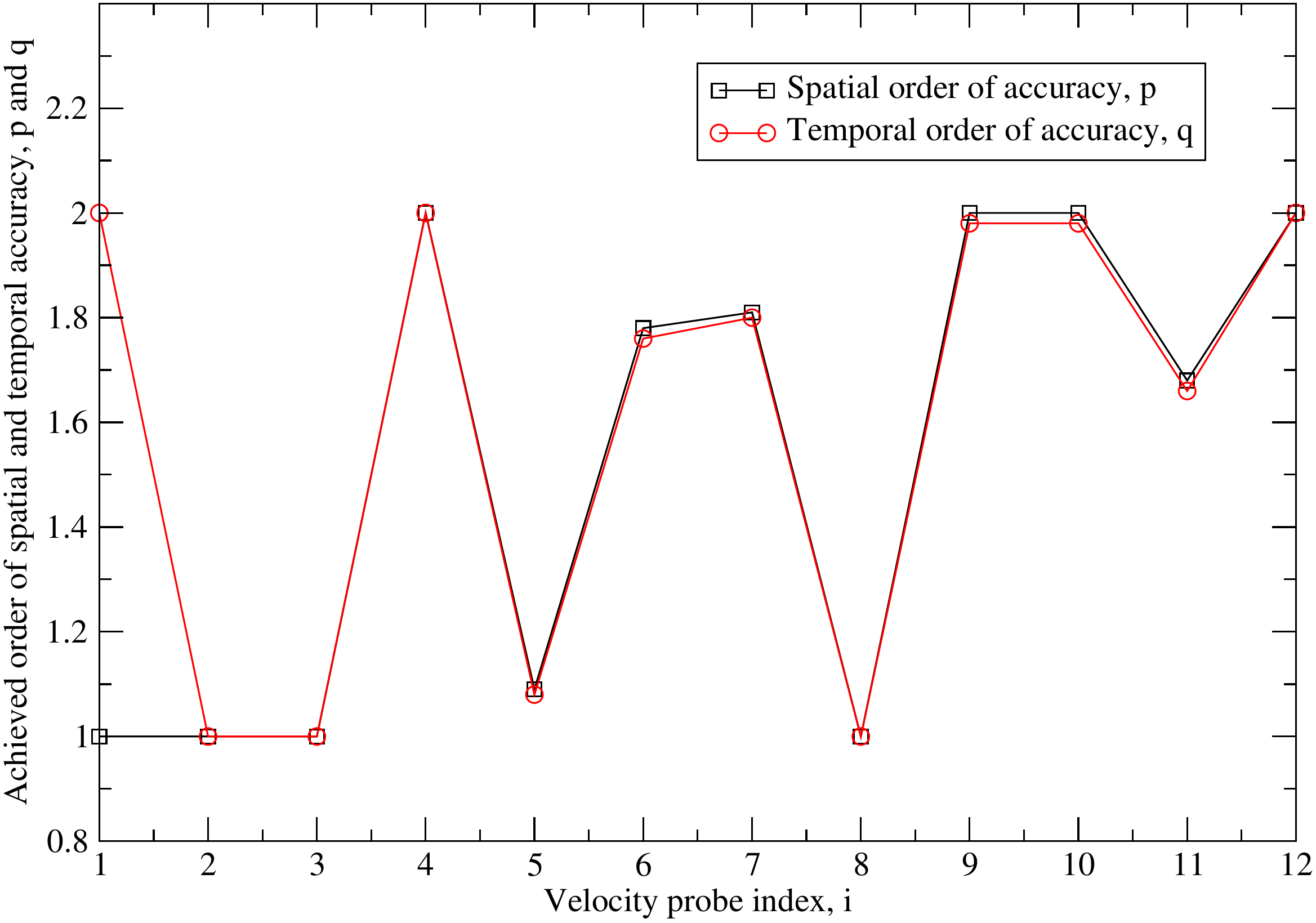}
  \caption{Achieved order of convergence for all probes.}
  \label{fig:velocityPhasesXAccuracy}
  \end{subfigure}%
\end{center}
\caption{Uncertainty analysis for $x$ (horizontal) component of the velocity
field: first order phases.}
\label{tab:uncertaintyVelocityXPhases}
\end{figure}
\begin{figure}[b!]
\begin{center}
  \begin{subfigure}[b]{0.45\textwidth}
    \begin{tabular}{l c c c c}
      \hline
      Item & $u_{yA}$, m/s & $U_{U_{yA}}\%$ & $p$ & $q$ \\
      \hline
      \ensuremath{u_{yAP1}} & $2.77 \times 10^{-1}$ & $1.1\%$ & $^{*~1, 2}$ & $1.00$ \\
      \ensuremath{u_{yAP2}} & $2.73 \times 10^{-1}$ & $0.3\%$ & $^{*~1, 2}$ & $^{*~1, 2}$ \\
      \ensuremath{u_{yAP3}} & $2.73 \times 10^{-1}$ & $1.2\%$ & $^{*~1, 2}$ & $1.00$ \\
      \ensuremath{u_{yAP4}} & $2.65 \times 10^{-1}$ & $1.5\%$ & $2.00$ & $1.99$ \\
      \ensuremath{u_{yAP5}} & $2.53 \times 10^{-1}$ & $6.8\%$ & $^{*~1, 2}$ & $^{*~1, 2}$ \\
      \ensuremath{u_{yAP6}} & $2.71 \times 10^{-1}$ & $2.4\%$ & $^{*~1, 2}$ & $1.00$ \\
      \ensuremath{u_{yAP7}} & $2.70 \times 10^{-1}$ & $2.0\%$ & $^{*~1, 2}$ & $1.00$ \\
      \ensuremath{u_{yAP8}} & $2.43 \times 10^{-1}$ & $10.5\%$ & $1.01$ & $1.00$ \\
      \ensuremath{u_{yAP9}} & $2.71 \times 10^{-1}$ & $2.8\%$ & $^{*~1, 2}$ & $^{*~1, 2}$ \\
      \ensuremath{u_{yAP10}} & $2.69 \times 10^{-1}$ & $1.9\%$ & $^{*~1, 2}$ & $1.00$ \\
      \ensuremath{u_{yAP11}} & $2.56 \times 10^{-1}$ & $4.6\%$ & $1.00$ & $0.99$ \\
      \ensuremath{u_{yAP12}} & $2.69 \times 10^{-1}$ & $4.9\%$ & $2.00$ & $1.21$ \\
      \hline
      Average & $2.66 \times 10^{-1}$ & $3.33$ & $1.17$ & $1.10$ \\
      \hline
      \multicolumn{5}{l}{\scriptsize $^{*~1, 2}$ Fit was made using first and
      second order exponents}\\
      \hline
    \end{tabular}
    \label{tab:uncertaintyVelocityYTable}
    \caption{Results for all probes.}
  \end{subfigure}%
  \begin{subfigure}[b]{0.35\textwidth}
  \includegraphics[width=7cm]{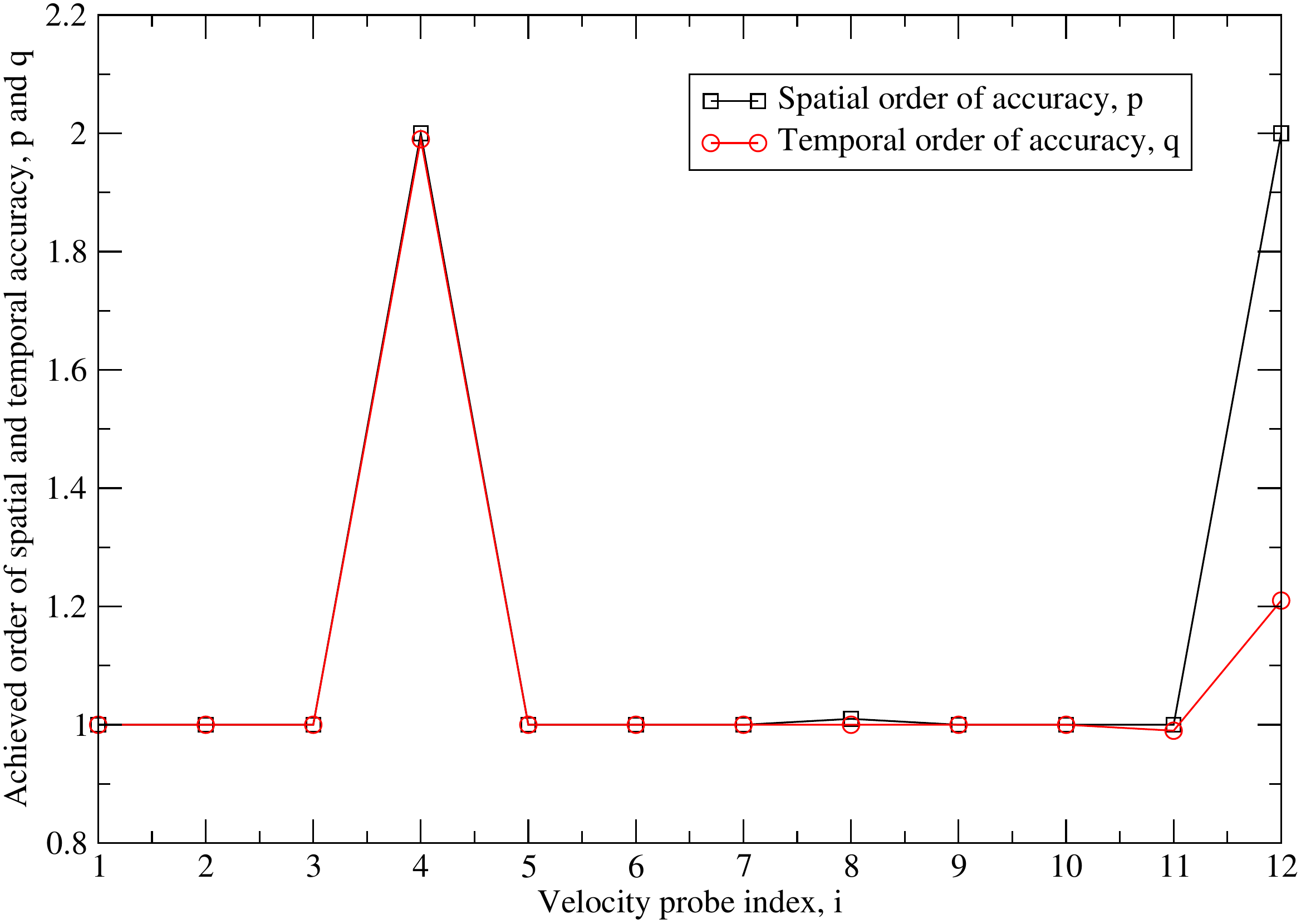}
  \caption{Achieved order of convergence for all probes.}
  \label{fig:velocityAmplitudesYAccuracy}
  \end{subfigure}%
\end{center}
\caption{Uncertainty analysis for $y$ (vertical) component of the velocity
field: first order amplitudes.}
\label{tab:uncertaintyVelocityYAmplitudes}
\end{figure}
\begin{figure}[t!]
\begin{center}
  \begin{subfigure}[b]{0.45\textwidth}
    \begin{tabular}{l c c c c}
      \hline
      Item & $u_{y\theta}$, $^{\circ}$ & $U_{U_{y\theta}}$, $^{\circ}$ & $p$ & $q$ \\
      \hline
      \ensuremath{u_{y\theta P1}} & $256$ & $11.3$  & $2.00$ & $^{*~1, 2}$ \\
      \ensuremath{u_{y\theta P2}} & $267$ & $3.6$   & $1.05$ & $1.00$ \\
      \ensuremath{u_{y\theta P3}} & $278$ & $9.2$   & $^{*~1, 2}$ & $2.00$ \\
      \ensuremath{u_{y\theta P4}} & $270$ & $2.9$   & $2.00$ & $2.00$ \\
      \ensuremath{u_{y\theta P5}} & $278$ & $11.7$  & $1.29$ & $1.28$ \\
      \ensuremath{u_{y\theta P6}} & $270$ & $8.5$   & $1.72$ & $1.71$ \\
      \ensuremath{u_{y\theta P7}} & $270$ & $7.9$   & $1.61$ & $1.59$ \\
      \ensuremath{u_{y\theta P8}} & $269$ & $7.6$   & $1.07$ & $1.06$ \\
      \ensuremath{u_{y\theta P9}} & $277$ & $14.0$  & $1.15$ & $1.13$ \\
      \ensuremath{u_{y\theta P10}} & $277$ & $13.3$ & $1.25$ & $1.24$ \\
      \ensuremath{u_{y\theta P11}} & $279$ & $14.2$ & $1.63$ & $1.62$ \\
      \ensuremath{u_{y\theta P12}} & $273$ & $3.8$  & $2.00$ & $2.00$ \\
      \hline
      Average & $272$ & $9.0$ & $1.48$ & $1.47$ \\
      \hline
      \multicolumn{5}{l}{\scriptsize $^{*~1, 2}$ Fit was made using first and
      second order exponents}\\
      \hline
    \end{tabular}
    \label{tab:uncertaintyVelocityYPhaseTable}
    \caption{Results for all probes.}
  \end{subfigure}%
  \begin{subfigure}[b]{0.35\textwidth}
  \includegraphics[width=7cm]{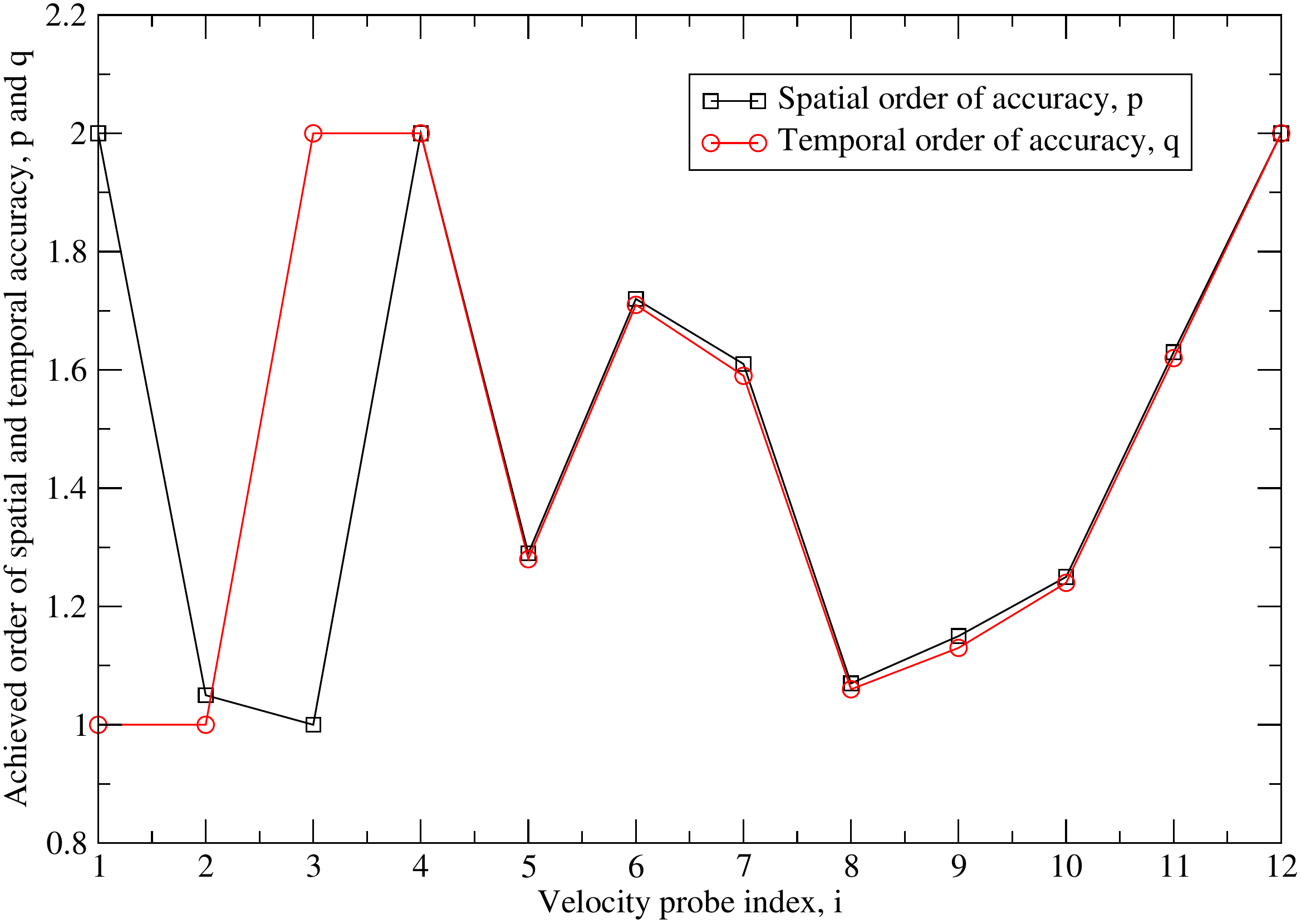}
  \caption{Achieved order of convergence for all probes.}
  \label{fig:velocityPhasesYAccuracy}
  \end{subfigure}%
\end{center}
\caption{Uncertainty analysis for $y$ (horizontal) component of the velocity
field: first order phases.}
\label{tab:uncertaintyVelocityYPhases}
\end{figure}

\noindent {\bfseries Verification study for first order effects.}
Achieved order of spatial $p$ and temporal $q$ convergence of first
order amplitude of wave elevation, $\eta_A$ is presented
in~\autoref{tab:uncertaintyAmplitudes} following the guidelines by E\c{c}a
and Hoekstra~\cite{ecaHoekstra2014} and using their open access
tool~\cite{uncertaintyReFrescoWebsite}. The achieved orders of spatial and
temporal accuracy range from 0.54 to 2.00, while the average value for all wave
gauges is 1.30. Average numerical uncertainty for the fine grid $\eta_U$ is
2.25\% considering all wave gauges. The achieved
order of accuracy is lower further downstream of the numerical wave tank, and
consequently numerical uncertainty for the fine grid $U_{\eta A}$ is larger.
This trend is presented in~\autoref{fig:amplitudesAccuracy}. From previous
experience with wave--related problems in ocean and marine engineering, it is
sufficient to consider only a few wave lengths in the computational
domain~\cite{vukcevicEtAl2016a,vukcevicPhD2016}. If only the first five wave
lengths were considered, the average order of accuracy would be 1.64 for space
and 1.63 for time. Consequently, the average numerical uncertainty for first
five wave lengths is 0.6\%, compared to 2.25\% for all wave lengths. In
addition, the second column of~\autoref{tab:uncertaintyAmplitudes} presents the
guessed asymptotic solution for each of wave gauges, where the average first
order amplitude is $0.142$ metres.\\
\indent~\autoref{tab:uncertaintyPhases} presents verification analysis
for first order harmonic phases, $\eta_{\theta}$. The achieved orders of
spatial and temporal accuracy range from 1.23 to 2.00, with average values 1.86
for space and 1.91 for time. Better convergence with refinement is thus obtained
for first order harmonic phases compared to first order harmonic amplitudes.
The dispersion error (error in phase) is well--behaved compared to the
dissipation error (error in amplitude), although more studies should be
carried out to draw more general conclusions. Achieved spatial and temporal
orders of convergence are well--behaved for all wave gauges as presented
in~\autoref{fig:phasesAccuracy}, except for the outlier at wave gauge 5 that
measured low order of spatial convergence.  Numerical uncertainties on Grid 6
for the phases are presented in degrees in the third column
of~\autoref{tab:uncertaintyPhases}, with average of approximately
$4{^\circ}$.\\
%
%
\indent~\autoref{tab:uncertaintyVelocityXAmplitudes} presents the results of the
verification analysis for the first order amplitude of $x$ (horizontal)
component of the velocity field, $u_{xA}$. Averaged for all probes, $u_{xA}$
exhibits convergence with grid and time step refinement of 1.69 and 1.62,
respectively. Here, a single ill--behaved convergence has been observed where a
fit was made using first and second order exponents following E\c{c}a and
Hoekstra~\cite{ecaHoekstra2014}. The convergence shows irregular behaviour for
different probes. The associated numerical uncertainties on the fine grid are
lower than 3.5\%, except for the two outliers at probes 3 and 4 with high
numerical uncertainties of $12\%$ and $15.5\%$, respectively.\\
\indent~\autoref{tab:uncertaintyVelocityXPhases} presents the results for the
first order phase of $x$ component of the velocity field, $u_{x\theta}$.
Achieved order of spatial accuracy and temporal accuracy is 1.53 and 1.61,
respectively, averaged over all probes. The associated numerical uncertainties
range from 2.2 to $28.8^{\circ}$. The analysis shows irregular behaviour across
all velocity probes.\\
%
%
\indent~\autoref{tab:uncertaintyVelocityYAmplitudes} presents the results for
the first order amplitude of $y$ (vertical) component of the velocity field,
$u_{yA}$. Achieved order of spatial accuracy and temporal accuracy is lower
compared to wave elevation and horizontal velocity component, with average
values of 1.17 and 1.10, respectively. The numerical uncertainty is lower than
$3\%$ for majority of the probes, with outliers up to 10.5\%.\\
\indent~\autoref{tab:uncertaintyVelocityYPhases} presents the results for
the first order phase of $y$ (vertical) component of the velocity field,
$u_{y\theta}$. Achieved orders of spatial accuracy and temporal accuracy are
1.48 and 1.47, respectively, averaged over all probes. The numerical uncertainty
is on average $9^{\circ}$ .\\

\begin{figure}[b!]
\begin{center}
  \begin{subfigure}[b]{0.45\textwidth}
    \begin{tabular}{l c c c c}
      \hline
      Item & $\eta_{A}$, m & $U_{\eta A}$ & $p$ & $q$ \\
      \hline
      \ensuremath{\eta_{AG1}} & $2.64 \times 10^{-2}$ & $0.1\%$ & $2.00$ & $1.96$ \\
      \ensuremath{\eta_{AG2}} & $2.45 \times 10^{-2}$ & $8.5\%$ & $^{*~1, 2}$ & $^{*~1, 2}$ \\
      \ensuremath{\eta_{AG3}} & $2.23 \times 10^{-2}$ & $18.8\%$ & $1.09$ & $1.08$ \\
      \ensuremath{\eta_{AG4}} & $2.10 \times 10^{-2}$ & $21.3\%$ & $^{*~1, 2}$ & $^{*~1, 2}$ \\
      \ensuremath{\eta_{AG5}} & $2.56 \times 10^{-2}$ & $7.9\%$ & $2.00$ & $2.00$ \\
      \ensuremath{\eta_{AG6}} & $2.30 \times 10^{-2}$ & $8.8\%$ & $1.17$ & $1.16$ \\
      \ensuremath{\eta_{AG7}} & $2.48 \times 10^{-2}$ & $1.8\%$ & $1.85$ & $1.85$ \\
      \ensuremath{\eta_{AG8}} & $2.71 \times 10^{-2}$ & $13.4\%$ & $0.67$ & $2.00$ \\
      \ensuremath{\eta_{AG9}} & $2.52 \times 10^{-2}$ & $3.4\%$ & $1.68$ & $1.68$ \\
      \ensuremath{\eta_{AG10}} & $2.35 \times 10^{-2}$ & $5.9\%$ & $1.15$ & $1.14$ \\
      \ensuremath{\eta_{AG11}} & $2.43 \times 10^{-2}$ & $1.6\%$ & $1.78$ & $1.77$ \\
      \ensuremath{\eta_{AG12}} & $2.41 \times 10^{-2}$ & $2.3\%$ & $1.60$ & $1.59$ \\
      \hline
      Average & $2.43 \times 10^{-2}$ & $7.81\%$ & $1.42$ & $1.52$ \\
      \hline
      \multicolumn{5}{l}{\scriptsize $^{*~1, 2}$ Fit was made using first and
      second order exponents}\\
      \hline
    \end{tabular}
    \label{tab:uncertaintyAmplitudeTable2}
    \caption{Results for all wave gauges.}
  \end{subfigure}%
  \begin{subfigure}[b]{0.35\textwidth}
  \includegraphics[width=7cm]{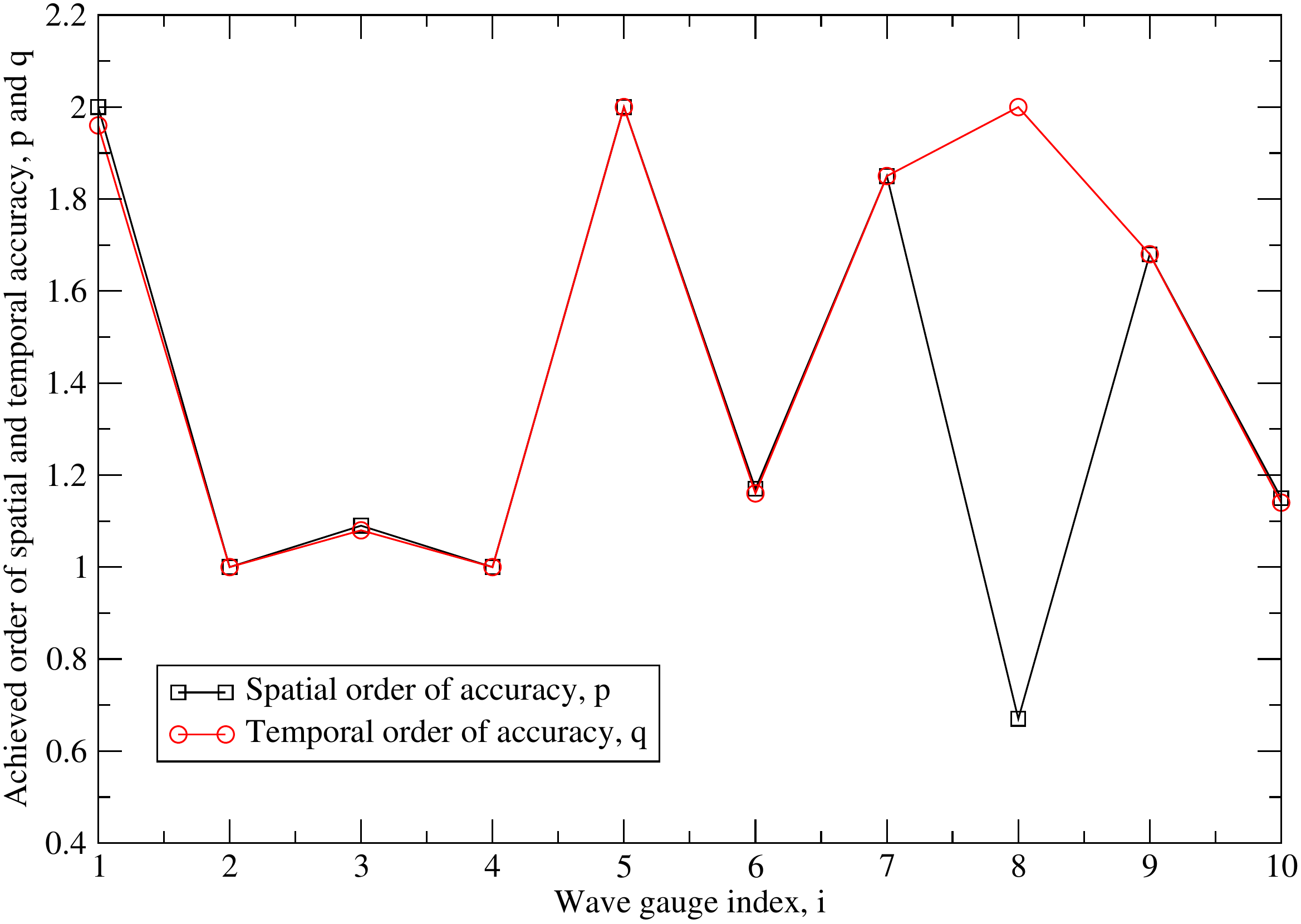}
  \caption{Achieved order of convergence for all wave gauges.}
  \label{fig:amplitudesAccuracy2}
  \end{subfigure}%
\end{center}
\caption{Uncertainty analysis for wave elevation: second order amplitudes.}
\label{tab:uncertaintyAmplitudes2}
\end{figure}

\begin{figure}[t!]
\begin{center}
  \begin{subfigure}[b]{0.45\textwidth}
    \begin{tabular}{l c c c c}
      \hline
      Item & $\eta_{\theta}$, $^{\circ}$ & $U_{\eta_{\theta}}$, $^{\circ}$ & $p$ & $q$ \\
      \hline
      \ensuremath{\eta_{\theta G1}} & $184$ & $2.3$ & $0.93$ & $0.92$ \\
      \ensuremath{\eta_{\theta G2}} & $182$ & $1.8$ & $2.00$ & $1.98$ \\
      \ensuremath{\eta_{\theta G3}} & $181$ & $1.3$ & $2.00$ & $2.00$ \\
      \ensuremath{\eta_{\theta G4}} & $178$ & $3.2$ & $2.00$ & $2.00$ \\
      \ensuremath{\eta_{\theta G5}} & $176$ & $5.4$ & $2.00$ & $2.00$ \\
      \ensuremath{\eta_{\theta G6}} & $175$ & $4.9$ & $1.87$ & $2.00$ \\
      \ensuremath{\eta_{\theta G7}} & $172$ & $10.2$& $2.00$ & $2.00$ \\
      \ensuremath{\eta_{\theta G8}} & $174$ & $7.9$ & $2.00$ & $1.98$ \\
      \ensuremath{\eta_{\theta G9}} & $173$ & $11.0$& $1.91$ & $1.89$ \\
      \ensuremath{\eta_{\theta G10}} & $174$ & $10.1$& $2.00$ & $1.98$ \\
      \ensuremath{\eta_{\theta G11}} & $172$ & $8.3$ & $1.81$ & $2.00$ \\
      \ensuremath{\eta_{\theta G12}} & $173$ & $10.1$& $2.00$ & $1.98$ \\
      \hline
      Average & $176$ & $6.37$ & $1.88$ & $1.89$ \\
    \end{tabular}
    \label{tab:uncertaintyPhaseTable2}
    \caption{Results for all wave gauges.}
  \end{subfigure}%
  \begin{subfigure}[b]{0.35\textwidth}
  \includegraphics[width=7cm]{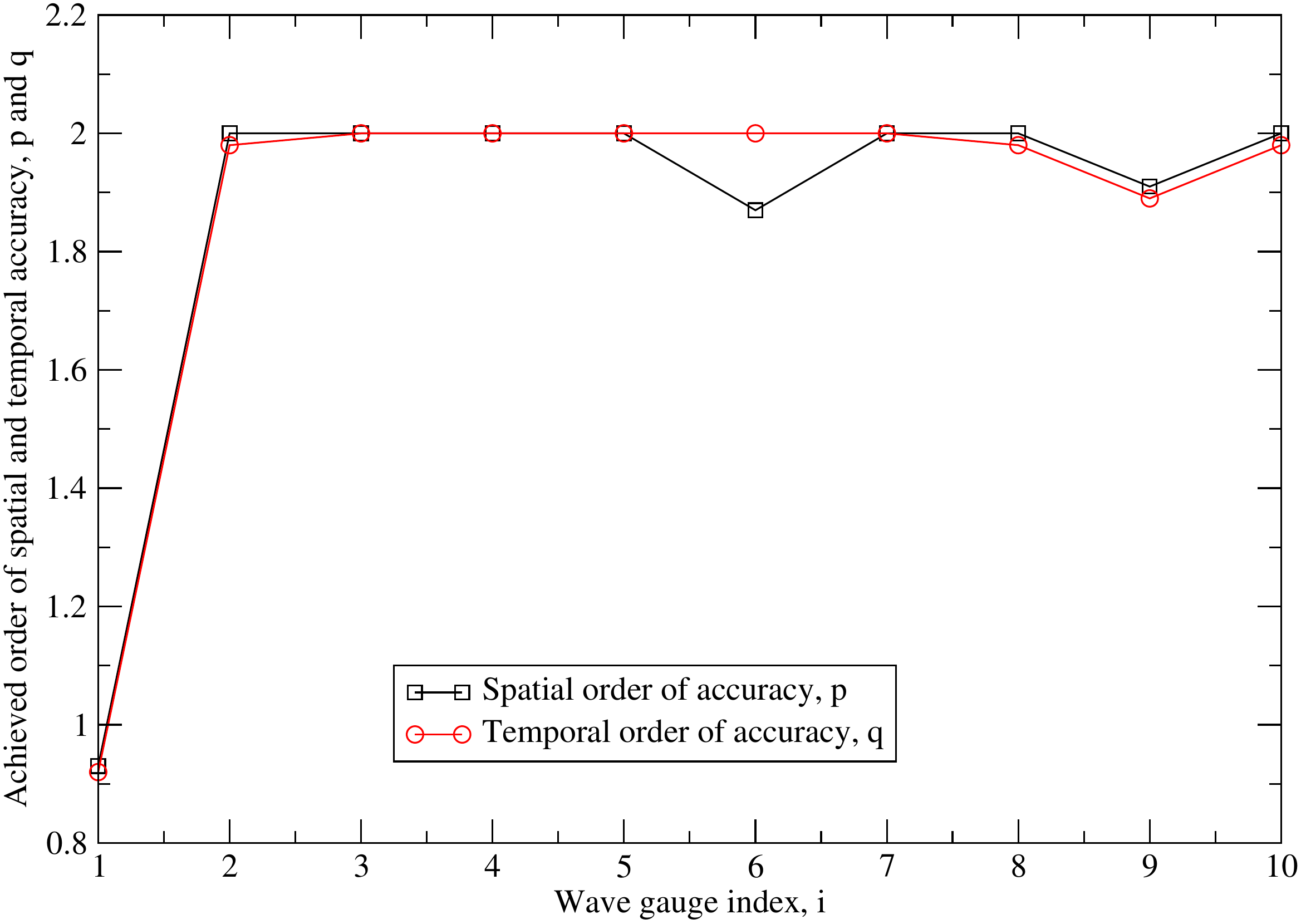}
  \caption{Achieved order of convergence for all wave gauges.}
  \label{fig:phasesAccuracy2}
  \end{subfigure}%
\end{center}
\caption{Uncertainty analysis for wave elevation: second order phases.}
\label{tab:uncertaintyPhases2}
\end{figure}

\noindent {\bfseries Verification study for second order effects.}
The verification study is continued by performing the same analysis for second
order effects since the wave is moderately nonlinear ($kH/2 =
0.174$).~\autoref{tab:uncertaintyAmplitudes2} presents verification results for
second order wave elevation amplitudes. The average asymptotic solution over all
wave gauges is $0.0243$ metres, which is approximately five times smaller
than the average first order amplitude ($0.142$ metres). The average
uncertainty is $7.81\%$, ranging from 0.1\% to 21.3\%. Achieved orders of
spatial and temporal accuracy range from $0.67$ to 2, with average values of
1.42 and 1.52, respectively.\\
\indent \autoref{tab:uncertaintyPhases2} presents the results for second order
wave elevation phases. The asymptotic solution for all wave gauges is
approximately 176$^{\circ}$, while the corresponding numerical uncertainty is
$6.4^{\circ}$ on average. The achieved orders of spatial and temporal accuracy
range from 0.92 to 2.00, with the average values of 1.88 and 1.89,
respectively.\\

\begin{figure}[t!]
\begin{center}
  \begin{subfigure}[b]{0.45\textwidth}
    \begin{tabular}{l c c c c}
      \hline
      Item & $u_{xA}$, m/s & $U_{U_{xA}}\%$ & $p$ & $q$ \\
      \hline
      \ensuremath{u_{xAP1}} & $3.89 \times 10^{-2}$ & $8.5\%$ & $1.01$ & $1.00$ \\
      \ensuremath{u_{xAP2}} & $2.82 \times 10^{-2}$ & $27.0\%$ & $^{*~1, 2}$ & $2.00$ \\
      \ensuremath{u_{xAP3}} & $3.09 \times 10^{-2}$ & $17.7\%$ & $1.63$ & $1.62$ \\
      \ensuremath{u_{xAP4}} & $3.42 \times 10^{-2}$ & $0.1\%$ & $2.00$ & $1.99$ \\
      \ensuremath{u_{xAP5}} & $3.59 \times 10^{-2}$ & $17.8\%$ & $2.00$ & $2.00$ \\
      \ensuremath{u_{xAP6}} & $3.26 \times 10^{-2}$ & $6.5\%$ & $1.99$ & $1.98$ \\
      \ensuremath{u_{xAP7}} & $3.62 \times 10^{-2}$ & $18.1\%$ & $2.00$ & $2.00$ \\
      \ensuremath{u_{xAP8}} & $4.88 \times 10^{-2}$ & $141.8\%$ & $1.64$ & $0.72$ \\
      \ensuremath{u_{xAP9}} & $2.96 \times 10^{-2}$ & $17.9\%$ & $1.53$ & $1.52$ \\
      \ensuremath{u_{xAP10}} & $3.34 \times 10^{-2}$&  $6.5\%$ & $1.95$ & $1.94$ \\
      \ensuremath{u_{xAP11}} & $3.19 \times 10^{-2}$&  $9.7\%$ & $2.00$ & $1.99$ \\
      \ensuremath{u_{xAP12}} & $2.95 \times 10^{-2}$&  $29.5\%$ & $1.00$ & $0.99$ \\
      \hline
      Average & $3.42 \times 10^{-2}$ & $25.09$ & $1.65$ & $1.65$ \\
      \hline
      \multicolumn{5}{l}{\scriptsize $^{*~1, 2}$ Fit was made using first and
      second order exponents}\\
      \hline
    \end{tabular}
    \label{tab:uncertaintyVelocityXAmplitudesTable2}
    \caption{Results for all probes.}
  \end{subfigure}%
  \begin{subfigure}[b]{0.35\textwidth}
  \includegraphics[width=7cm]{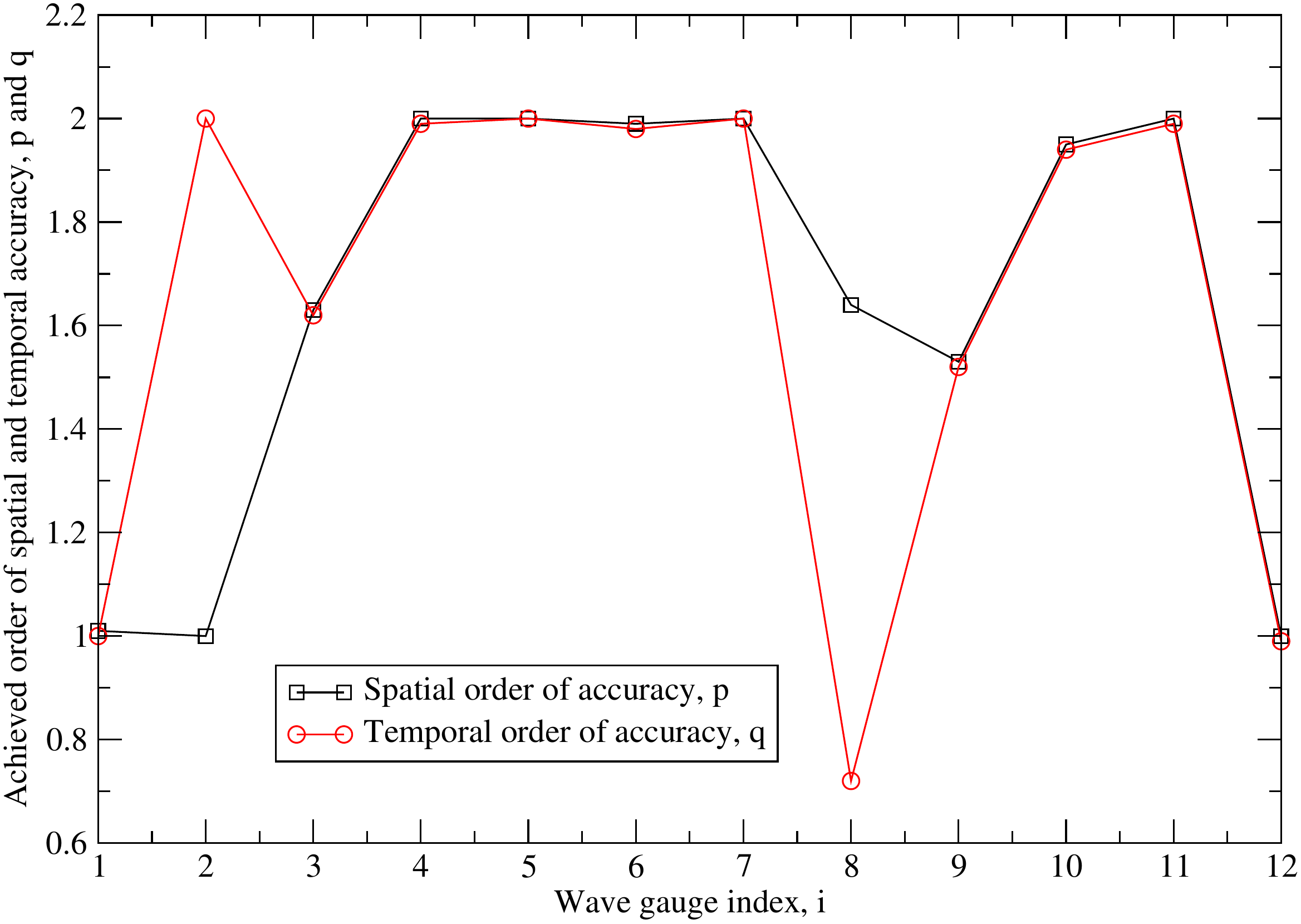}
  \caption{Achieved order of convergence for all probes.}
  \label{fig:velocityAmplitudesXAccuracy2}
  \end{subfigure}%
\end{center}
\caption{Uncertainty analysis for $x$ (horizontal) component of the velocity
field: second order amplitudes.}
\label{tab:uncertaintyVelocityXAmplitudes2}
\end{figure}

\begin{figure}[t!]
\begin{center}
  \begin{subfigure}[b]{0.45\textwidth}
    \begin{tabular}{l c c c c}
      \hline
      Item & $u_{x\theta}$, $^{\circ}$ & $U_{U_{x\theta}}$, $^{\circ}$ & $p$ & $q$ \\
      \hline
      \ensuremath{u_{x\theta P1}} & $130 $  & $61.4$ & $^{*~1, 2}$ & $2.00$ \\
      \ensuremath{u_{x\theta P2}} & $175 $  & $26.8$ & $2.00$ & $1.98$ \\
      \ensuremath{u_{x\theta P3}} & $177 $  & $3.4$ & $2.00$ & $1.99$ \\
      \ensuremath{u_{x\theta P4}} & $179 $  & $2.9$ & $2.00$ & $2.00$ \\
      \ensuremath{u_{x\theta P5}} & $192 $  & $29.3$ & $1.40$ & $1.39$ \\
      \ensuremath{u_{x\theta P6}} & $166 $  & $12.8$ & $2.00$ & $2.00$ \\
      \ensuremath{u_{x\theta P7}} & $172 $  & $9.4$ & $2.00$ & $1.98$ \\
      \ensuremath{u_{x\theta P8}} & $171 $  & $16.7$ & $1.40$ & $1.38$ \\
      \ensuremath{u_{x\theta P9}} & $165 $  & $8.3$ & $2.00$ & $2.00$ \\
      \ensuremath{u_{x\theta P10}} & $197$ & $45.4$ & $1.20$ & $1.19$ \\
      \ensuremath{u_{x\theta P11}} & $193$ & $39.2$ & $1.75$ & $1.73$ \\
      \ensuremath{u_{x\theta P12}} & $200$ & $26.6$ & $0.84$ & $2.00$ \\
      \hline
      Average & $176$ & $23.5$ & $1.63$ & $1.80$ \\
      \hline
      \multicolumn{5}{l}{\scriptsize $^{*~1, 2}$ Fit was made using first and
      second order exponents}\\
      \hline
    \end{tabular}
    \label{tab:uncertaintyVelocityXPhaseTable2}
    \caption{Results for all probes.}
  \end{subfigure}%
  \begin{subfigure}[b]{0.35\textwidth}
  \includegraphics[width=7cm]{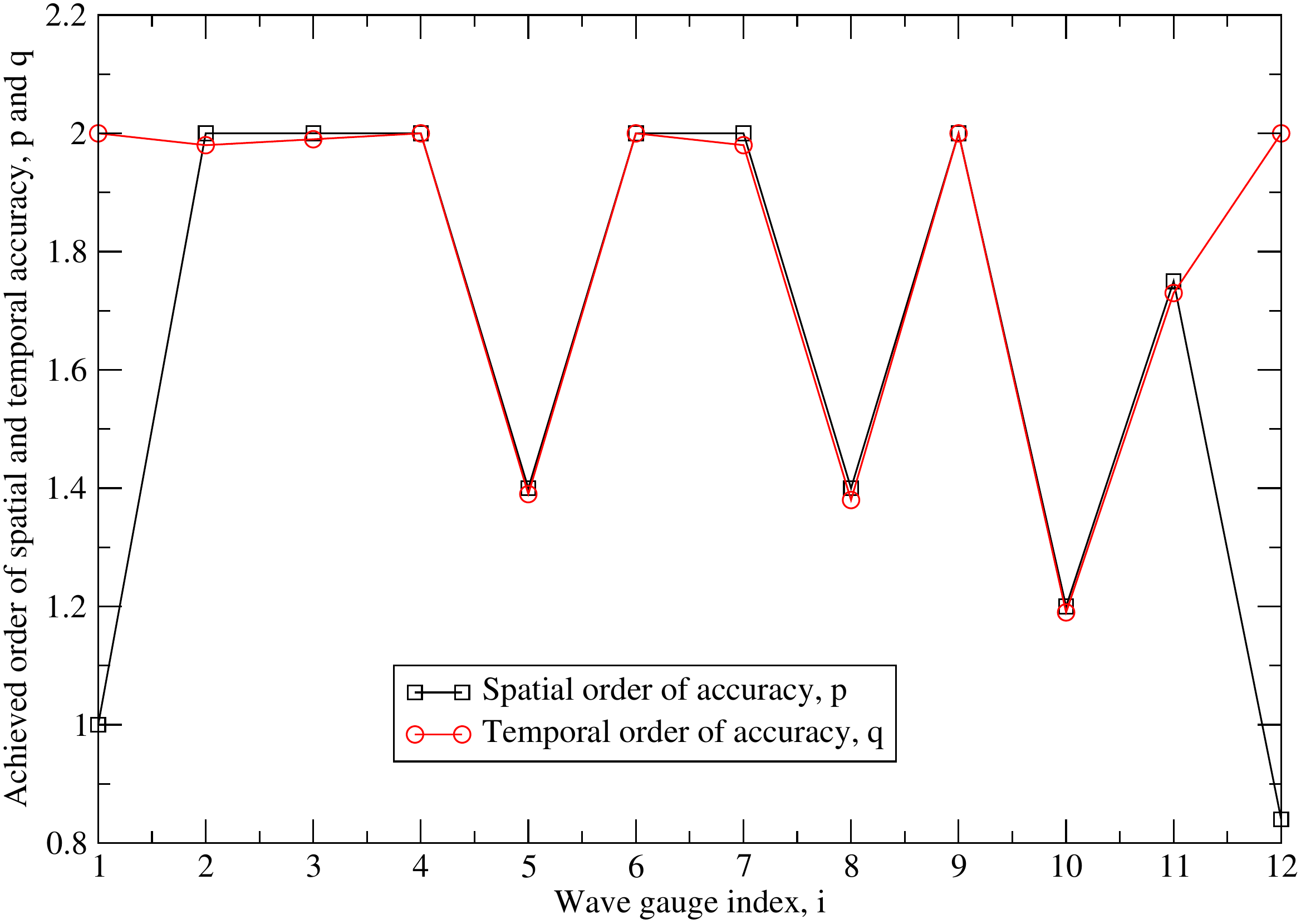}
  \caption{Achieved order of convergence for all probes.}
  \label{fig:velocityPhasesXAccuracy2}
  \end{subfigure}%
\end{center}
\caption{Uncertainty analysis for $x$ (horizontal) component of the velocity
field: second order phases.}
\label{tab:uncertaintyVelocityXPhases2}
\end{figure}

\indent \autoref{tab:uncertaintyVelocityXAmplitudes2} presents the results for
second order horizontal ($x$ component) velocity amplitudes, where the average
amplitude of 0.0342 m/s is obtained over all wave lengths, although with quite
irregular results for different wave length. The numerical uncertainty ranges
from 0.1\% to very high 141.8\%, which can be considered an outlier, increasing
the average numerical uncertainty to 25.1\%. Achieved orders of spatial and
temporal accuracy range from 0.72 to 2.00, with averages values of 1.65 for both
space and time.\\
\indent \autoref{tab:uncertaintyVelocityXPhases2} presents the results for
second order horizontal ($x$ component) velocity phases, which shows quite
irregular results for different wave gauges. The asymptotic solution ranges
from 130$^{\circ}$ for the first wave probe to 200$^{\circ}$ for the last probe.
On average, the guessed asymptotic solution is $176^{\circ}$, with the average
uncertainty of 13.1$^{\circ}$. The achieved order of spatial and temporal
accuracy is 1.63 and 1.80, respectively, on average for all velocity probes.\\

\begin{figure}[t!]
\begin{center}
  \begin{subfigure}[b]{0.45\textwidth}
    \begin{tabular}{l c c c c}
      \hline
      Item & $u_{yA}$, m/s & $U_{U_{yA}}\%$ & $p$ & $q$ \\
      \hline
      \ensuremath{u_{yAP1}} & $3.12 \times 10^{-2}$ & $9.2\%$ & $2.00$ & $1.98$ \\
      \ensuremath{u_{yAP2}} & $2.88 \times 10^{-2}$ & $17.5\%$ & $^{*~1, 2}$ & $2.00$ \\
      \ensuremath{u_{yAP3}} & $2.87 \times 10^{-2}$ & $15.4\%$ & $1.64$ & $1.63$ \\
      \ensuremath{u_{yAP4}} & $3.18 \times 10^{-2}$ & $0.4\%$ & $2.00$ & $1.99$ \\
      \ensuremath{u_{yAP5}} & $4.39 \times 10^{-2}$ & $76.2\%$ & $1.51$ & $0.60$ \\
      \ensuremath{u_{yAP6}} & $3.77 \times 10^{-2}$ & $52.9\%$ & $1.59$ & $0.69$ \\
      \ensuremath{u_{yAP7}} & $3.05 \times 10^{-2}$ & $4.5\%$ & $2.00$ & $1.99$ \\
      \ensuremath{u_{yAP8}} & $3.18 \times 10^{-2}$ & $1.0\%$ & $2.00$ & $1.99$ \\
      \ensuremath{u_{yAP9}} & $3.00 \times 10^{-2}$ & $19.3\%$ & $2.00$ & $1.99$ \\
      \ensuremath{u_{yAP10}} & $3.12 \times 10^{-2}$ & $4.5\%$ & $2.00$ & $1.99$ \\
      \ensuremath{u_{yAP11}} & $3.07 \times 10^{-2}$ & $6.1\%$ & $2.00$ & $1.99$ \\
      \ensuremath{u_{yAP12}} & $3.30 \times 10^{-2}$ & $3.9\%$ & $2.00$ & $1.98$ \\
      \hline
      Average & $3.24 \times 10^{-2}$ & $17.58$ & $1.82$ & $1.75$ \\
      \hline
      \multicolumn{5}{l}{\scriptsize $^{*~1, 2}$ Fit was made using first and
      second order exponents}\\
      \hline
    \end{tabular}
    \label{tab:uncertaintyVelocityYTable2}
    \caption{Results for all probes.}
  \end{subfigure}%
  \begin{subfigure}[b]{0.35\textwidth}
  \includegraphics[width=7cm]{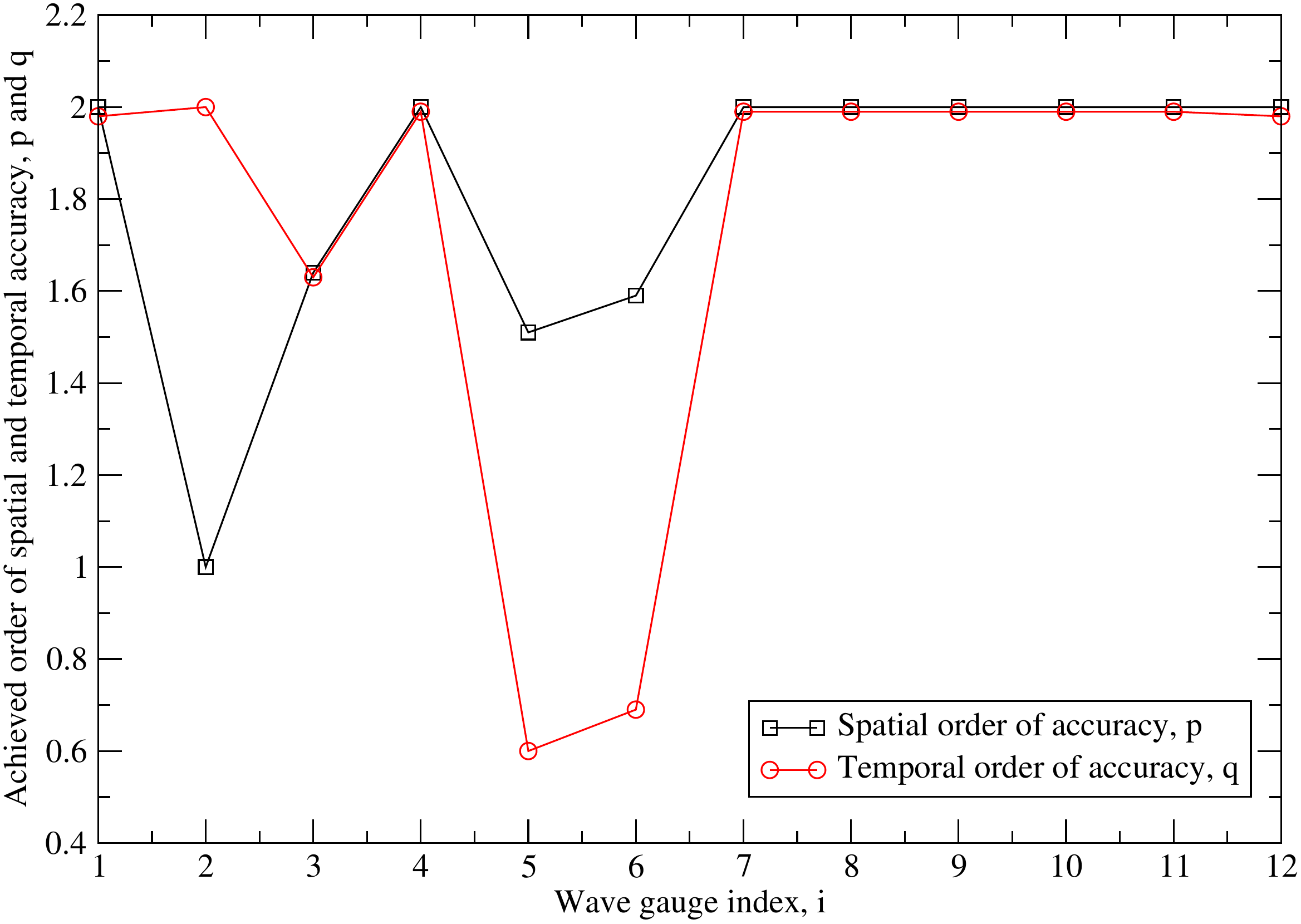}
  \caption{Achieved order of convergence for all probes.}
  \label{fig:velocityAmplitudesYAccuracy2}
  \end{subfigure}%
\end{center}
\caption{Uncertainty analysis for $y$ (vertical) component of the velocity
field: second order amplitudes.}
\label{tab:uncertaintyVelocityYAmplitudes2}
\end{figure}

\begin{figure}[t!]
\begin{center}
  \begin{subfigure}[b]{0.45\textwidth}
    \begin{tabular}{l c c c c}
      \hline
      Item & $u_{y\theta}$, $^{\circ}$ & $U_{U_{y\theta}}$, $^{\circ}$ & $p$ & $q$ \\
      \hline
      \ensuremath{u_{y\theta P1}} & $262 $ & $7.0$ & $^{*~1, 2}$ & $1.00$ \\
      \ensuremath{u_{y\theta P2}} & $264 $ & $18.9$ & $2.00$ & $1.98$ \\
      \ensuremath{u_{y\theta P3}} & $267 $ & $2.7$ & $2.00$ & $1.99$ \\
      \ensuremath{u_{y\theta P4}} & $269 $ & $2.3$ & $2.00$ & $2.00$ \\
      \ensuremath{u_{y\theta P5}} & $298 $ & $32.4$ & $1.07$ & $1.06$ \\
      \ensuremath{u_{y\theta P6}} & $256 $ & $7.9$ & $2.00$ & $2.00$ \\
      \ensuremath{u_{y\theta P7}} & $263 $ & $5.9$ & $2.00$ & $1.98$ \\
      \ensuremath{u_{y\theta P8}} & $262 $ & $10.4$ & $1.22$ & $1.21$ \\
      \ensuremath{u_{y\theta P9}} & $257 $ & $5.0$ & $2.00$ & $2.00$ \\
      \ensuremath{u_{y\theta P10}} & $306$ & $44.1$ & $0.99$ & $0.98$ \\
      \ensuremath{u_{y\theta P11}} & $283$ & $24.1$ & $1.73$ & $1.71$ \\
      \ensuremath{u_{y\theta P12}} & $300$ & $25.7$ & $2.00$ & $0.54$ \\
      \hline
      Average & $274$ & $15.6$ & $1.66$ & $1.54$ \\
      \hline
      \multicolumn{5}{l}{\scriptsize $^{*~1, 2}$ Fit was made using first and
      second order exponents}\\
      \hline
    \end{tabular}
    \label{tab:uncertaintyVelocityYPhaseTable2}
    \caption{Results for all probes.}
  \end{subfigure}%
  \begin{subfigure}[b]{0.35\textwidth}
  \includegraphics[width=7cm]{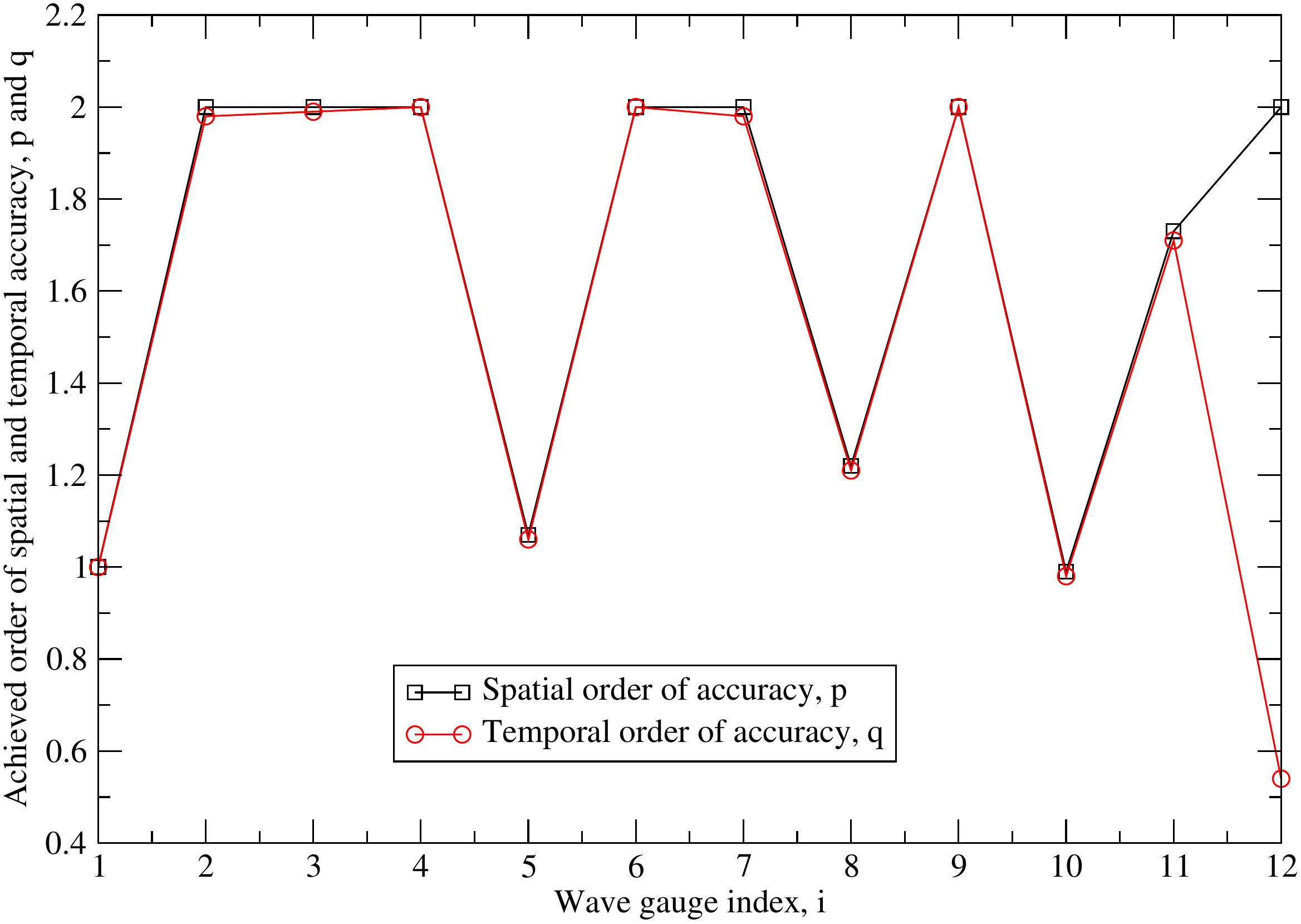}
  \caption{Achieved order of convergence for all probes.}
  \label{fig:velocityPhasesYAccuracy2}
  \end{subfigure}%
\end{center}
\caption{Uncertainty analysis for $y$ (horizontal) component of the velocity
field: second order phases.}
\label{tab:uncertaintyVelocityYPhases2}
\end{figure}

\indent \autoref{tab:uncertaintyVelocityYAmplitudes2} presents the results for
second order vertical ($y$ component) velocity amplitudes. Similarly to second
order horizontal velocity amplitudes, the asymptotic solutions are quite
irregular for different wave probes, with the average value of 0.0324 m/s. The
numerical uncertainty is 17.58\% on average, where the largest uncertainty is
obtained for velocity probe 5 with 76.2\%, where a low order of temporal
accuracy is achieved, $q = 0.60$. Averaged over all probes, achieved orders of
spatial and temporal accuracy are 1.82 and 1.75, respectively.\\
\indent \autoref{tab:uncertaintyVelocityYPhases2} presents the results for
second order vertical ($y$ component) velocity phases. The guessed asymptotic
solution for the phases varies from 262$^{\circ}$ to 306$^{\circ}$, with the
average value of 274$^{\circ}$. The corresponding numerical uncertainty
ranges from 2.7$^{\circ}$ to 44.1$^{\circ}$ obtained at velocity probe 10.
The average numerical uncertainty is 15.6$^{\circ}$, while the achieved orders
of spatial and temporal accuracy are 1.66 and 1.54, respectively, averaged over
all probes.\\

\subsubsection{Validation study for the progressive wave}
\label{sec:validation}

\begin{figure}[!p]
\begin{center}
  \begin{subfigure}[b]{0.45\textwidth}
  \includegraphics[width=\textwidth]{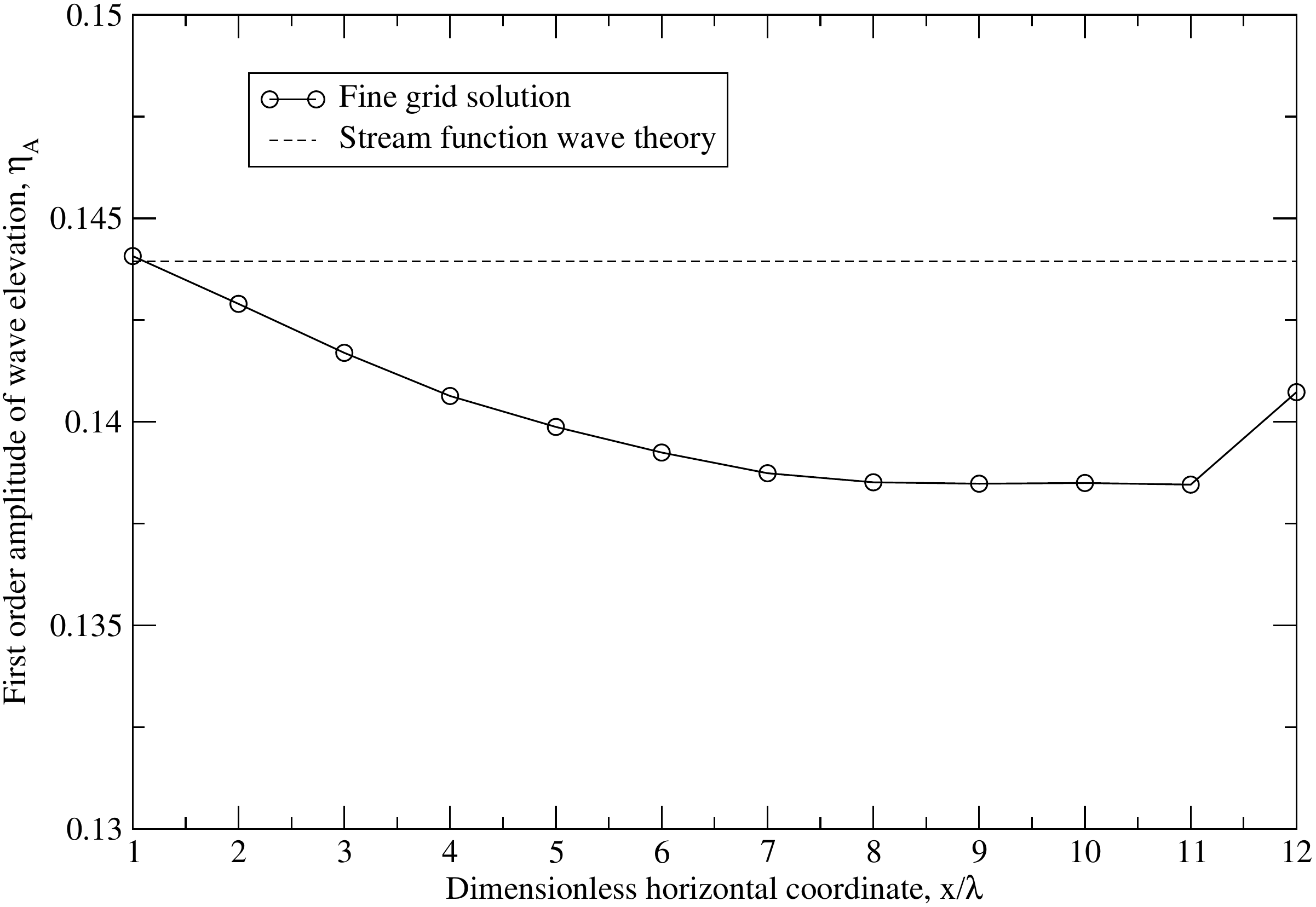}
  \caption{First order amplitudes.}
  \label{fig:etaAmplitudesValidation}
  \end{subfigure}%
  \begin{subfigure}[b]{0.45\textwidth}
  \includegraphics[width=\textwidth]{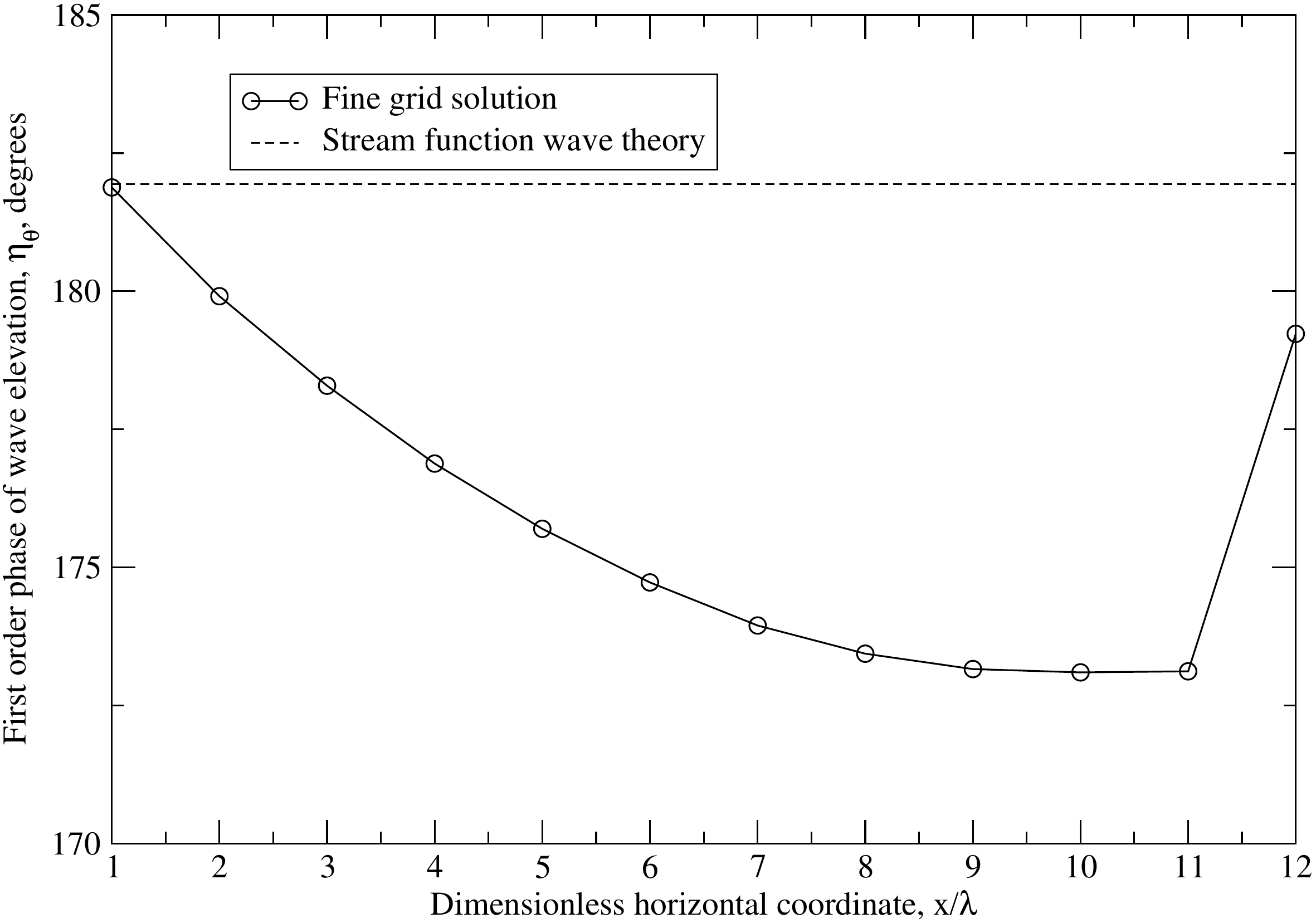}
  \caption{First order phases.}
  \label{fig:etaPhasesValidation}
  \end{subfigure}%
\end{center}
\caption{Comparison of fine grid (Grid 6) first order wave elevations with the
stream function wave theory~\cite{rieneckerFenton1981} for all wave gauges.}
\label{fig:etaValidation}
\end{figure}

\noindent The remainder of the analysis is dedicated to validation by comparing
the results obtained with the present model with fully nonlinear, potential flow
based stream function wave
theory~\cite{rieneckerFenton1981}.~\autoref{fig:etaValidation} presents the
comparison of first order wave elevations obtained with the fine grid (Grid 6)
for all wave gauges ranging from $x = 1\lambda$ to $x = 12\lambda$.\\
%
%

\begin{figure}[!p]
\begin{center}
  \begin{subfigure}[b]{0.45\textwidth}
  \includegraphics[width=\textwidth]{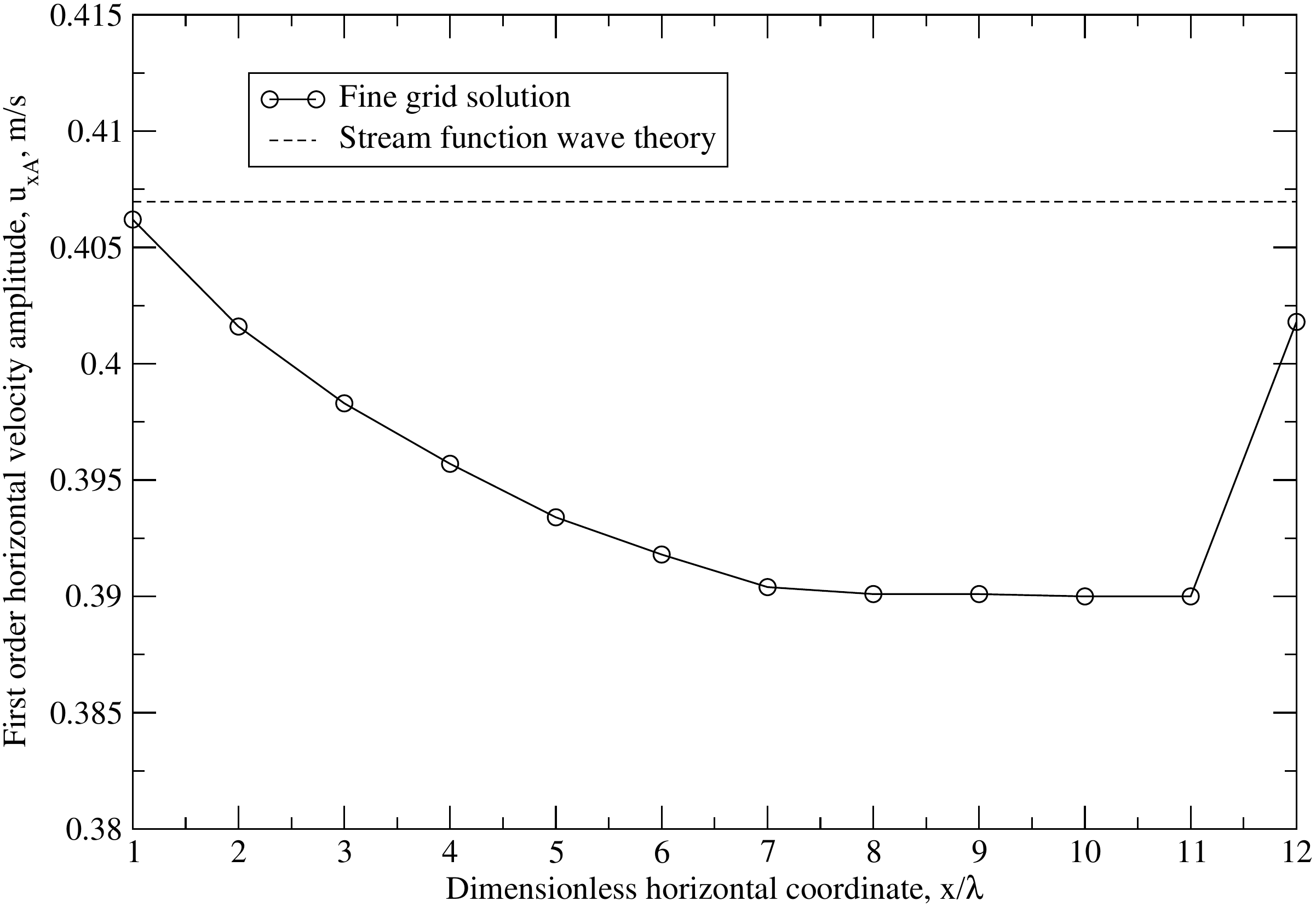}
  \caption{First order amplitudes.}
  \label{fig:UxAmplitudesValidation}
  \end{subfigure}%
  \begin{subfigure}[b]{0.45\textwidth}
  \includegraphics[width=\textwidth]{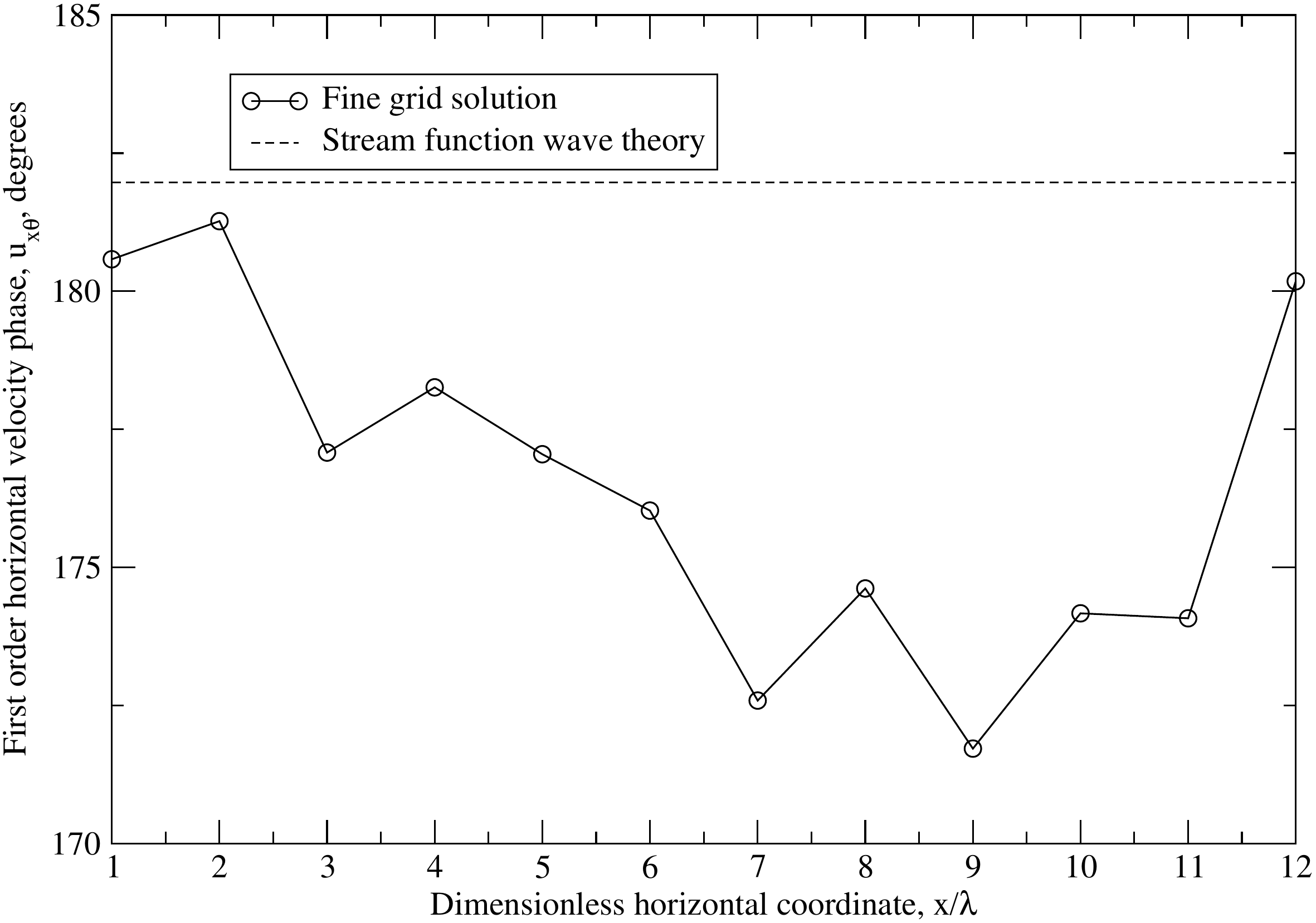}
  \caption{First order phases.}
  \label{fig:UxPhasesValidation}
  \end{subfigure}%
\end{center}
\caption{Comparison of fine grid (Grid 6) first order horizontal velocities with
the stream function wave theory~\cite{rieneckerFenton1981} for all probes.}
\label{fig:UxValidation}
\end{figure}

\begin{figure}[!p]
\begin{center}
  \begin{subfigure}[b]{0.45\textwidth}
  \includegraphics[width=\textwidth]{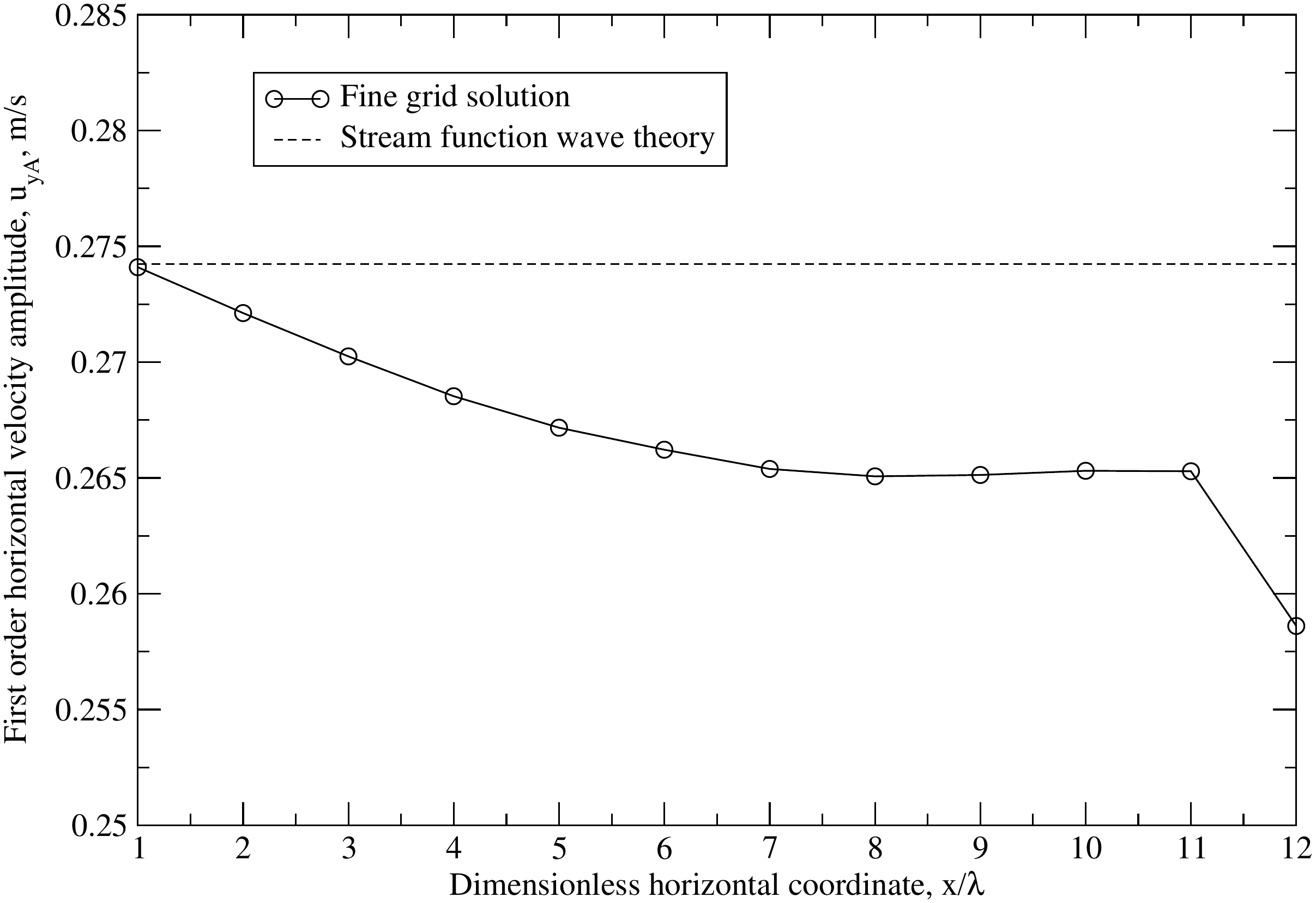}
  \caption{First order amplitudes.}
  \label{fig:UyAmplitudesValidation}
  \end{subfigure}%
  \begin{subfigure}[b]{0.45\textwidth}
  \includegraphics[width=\textwidth]{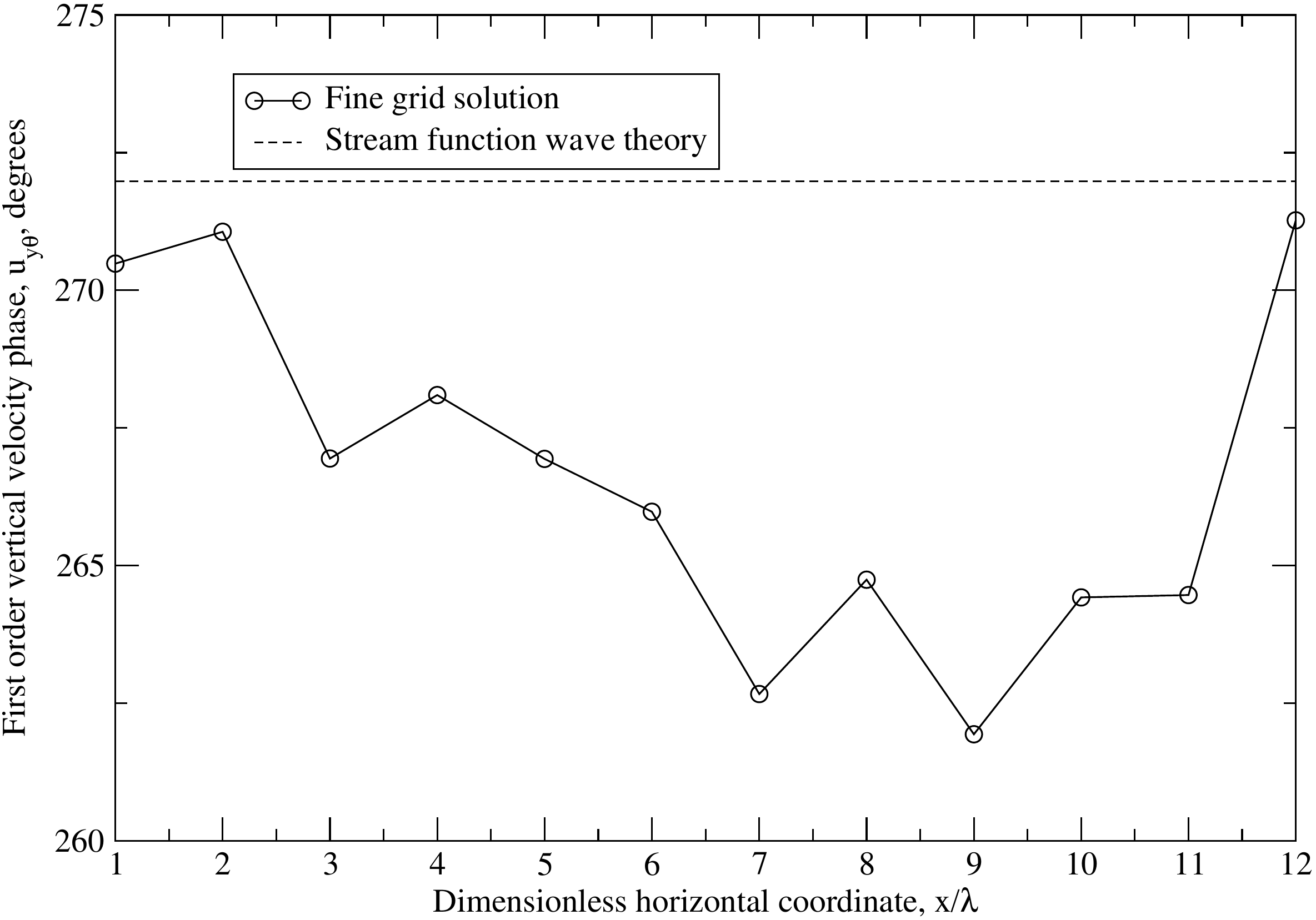}
  \caption{First order phases.}
  \label{fig:UyPhasesValidation}
  \end{subfigure}%
\end{center}
\caption{Comparison of fine grid (Grid 6) first order vertical velocities with
the stream function wave theory~\cite{rieneckerFenton1981} for all probes.}
\label{fig:UyValidation}
\end{figure}

\begin{figure}[!p]
\begin{center}
  \begin{subfigure}[b]{0.45\textwidth}
  \includegraphics[width=\textwidth]{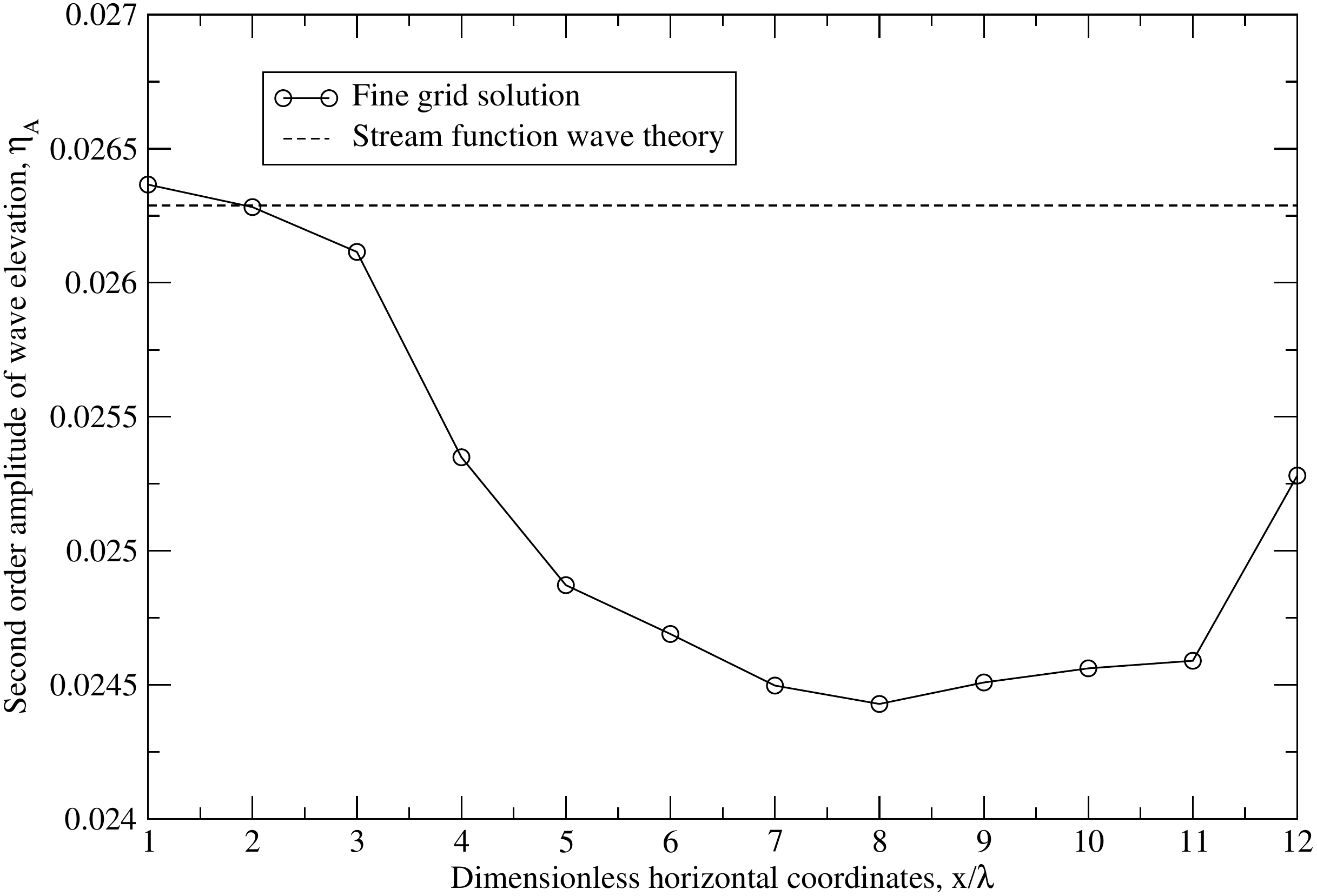}
  \caption{Second order amplitudes.}
  \label{fig:etaAmplitudes2Validation}
  \end{subfigure}%
  \begin{subfigure}[b]{0.45\textwidth}
  \includegraphics[width=\textwidth]{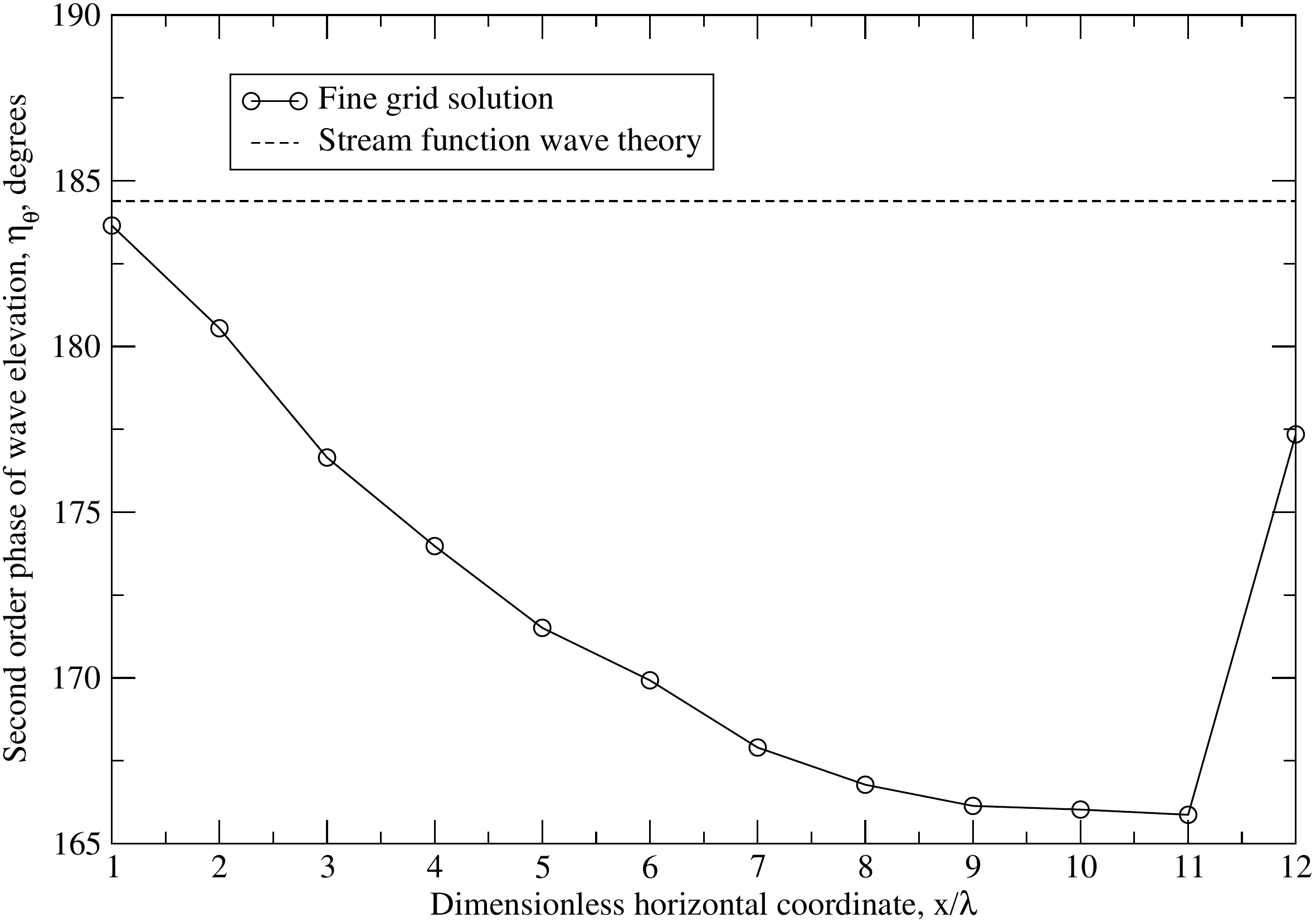}
  \caption{Second order phases.}
  \label{fig:etaPhases2Validation}
  \end{subfigure}%
\end{center}
\caption{Comparison of fine grid (Grid 6) second order wave elevations with the
stream function wave theory~\cite{rieneckerFenton1981} for all wave gauges.}
\label{fig:eta2Validation}
\end{figure}

\begin{figure}[!p]
\begin{center}
  \begin{subfigure}[b]{0.45\textwidth}
  \includegraphics[width=\textwidth]{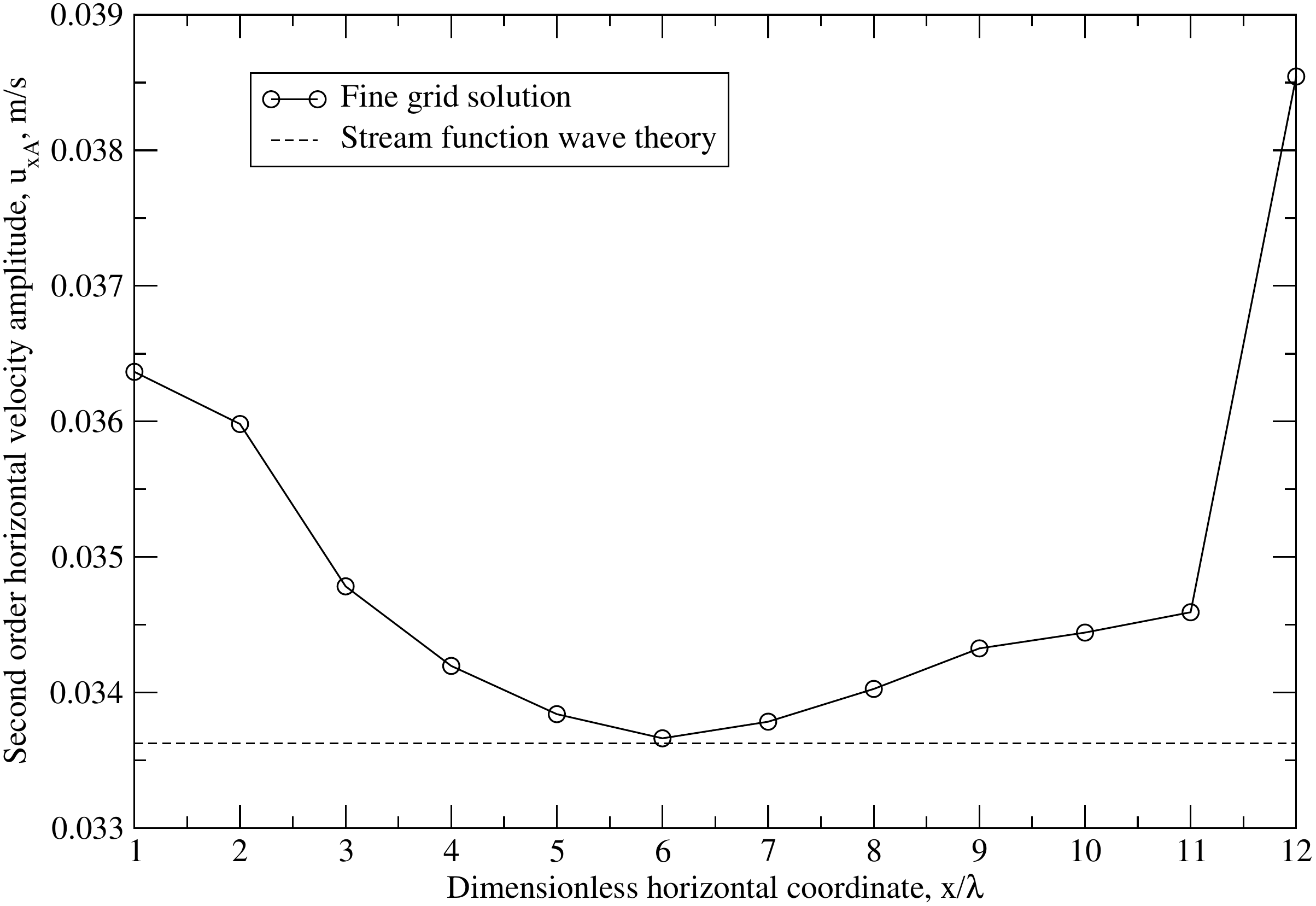}
  \caption{Second order amplitudes.}
  \label{fig:UxAmplitudes2Validation}
  \end{subfigure}%
  \begin{subfigure}[b]{0.45\textwidth}
  \includegraphics[width=\textwidth]{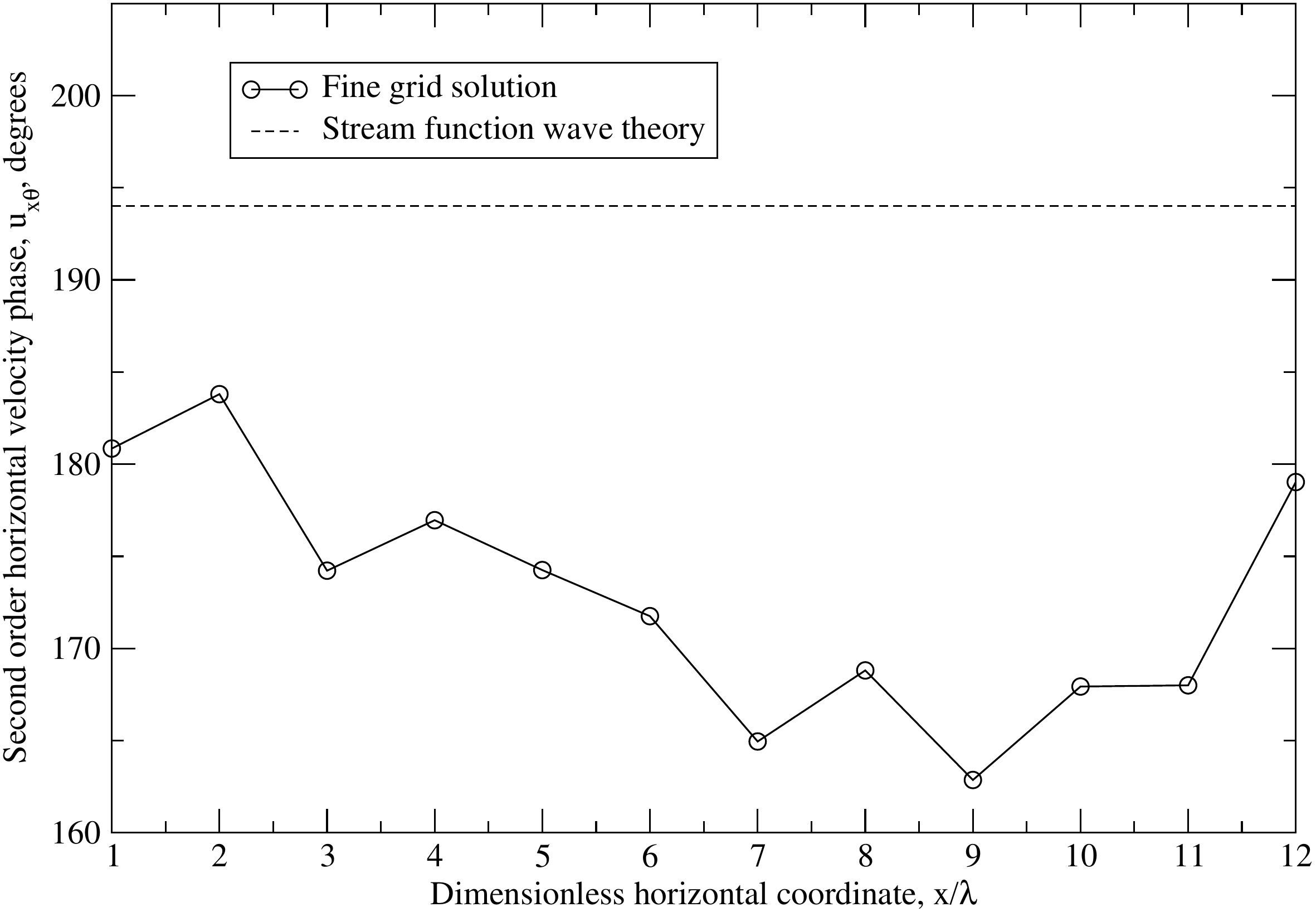}
  \caption{Second order phases.}
  \label{fig:UxPhases2Validation}
  \end{subfigure}%
\end{center}
\caption{Comparison of fine grid (Grid 6) second order horizontal velocities
with the stream function wave theory~\cite{rieneckerFenton1981} for all probes.}
\label{fig:Ux2Validation}
\end{figure}

\begin{figure}[!p]
\begin{center}
  \begin{subfigure}[b]{0.45\textwidth}
  \includegraphics[width=\textwidth]{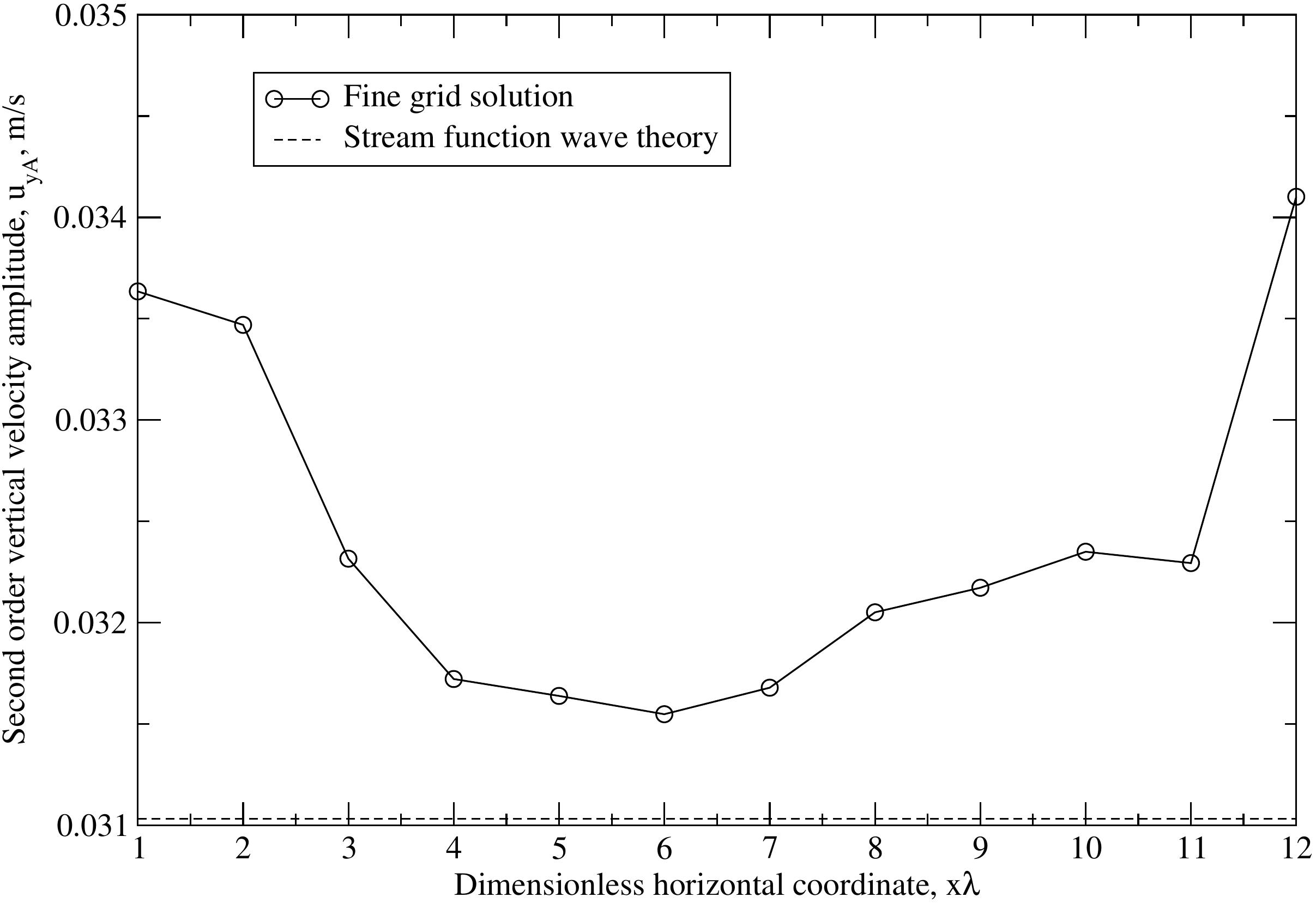}
  \caption{Second order amplitudes.}
  \label{fig:UyAmplitudes2Validation}
  \end{subfigure}%
  \begin{subfigure}[b]{0.45\textwidth}
  \includegraphics[width=\textwidth]{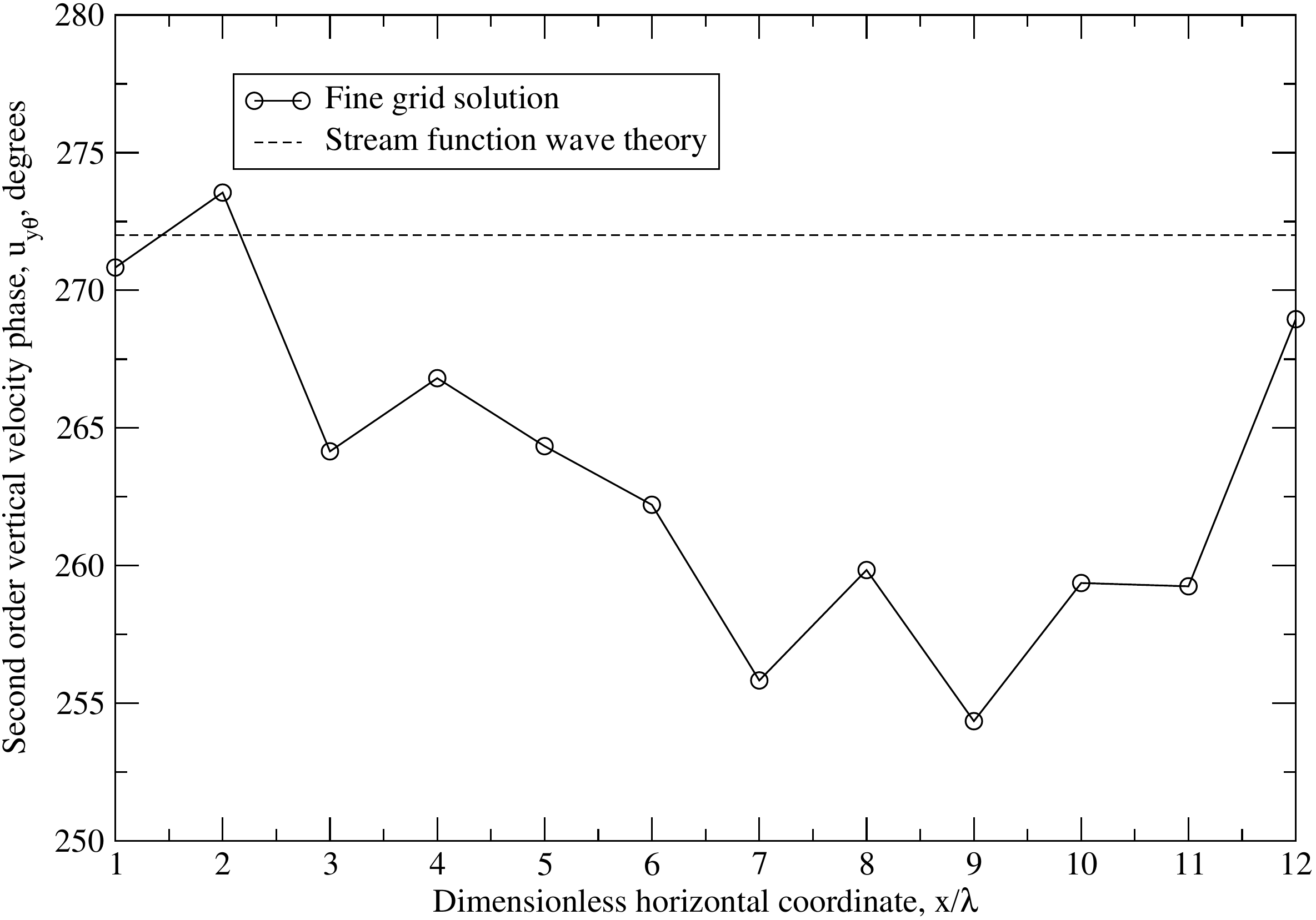}
  \caption{Second order phases.}
  \label{fig:UyPhases2Validation}
  \end{subfigure}%
\end{center}
\caption{Comparison of fine grid (Grid 6) second order vertical velocities with
the stream function wave theory~\cite{rieneckerFenton1981} for all probes.}
\label{fig:Uy2Validation}
\end{figure}

\noindent {\bfseries Validation study for first order effects.}
The dissipation error associated with~\autoref{fig:etaAmplitudesValidation}
increases downstream of the numerical wave tank and stabilises at approximately
$3.81\%$ at $x = 8\lambda$. The dissipation error is lowered at the last wave
gauge since the relaxation zone starts to force the solution towards the target
solution as discussed in~\autoref{sec:numericalModel}. The dispersion error
presented in~\autoref{fig:etaPhasesValidation} shows similar trend, with the
largest difference being slightly less than $9^{\circ}$.\\
\indent~\autoref{fig:UxValidation} presents the comparison of first order
horizontal velocities obtained with the fine grid for all probes. Similarly to
wave elevation, the dissipation causes the fluid to lose linear momentum
downstream of the inlet. The error stabilises at approximately $x = 7\lambda$
at $4.2\%$ before it starts to tend to the stream function solution within the
relaxation zone at $x = 12\lambda$. The phase difference presented
in~\autoref{fig:UxPhasesValidation} shows increasing trend downstream of the
inlet, although with irregular and oscillatory behaviour. The largest
dissipation error is found at $x = 9\lambda$ with approximately $10.3^{\circ}$,
which is comparable to $9^{\circ}$ for wave elevation phase.\\
\indent~\autoref{fig:UyValidation} presents the comparison of first order
vertical velocities obtained with the fine grid for all probes. The trend for
the vertical velocity is similar to the one obtained for the horizontal
velocity. The dissipation error reaches a local maximum of approximately
$3.3\%$ at $x = 8\lambda$, suddenly growing to $5.7\%$ within the start of the
outlet relaxation zone at $x = 12\lambda$. This trend is different than the one
observed for horizontal velocity component and wave elevation and needs to be
investigated further. The dispersion of the vertical velocity field as presented
in~\autoref{fig:UyPhasesValidation} has exactly the same irregular trend as the
dispersion of the horizontal velocity field. The maximum dispersion error is
found at $x = 9\lambda$ and the difference in phase is approximately
$10^{\circ}$ (compared to $10.3^{\circ}$ for the horizontal velocity).\\
%
%
%
\noindent{\bfseries Validation study for second order effects.}
Compared to the first order wave elevation, the second order wave elevation exhibits
a similar trend of increasing dissipation and dispersion errors downstream of the
wave tank, as seen in~\autoref{fig:eta2Validation}.
\autoref{fig:etaAmplitudes2Validation} shows that the relative error of
the second order wave amplitude increases up to approximately 7\% at $x = 8\lambda$,
while it is less than 1\% until $x = 4\lambda$. The dispersion error grows
continuously up to $18^{\circ}$ at $x = 9\lambda$, where it stabilises.
Similarly to first order effects, the solution starts to be gradually forced
to the stream function solution in the last wave gauge at $x = 12\lambda$ due to
the relaxation zone.\\
\indent \autoref{fig:UxAmplitudes2Validation} shows the second order
amplitude of the horizontal velocity component, where it is interesting to note that
the solution obtained with the present model over--predicts the solution
obtained with stream function wave theory. The maximum differences compared to
stream function wave theory are obtained for probes that are within relaxation
zones $x = \lambda$ and $x = 12\lambda$. For probes not affected by relaxation
zones ($x = 2\lambda$ to $x = 11\lambda$), the maximum difference is
approximately 7\%. The comparison of phases for the second order horizontal
velocity is presented in~\autoref{fig:UxPhases2Validation}. As for the first
order phases, the behaviour is quite irregular for different longitudinal
locations, where the phase calculated with the present model lags behind from
$9^{\circ}$ at $x = 2\lambda$ to $32^{\circ}$ at $x = 9\lambda$.\\
\indent The second order of vertical velocity presented
in~\autoref{fig:Uy2Validation} shows similar behaviour as second order of
horizontal velocity. The amplitude is over--predicted compared to stream
function wave theory, as shown in~\autoref{fig:UyAmplitudes2Validation}, where
the maximum difference is approximately 10\% obtained for probe at $x =
12\lambda$, while the minimum difference is 2\% for probe at $x = 6\lambda$. The
second order phases show irregular behaviour on different longitudinal
locations, as shown in~\autoref{fig:UyPhases2Validation}. The difference ranges
from less than $2^{\circ}$ for probe $x = 2\lambda$ to approximately
$18^{\circ}$ for $x = 9\lambda$.\\

%

\begin{figure}[!h]
\begin{center}
  \begin{subfigure}[b]{0.4\textwidth}
  \includegraphics[width=\textwidth]{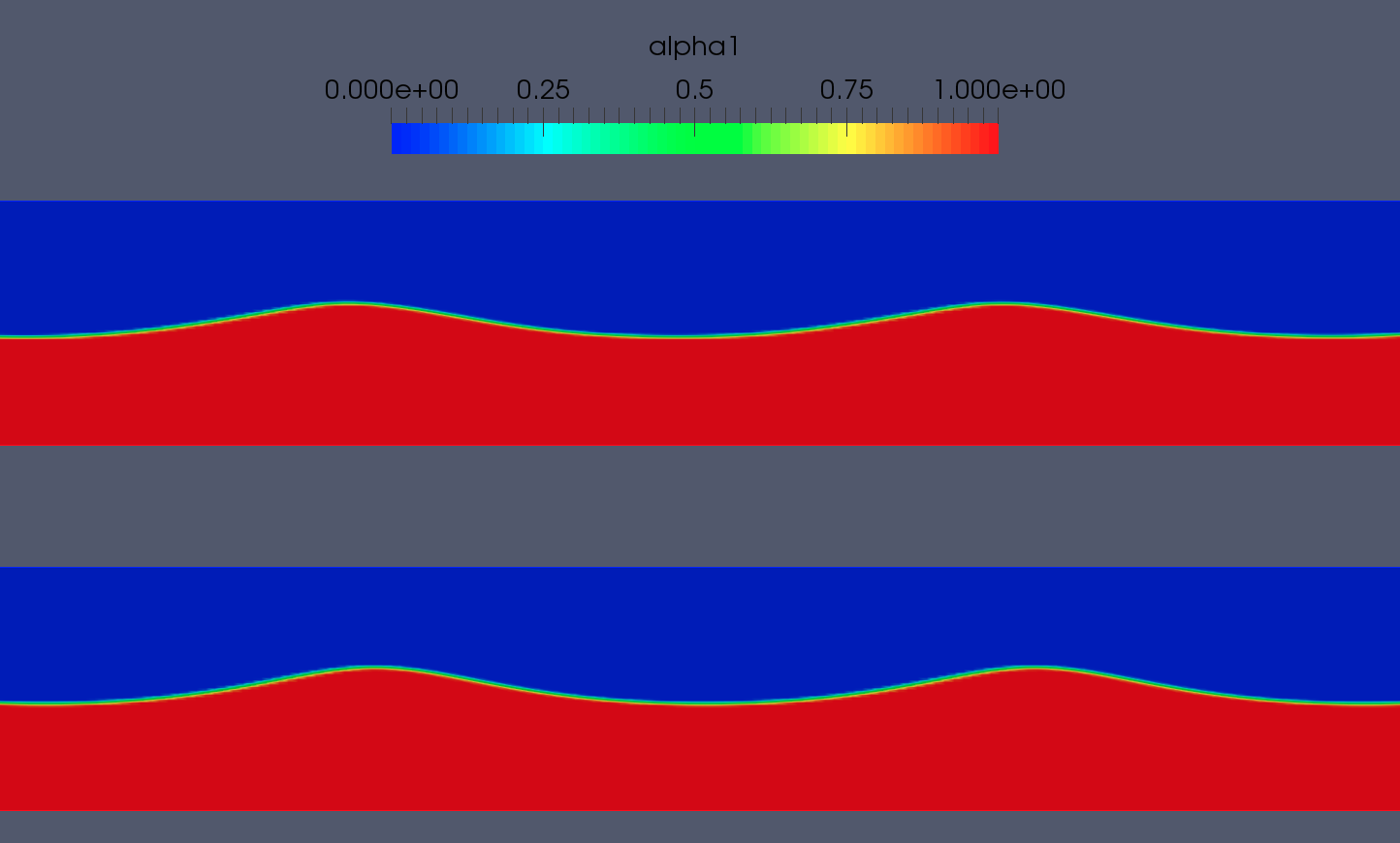}
  \caption{Volume fraction field.}
  \label{fig:alphaFieldComparison}
  \end{subfigure}%
  \begin{subfigure}[b]{0.4\textwidth}
  \includegraphics[width=0.985\textwidth]{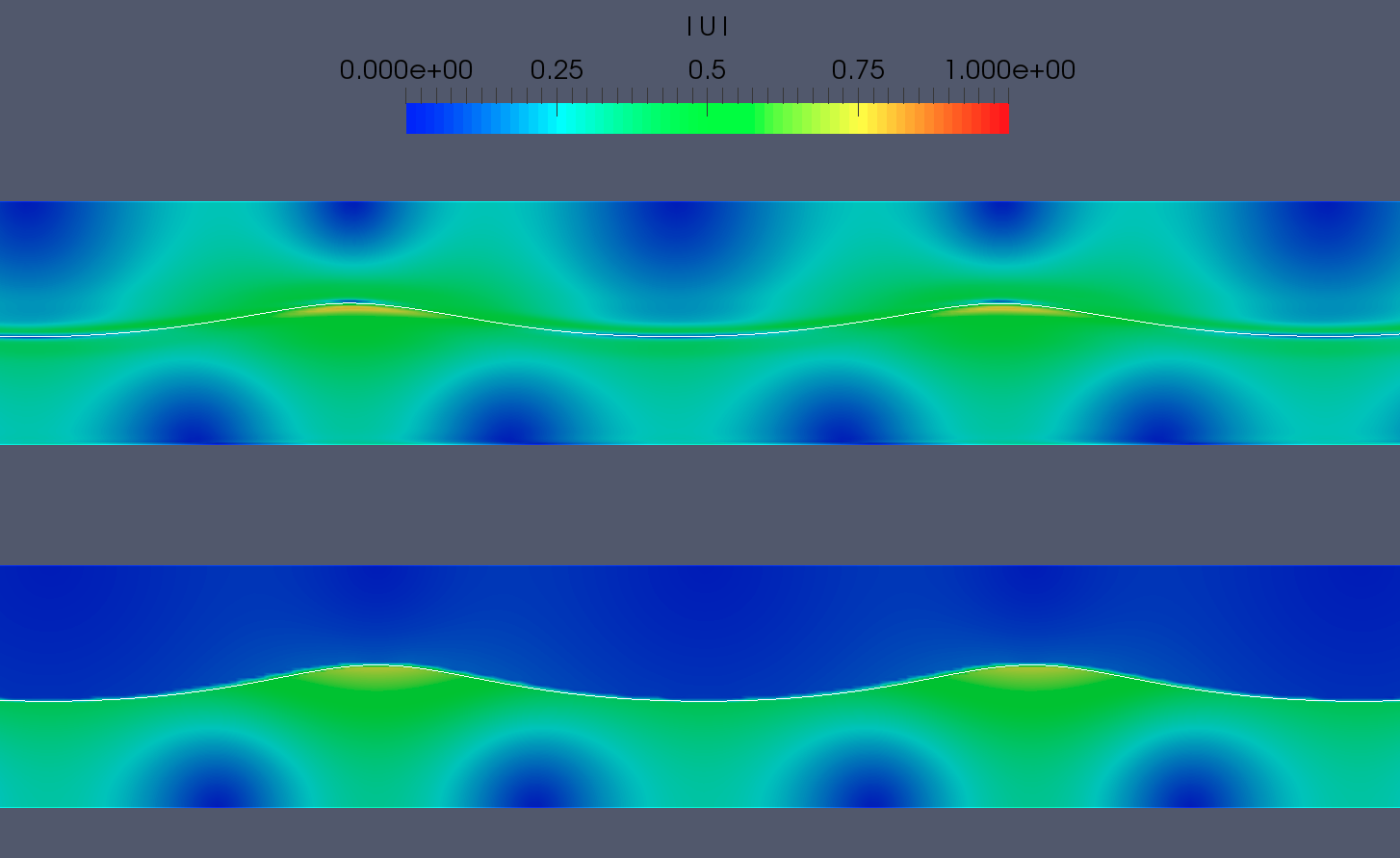}
  \caption{Velocity field.}
  \label{fig:velocityFieldComparison}
  \end{subfigure}%
\end{center}
\caption{Volume fraction and velocity fields obtained with the present model (upper
part) compared to stream function wave theory (lower part) in part of the domain
between $x \approx 5.5\lambda$ and $x \approx 7.5\lambda$ at $t = 10T$.}
\label{fig:alphaAndUComparison}
\end{figure}

\noindent {\bfseries Field data comparison.}
In addition to detailed data obtained with wave gauges and probes, we present a
qualitative comparison of volume fraction and velocity fields
in~\autoref{fig:alphaAndUComparison} for a part of the domain between $x \approx
5.5\lambda$ and $x \approx 7.5\lambda$ at $t = 10T$ (end of simulation). Top
figures present the solution obtained with the present model, while lower figures
present the solution obtained using the stream function wave theory.
\autoref{fig:alphaFieldComparison} shows that the free surface is well preserved
without numerical diffusion of the volume fraction field $\alpha$. In the stream
function solution, the velocity field is defined only up to the free
surface,\autoref{fig:velocityFieldComparison}. The present model takes
into account two--phase effects, thus producing an asymmetric velocity field
above the free surface. The velocity field in the water compares well with the
stream function wave theory.\\

\begin{figure}[!t]
\begin{center}
  \begin{subfigure}[b]{0.4\textwidth}
  \includegraphics[width=\textwidth]{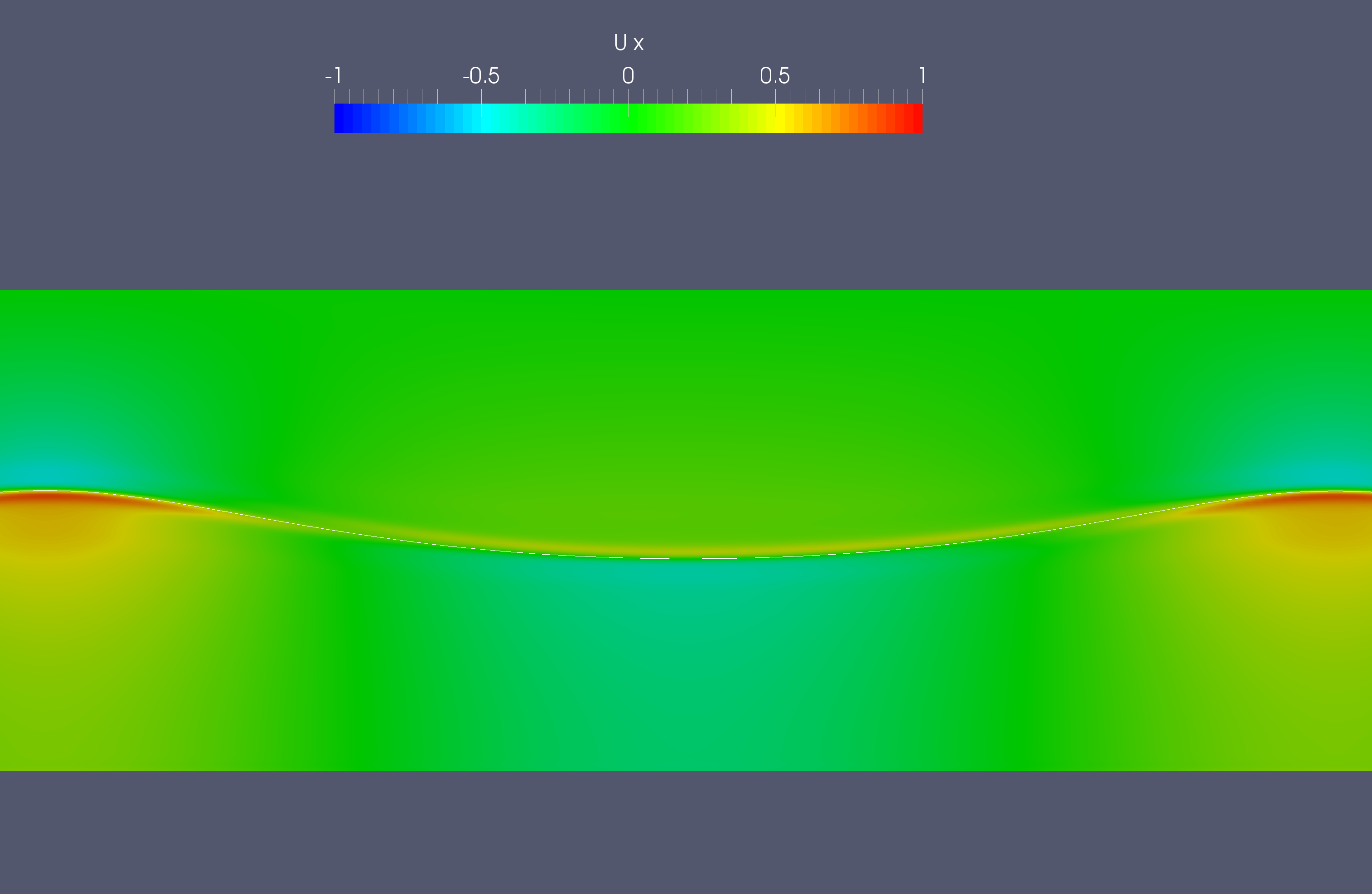}
  \caption{Horizontal component of the velocity field, $u_x$,}
  \label{fig:velocityFieldX}
  \end{subfigure}%
  \begin{subfigure}[b]{0.4\textwidth}
  \includegraphics[width=\textwidth]{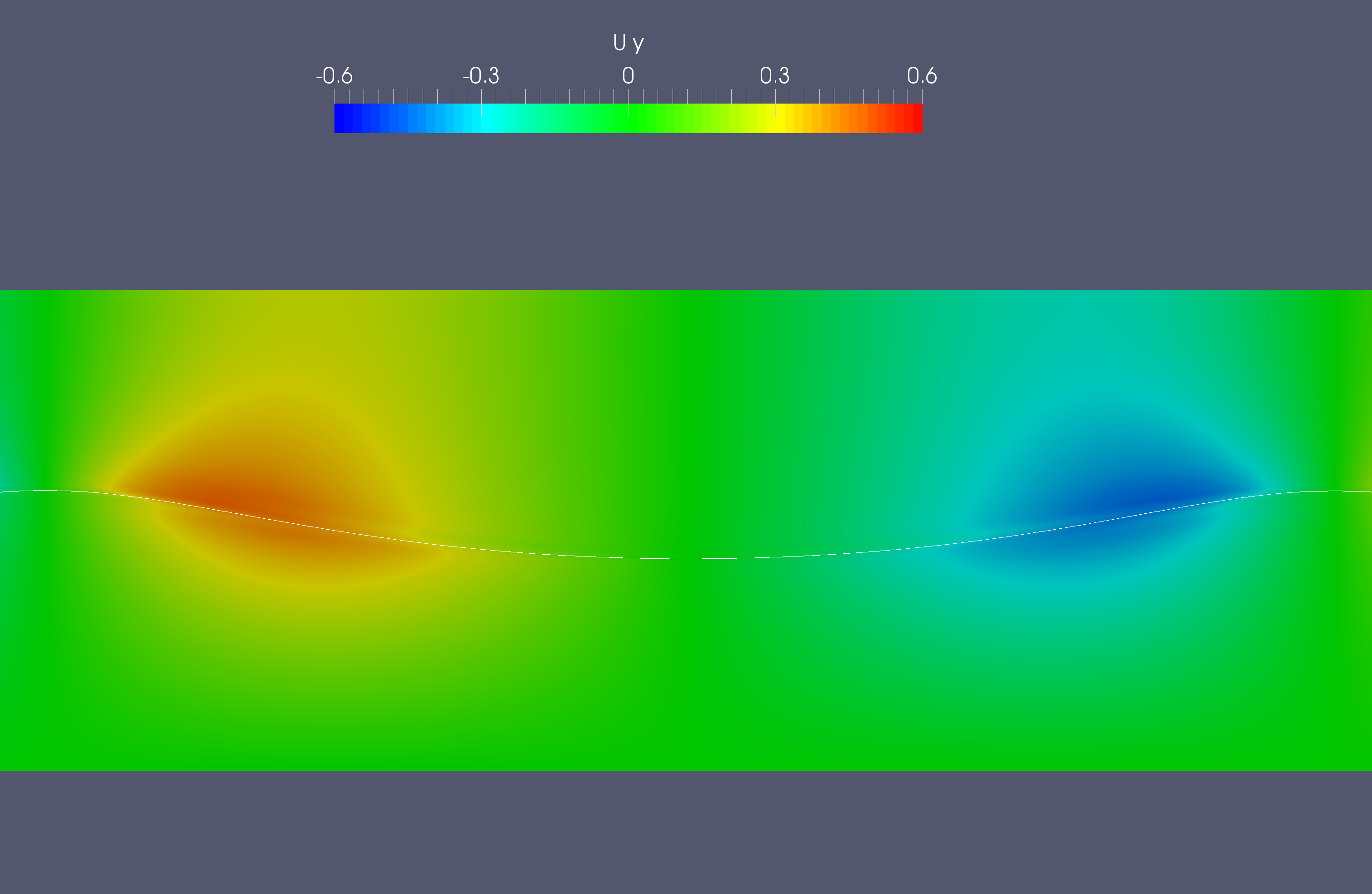}
  \caption{Vertical component of the velocity field, $u_y$,}
  \label{fig:velocityFieldY}
  \end{subfigure}%
  
  \begin{subfigure}[b]{0.4\textwidth}
  \includegraphics[width=\textwidth]{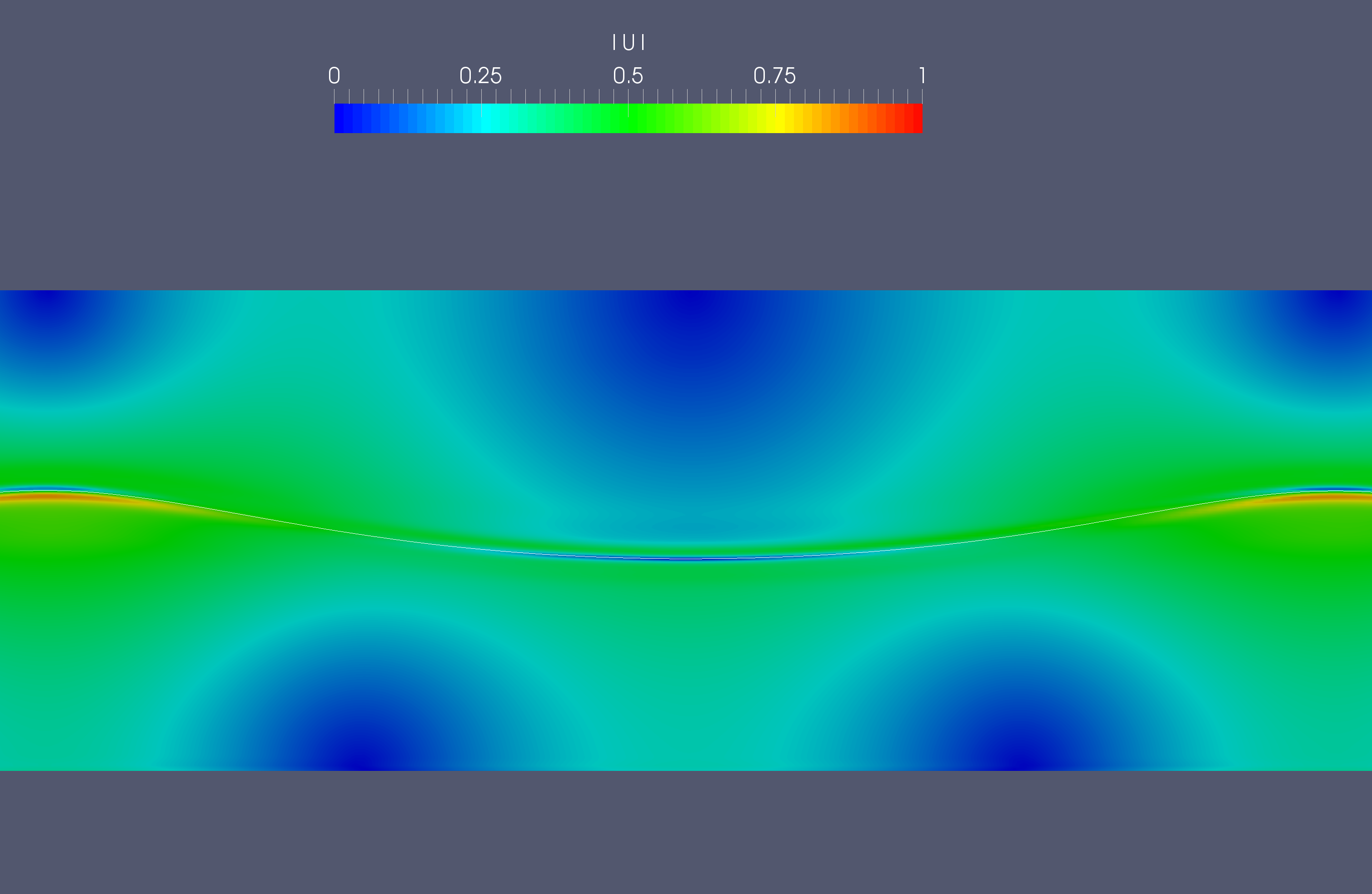}
  \caption{Velocity field magnitude, $|\U|$,}
  \label{fig:velocityFieldMagnitude}
  \end{subfigure}%
  \begin{subfigure}[b]{0.4\textwidth}
  \includegraphics[width=\textwidth]{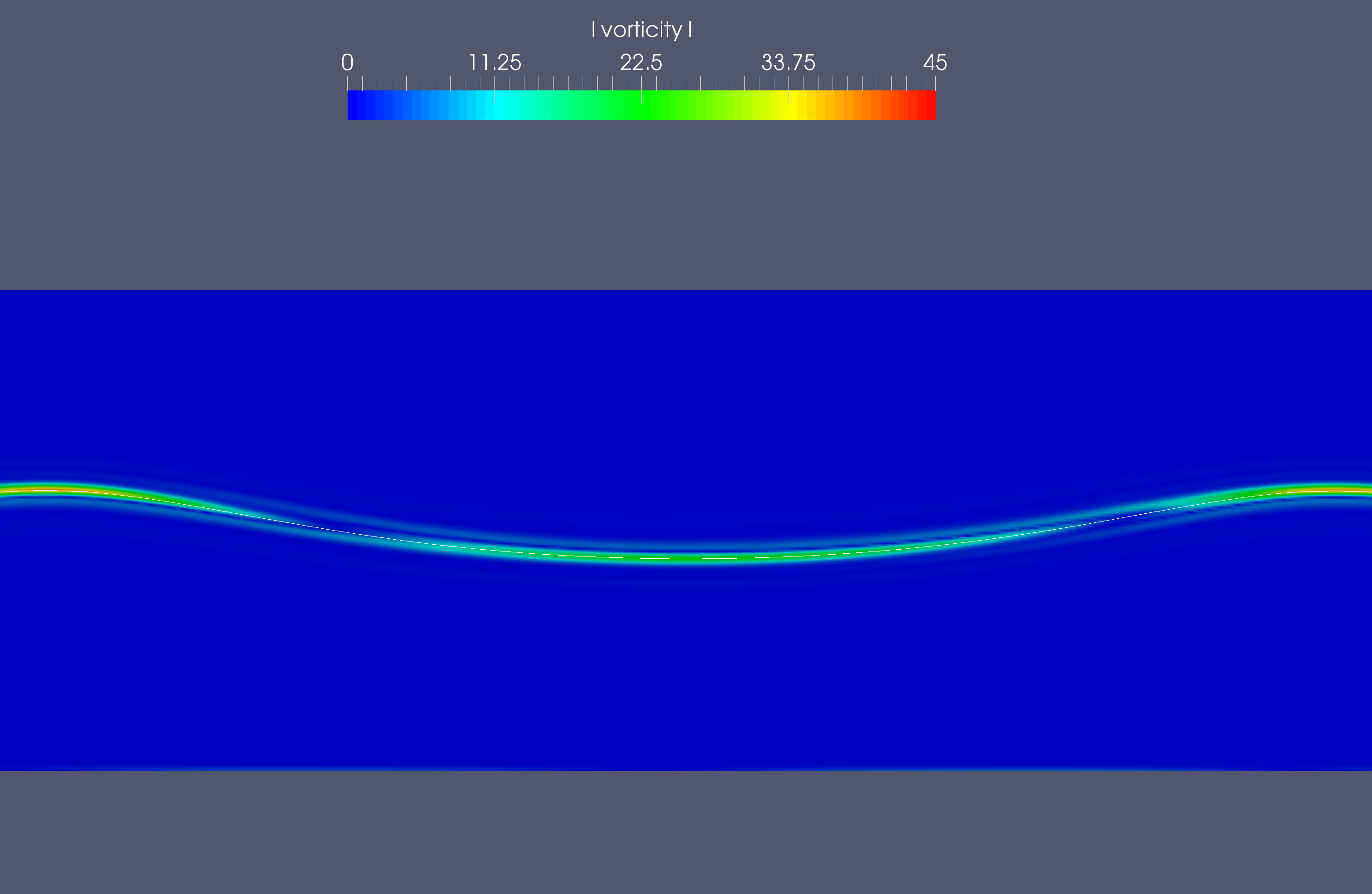}
  \caption{Vorticity field magnitude, $|\vec{\omega}|$}
  \label{fig:vorticityFieldMagnitude}
  \end{subfigure}%
\end{center}
\caption{Velocity and vorticity fields in part of the domain between $x =
\approx 6\lambda$ and $x = \approx 7\lambda$ at $t = 10T$.}
\label{fig:velocityAndVorticity}
\end{figure}

\noindent {\bfseries Note on two--phase effects and vorticity.}
It is important to clearly state the differences between the two computational
models compared within this study. Although the stream function wave theory
accounts for nonlinearity in the potential flow context, it does not take into
account two--phase, vorticity and viscosity effects. The present model takes
into account all of these effects and can predict the velocity in the air phase
as well, accounting for relevant jump conditions at the free
surface.~\autoref{fig:velocityAndVorticity} presents the velocity field obtained
with the present model in the part of the domain between $x \approx 6\lambda$ and $x
\approx 7\lambda$ at $t = 10T$, using the fine grid.  The horizontal velocity
component shown in~\autoref{fig:velocityFieldX} changes sign across the free
surface, \eg below the wave crest, the horizontal velocity is positive (in the
direction of wave propagation), while above the crest, it is negative (in the
direction opposite of wave propagation). Along with the vertical velocity field
presented in~\autoref{fig:velocityFieldY}, this leads to significant vorticity
effects near the free surface, as presented
in~\autoref{fig:vorticityFieldMagnitude}. Note that the vorticity field defined
as $\vec{\omega} = \curl \U$ vanishes far from the free surface, where the
potential flow model is therefore justified. Further investigation of the
combined vorticity and viscosity effects on propagation of steep waves shall be
the topic for future work.

\subsection{Wave breaking on a simplified ship model}
\label{sec:shipModel}

\begin{figure}[h!]
\begin{center}
  \begin{subfigure}[!h]{\textwidth}
  \centering
  \includegraphics[clip=true, trim = 2cm 17.7cm 10cm 3.1cm, width =0.65\textwidth]
      {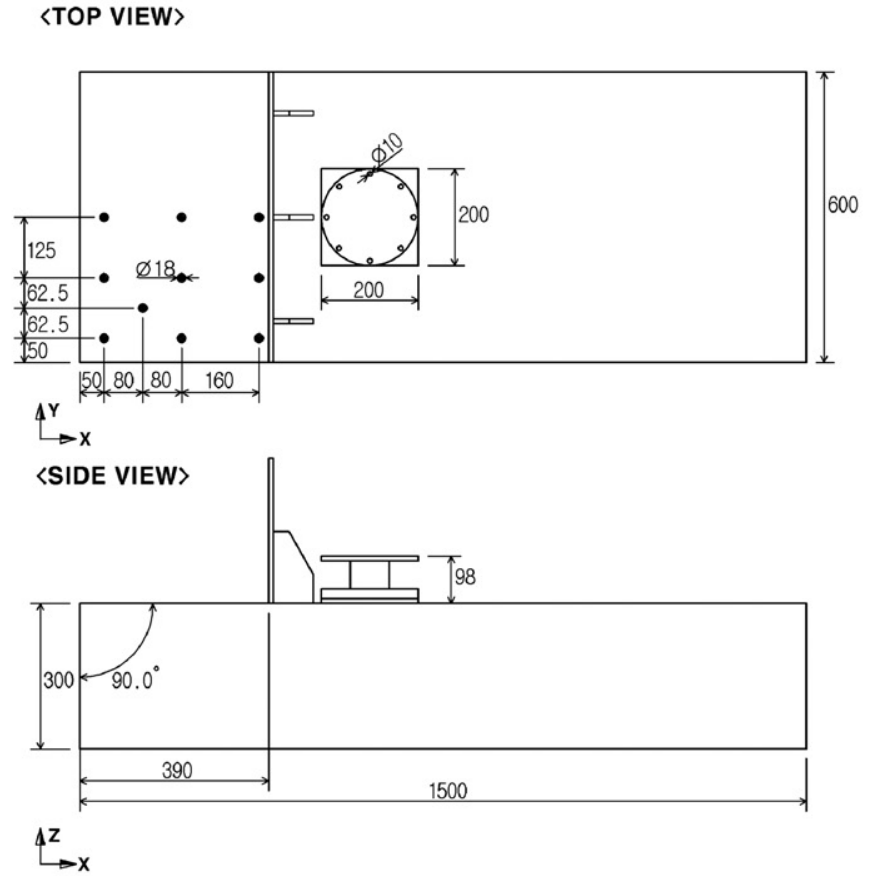}
  \caption{Dimensions of the model (in millimetres) and pressure gauge
  locations,}
  \label{fig:geometryStructure}
  \end{subfigure}%

  \begin{subfigure}[!h]{\textwidth}
  \centering
  \includegraphics[clip=true, trim = 1cm 2cm 1cm 2cm, width =0.4\textwidth]
      {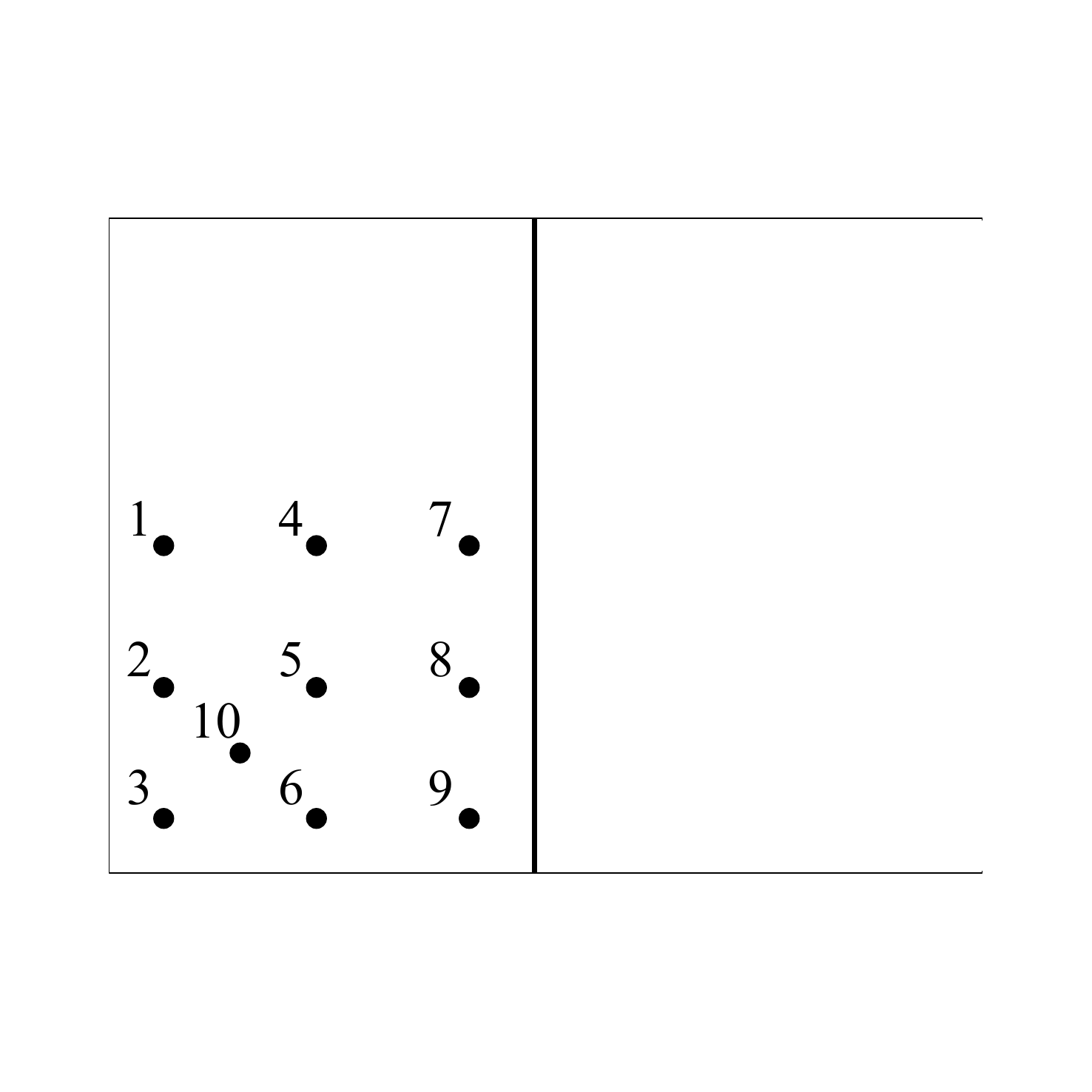}
  \caption{Indexing of pressure gauges.}
  \label{fig:probesStructure}
  \end{subfigure}
\end{center}
\caption{Geometry of the ship--like structure, courtesy of
Lee~\etal~\cite{leeEtAl2012}}
\label{fig:structure}
\end{figure}

\noindent For the second test case, we investigate a violent free surface flow
on a deck of a simplified ship--like structure in model scale. The structure
presented in~\autoref{fig:structure} is taken from Lee~\etal~\cite{leeEtAl2012},
who carried out comprehensive experimental analysis specifically for validation
of various CFD codes. They performed comprehensive studies including three
geometries, multiple regular waves impacting the structure and have measured
pressure at ten locations on the deck in front of a breakwater, as depicted
in~\autoref{fig:probesStructure}. In this study, we consider a rectangular
structure as presented in~\autoref{fig:geometryStructure} in a regular wave with
parameters given in~\autoref{tab:waveParametersStructure}. The wave is
moderately steep with $kH/2 = 0.188$, which is 7.5\% steeper than the wave
considered in~\autoref{sec:wavePropagation}.
\begin{table}[b!]
\caption{Wave parameters for the simplified ship model.}
\begin{center}
\begin{tabular}{*{3}{l}}
\hline
    Wave height& $H$, m & 0.13 \\
    \hline
    Wave length & $\lambda$, m & 2.25 \\
    \hline
    Water depth & $d$, m & 1 \\
    \hline
    Wave steepness & $kH/2$ & 0.188 \\
\hline
\end{tabular}
\end{center}
\label{tab:waveParametersStructure}
\end{table}

\begin{figure}[p!]
\centering
\includegraphics[width =0.8\textwidth]{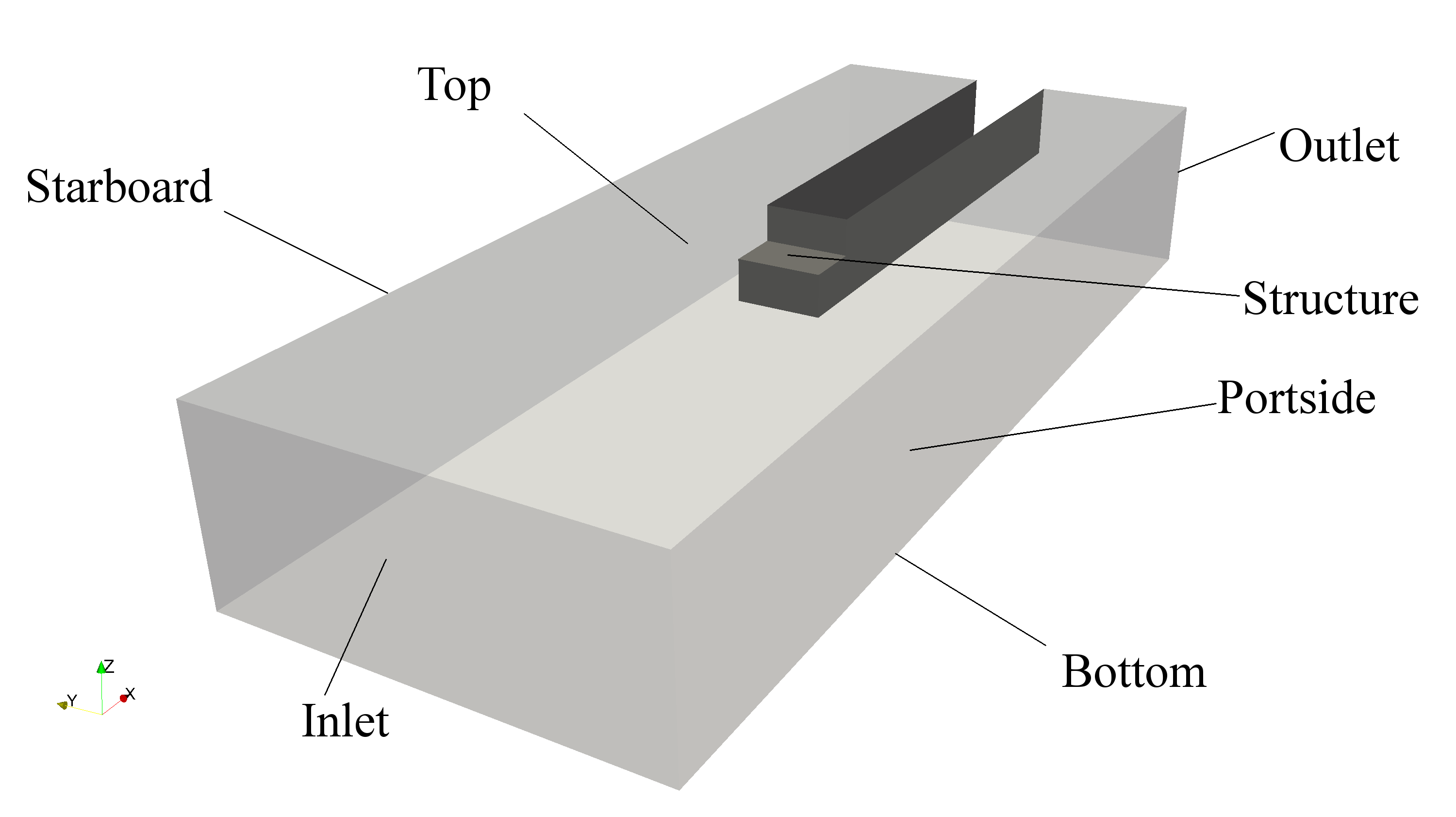}
\caption{Perspective view of the computational domain.}
\label{fig:domain}
\end{figure}

\begin{figure}[p!]
\centering
  \begin{subfigure}[!h]{0.4\textwidth}
  \includegraphics[width=\textwidth]{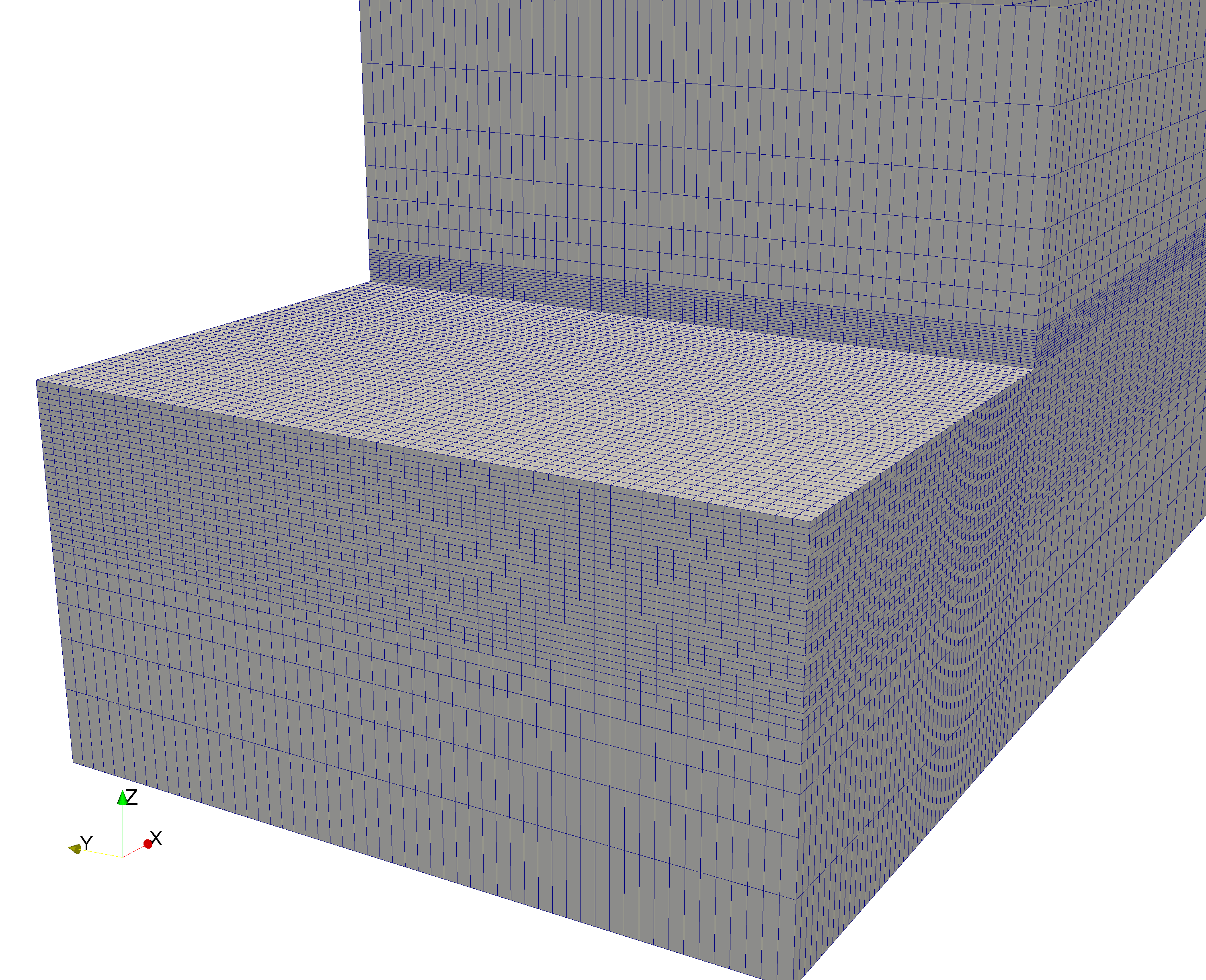}
  \caption{Surface grid of the structure,}
  \label{fig:gridSurface}
  \end{subfigure}~
  \begin{subfigure}[!h]{0.4\textwidth}
  \includegraphics[width=\textwidth]{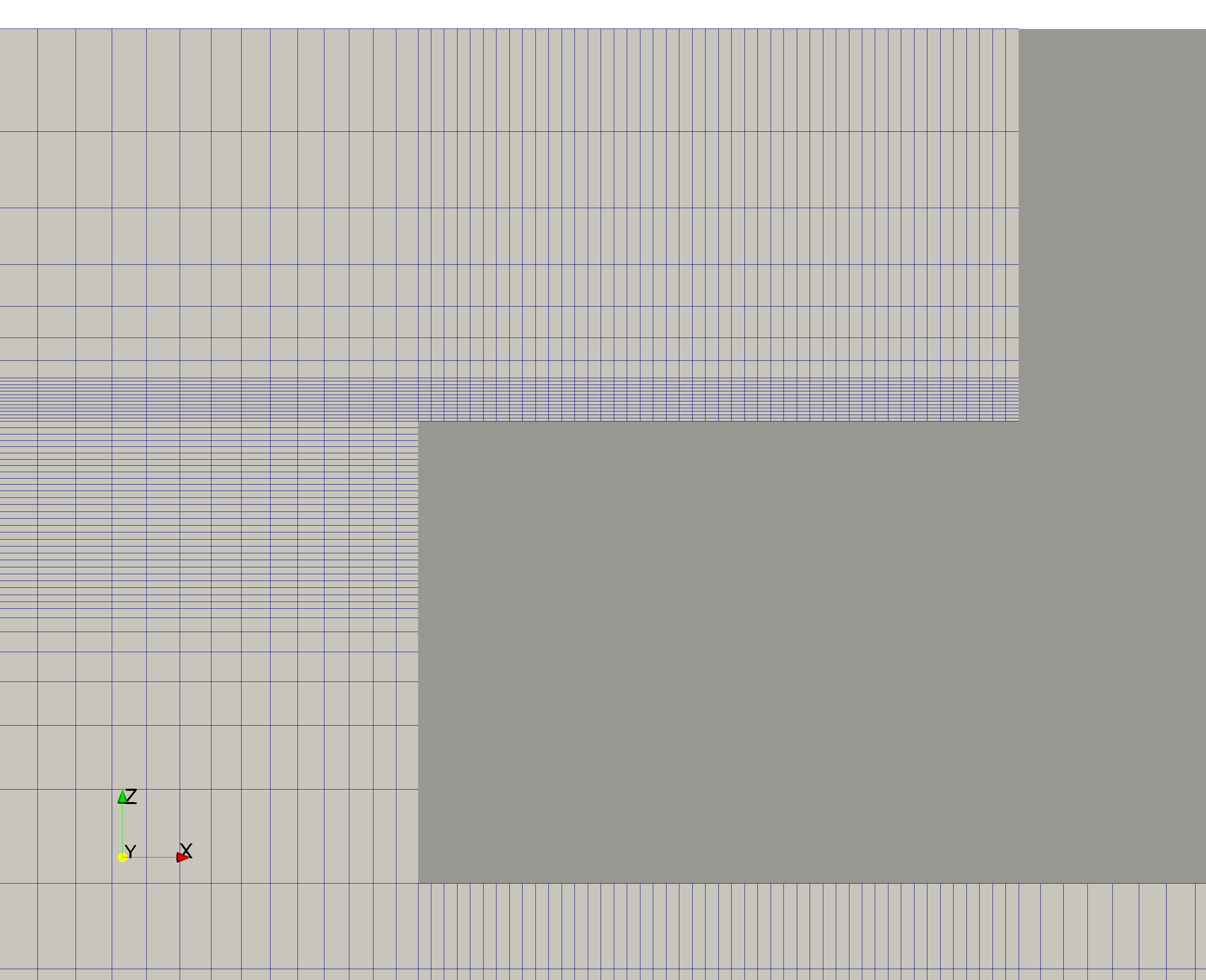}
  \caption{Longitudinal centre plane view.}
  \label{fig:gridLongitudinal}
  \end{subfigure}~
\caption{Coarse computational grid details.}
\label{fig:gridExample}
\end{figure}

\noindent The computational domain is identical to
Gatin~\etal~\cite{gatinEtAl2017a} and is presented in~\autoref{fig:domain}. The
relaxation zones are used at inlet, outlet, starboard and portside boundaries in
order to prevent wave reflection and reduce the size of the computational
domain. The domain is 6.5 metres long and the relaxation zones at inlet and
outlet are 2.5 metres long.  Height of the domain measured from the deck at
which the pressure probes are positioned is 0.3 meters. Depth of the domain from
the calm free surface to the bottom is 1 metre, while the width is 3 metres.
Simulations have been performed for 20 wave periods, while the time--step is
controlled with a maximum Courant--Friedrichs--Lewy (CFL) number of 0.75.\\
\indent The coarse computational grid is presented in~\autoref{fig:gridExample},
where~\autoref{fig:gridSurface} shows the surface grid at the structure,
and~\autoref{fig:gridLongitudinal} shows a slice through the longitudinal
centre plane. The grid is structured with grading towards the deck in the
vertical direction and towards the structure in the longitudinal direction.
The coarse grid is designed with approximately 23 cells per wave amplitude and
225 cells per wave length near the structure, and consists of 276\,699 cells.
In order to assess numerical uncertainty, three additional grids are used with a
constant refinement ratio of $r = \sqrt{2}$ for all three spatial dimensions and
time. Additional information regarding the grids is presented
in~\autoref{tab:gridsStructure}.
\begin{table}[h!]
\caption{Grids for the green--water case.}
\begin{center}
\begin{tabular}{*{5}{l}}
\hline
    Grid index & 1 & 2 & 3 & 4 \\
    \hline
    Number of cells & 276\,699 & 518\,476 & 1\,077\,515 &
        2\,181\,103 \\
    \hline
    Number of cells per wave length & 225 & 318 & 450 & 637 \\
    \hline
    Number of cells per wave height & 23 & 33 & 46 & 65 \\
\hline
\end{tabular}
\end{center}
\label{tab:gridsStructure}
\end{table}

\subsubsection{Verification study for the wave breaking case}
\label{sec:verificationBreaking}

\begin{figure}[h!]
\centering
\begin{subfigure}[!h]{0.3\textwidth}
\includegraphics[width=\textwidth]{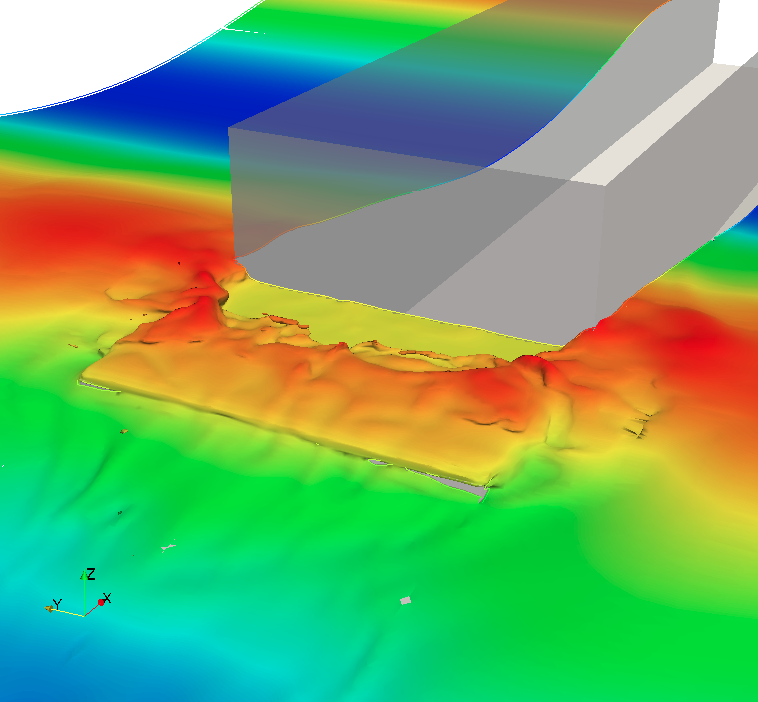}
\caption{t = 0,}
\end{subfigure}~
\begin{subfigure}[!h]{0.3\textwidth}
\includegraphics[width=\textwidth]{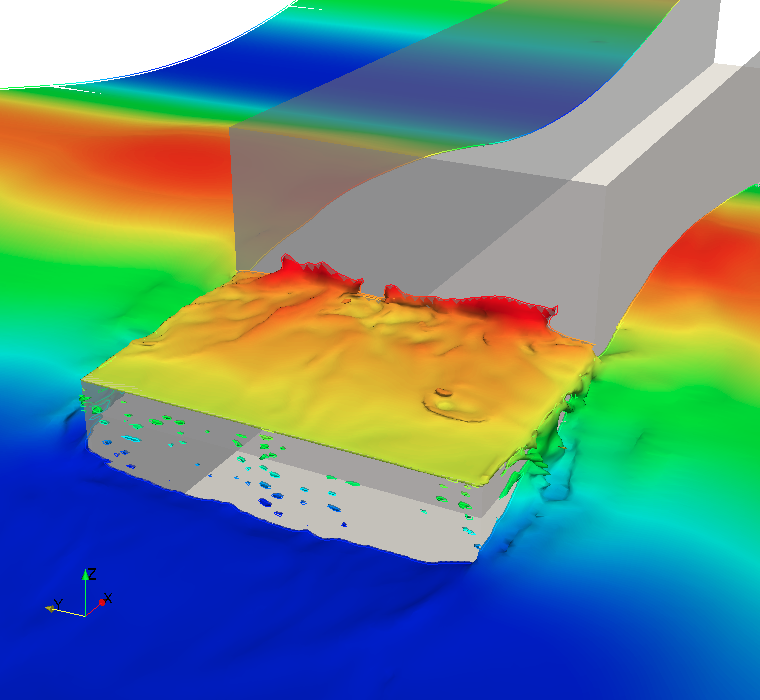}
\caption{t = T/6,}
\end{subfigure}~
\begin{subfigure}[!h]{0.3\textwidth}
\includegraphics[width=\textwidth]{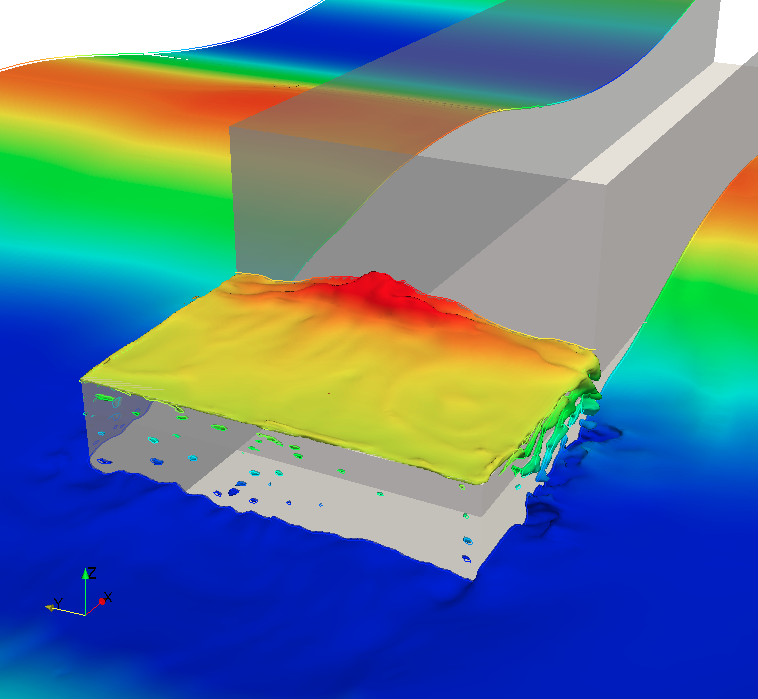}
\caption{t = 2T/6,}
\end{subfigure}%

\begin{subfigure}[!h]{0.3\textwidth}
\includegraphics[width=\textwidth]{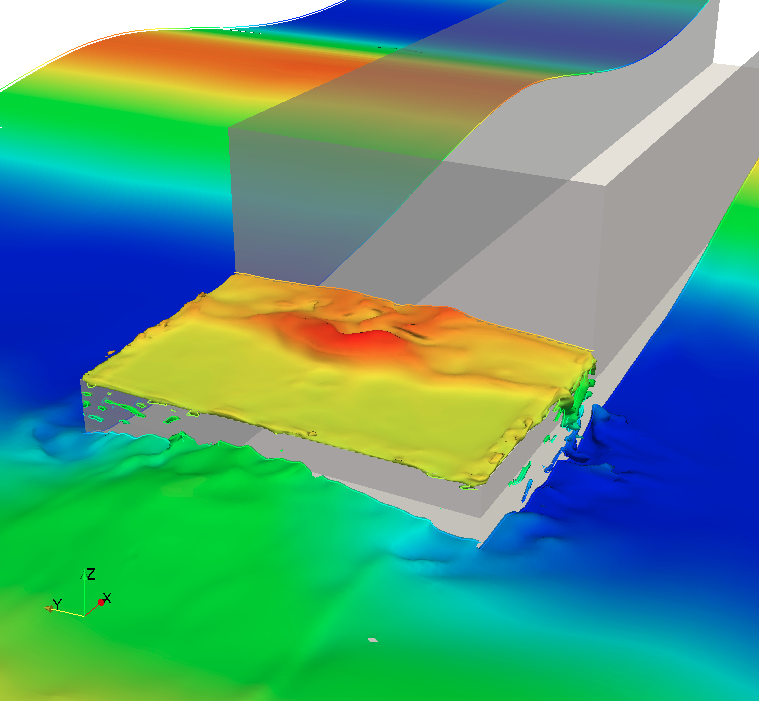}
\caption{t = 3T/6,}
\end{subfigure}~
\begin{subfigure}[!h]{0.3\textwidth}
\includegraphics[width=\textwidth]{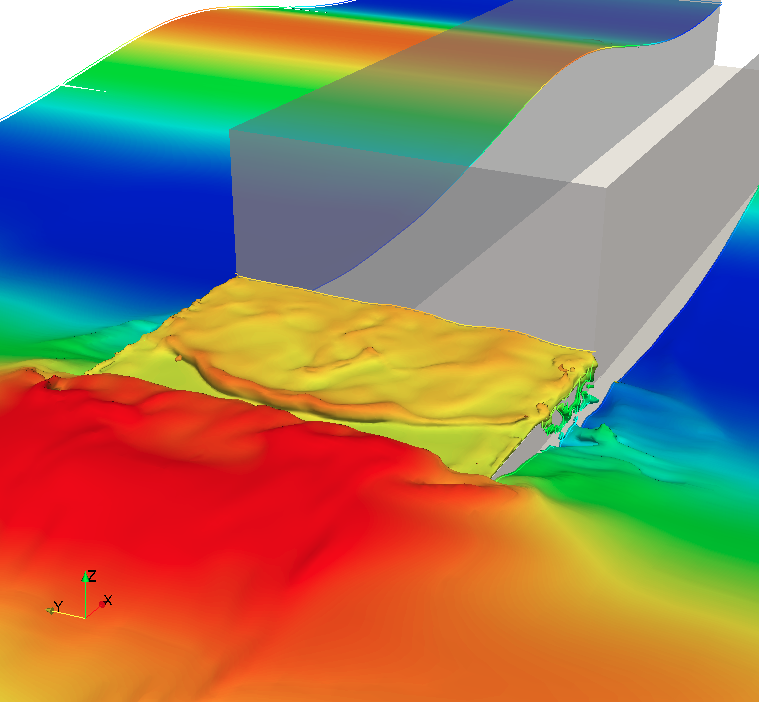}
\caption{t = 4T/6,}
\end{subfigure}~
\begin{subfigure}[!h]{0.3\textwidth}
\includegraphics[width=\textwidth]{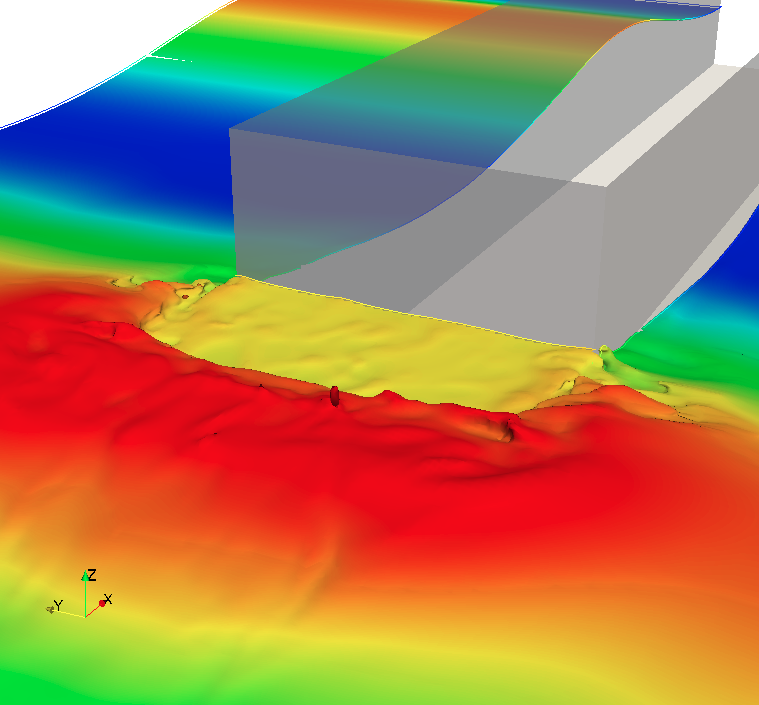}
\caption{t = 5T/6.}
\end{subfigure}%
\caption{Perspective view of the computational results during the last period.}
\label{fig:waveBreakingView}
\end{figure}

\begin{figure}[h!]
\centering
  \begin{subfigure}[!h]{0.45\textwidth}
  \includegraphics[width=0.95\textwidth]{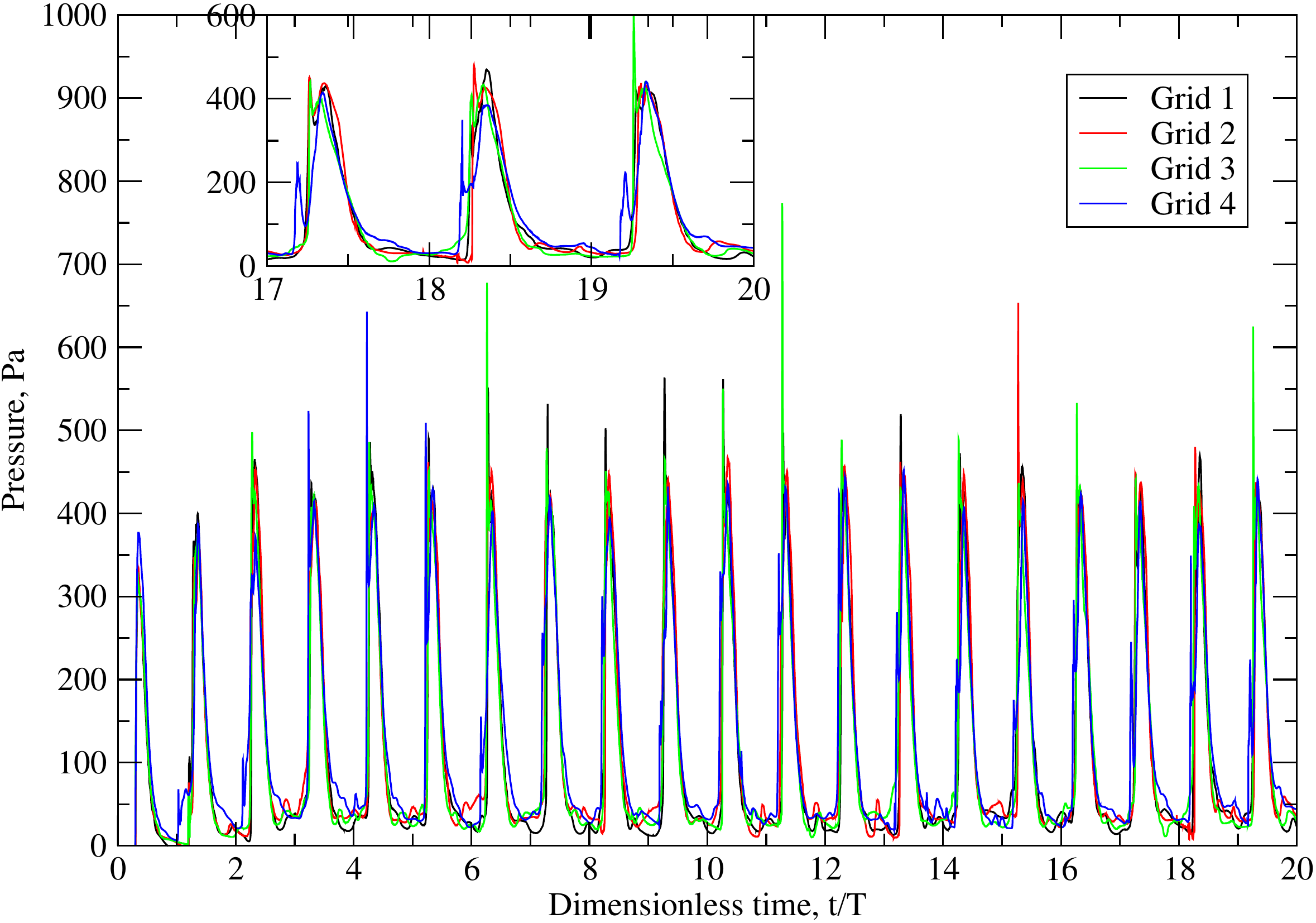}
  \caption{Pressure signal at pressure probe 3 obtained with all grids.}
  \label{fig:pressureSignal}
  \end{subfigure}~
  \begin{subfigure}[!h]{0.45\textwidth}
  \includegraphics[width=0.95\textwidth]{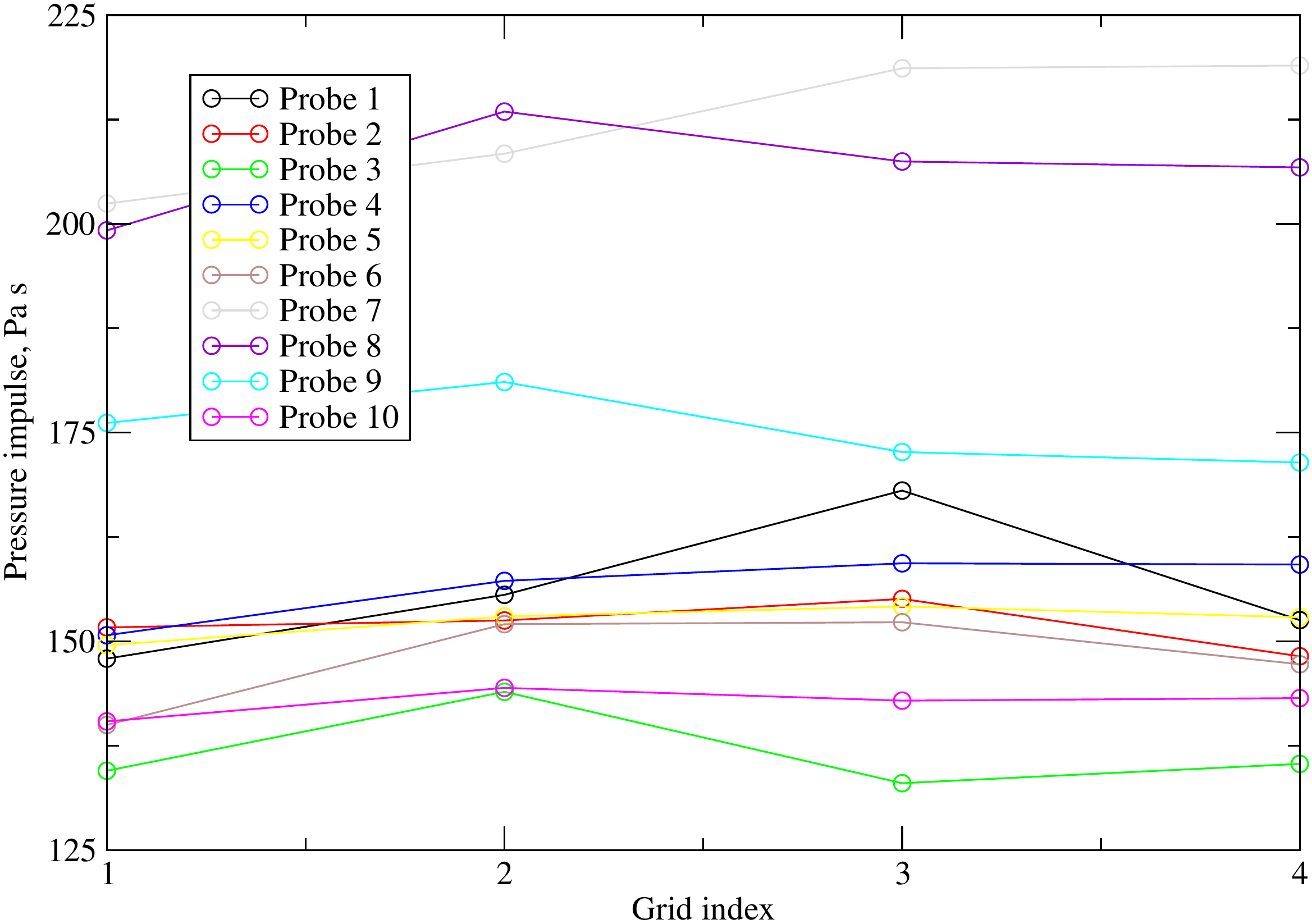}
  \caption{Average pressure impulses with all four grids for all probes.}
  \label{fig:pressureImpulses}
  \end{subfigure}%
\caption{Grid refinement study results for pressure and pressure impulses.}
\end{figure}

\begin{figure}[t!]
\caption{Uncertainty analysis for pressure impulses for all pressure probes.}
\begin{center}
  \begin{subfigure}[b]{0.45\textwidth}
    \begin{tabular}{l c c c c}
      \hline
      Item & $P$, Pas & $U_P, \%$ & $p$ & $q$ \\
      \hline
      \ensuremath{P_1}    & $156$ & $18.3\%$ & $1.00$ & $2.00$ \\
      \ensuremath{P_2}    & $152$ & $13.9\%$ & $1.39$ & $0.59$ \\
      \ensuremath{P_3}    & $137$ & $38.0\%$ & $1.00$ & $2.00$ \\
      \ensuremath{P_4}    & $156$ & $6.2\%$  & $^{*~1, 2}$ & $2.00$ \\
      \ensuremath{P_5}    & $152$ & $4.4\%$  & $2.00$ & $^{*~1, 2}$ \\
      \ensuremath{P_6}    & $148$ & $10.9\%$ & $2.00$ & $2.00$ \\
      \ensuremath{P_7}    & $212$ & $10.7\%$ & $1.00$ & $1.00$ \\
      \ensuremath{P_8}    & $207$ & $38.4\%$ & $^{*~1, 2}$ & $2.00$ \\
      \ensuremath{P_9}    & $175$ & $25.9\%$ & $1.00$ & $2.00$ \\
      \ensuremath{P_{10}} & $143$ & $9.5\%$  & $2.00$ & $^{*~1, 2}$ \\
      \hline
      Average & $164$ & $17.6$ & $1.34$ & $1.56$ \\
      \hline
      \multicolumn{5}{l}{\scriptsize $^{*~1, 2}$ Fit was made using first and
      second order exponents}\\
      \hline
    \end{tabular}
    \caption{Results for all pressure probes.}
    \label{tab:verificationWaveBreaking}
  \end{subfigure}%
  \begin{subfigure}[b]{0.35\textwidth}
  \includegraphics[width=7cm]{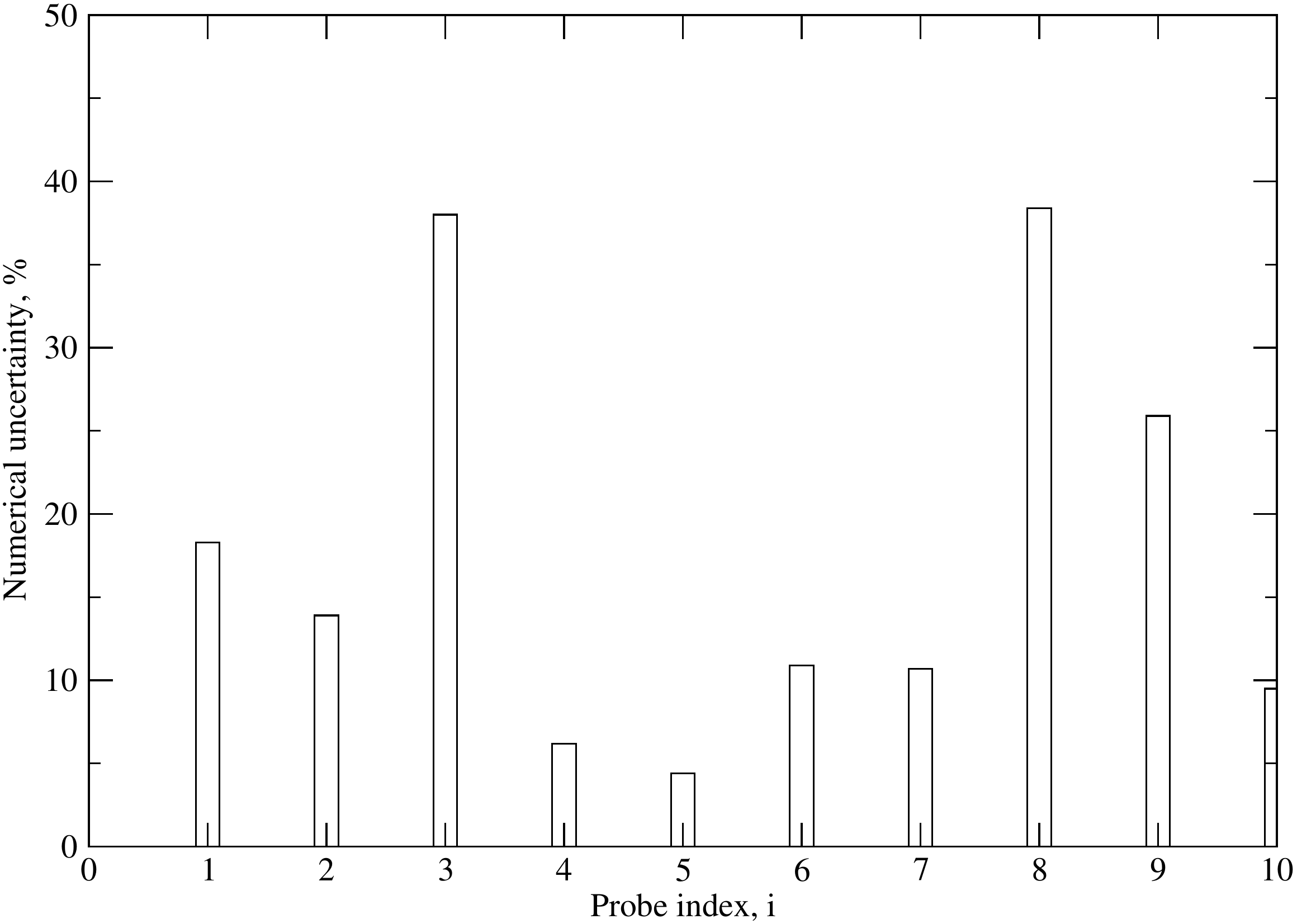}
  \caption{Numerical uncertainties.}
  \label{fig:uncertaintiesProbes}
  \end{subfigure}%
\end{center}
\label{tab:uncertaintyPressureImpulses}
\end{figure}

\noindent Prior to comparing the CFD results with experimental data, we perform
uncertainty assessment following the same guidelines by E\c{c}a and
Hoekstra~\cite{ecaHoekstra2014} as used for the wave propagation case. The focus
is placed on pressure at ten pressure gauges positioned at the deck.
The free surface flow at the deck exhibits complex flow
patterns, \autoref{fig:waveBreakingView}. Therefore higher numerical
uncertainties should be expected compared to wave propagation case. In
addition,~\autoref{fig:pressureSignal} presents the pressure signal at pressure
probe 3, indicating irregular behaviour in terms of pressure peaks. Due to
irregularity of pressure peaks, the uncertainty assessment is carried using an
average pressure impulse during last fifteen periods, defined as:
\begin{equation}
\label{eq:pImpulse}
    P =  \frac{\sum_{i=0}^{N} \int_{0}^{T} p_{i}\left(\tau\right) \d \tau}{N}
\end{equation}
\noindent Average pressure impulses for all wave gauges are presented
in~\autoref{fig:pressureImpulses} for all grids. Convergence with grid
refinement is not obtained for most of the pressure impulses. Nevertheless, the
procedure presented by E\c{c}a and Hoekstra~\cite{ecaHoekstra2014} yields
reasonable estimates of numerical uncertainty even when monotonic convergence is
not achieved. The verification results are summarized
in~\autoref{tab:uncertaintyPressureImpulses}, where the mean value of pressure
impulse $P$ for four grids is presented in the first column, while remaining
columns represent numerical uncertainty, achieved order of spatial convergence
and achieved order of temporal convergence, respectively. The numerical
uncertainty ranges from 4.4\% (pressure probe 5) to 38.4\% (pressure probe 8),
with the average value of 17.6\%, which is significantly higher than the
numerical uncertainties for the wave propagation case. As most of the pressure
impulses do not achieve convergence with grid refinement, the procedure by
E\c{c}a and Hoekstra~\cite{ecaHoekstra2014} estimates the order of convergence
between first and second order. Such irregular behaviour is expected,
considering the irregular behaviour of wave breaking during successive
periods (see~\autoref{fig:pressureSignal}).

\subsubsection{Validation study for the wave breaking case}
\label{sec:validationBreaking}

\begin{figure}[t!]
\centering
\includegraphics[width=0.6\textwidth]{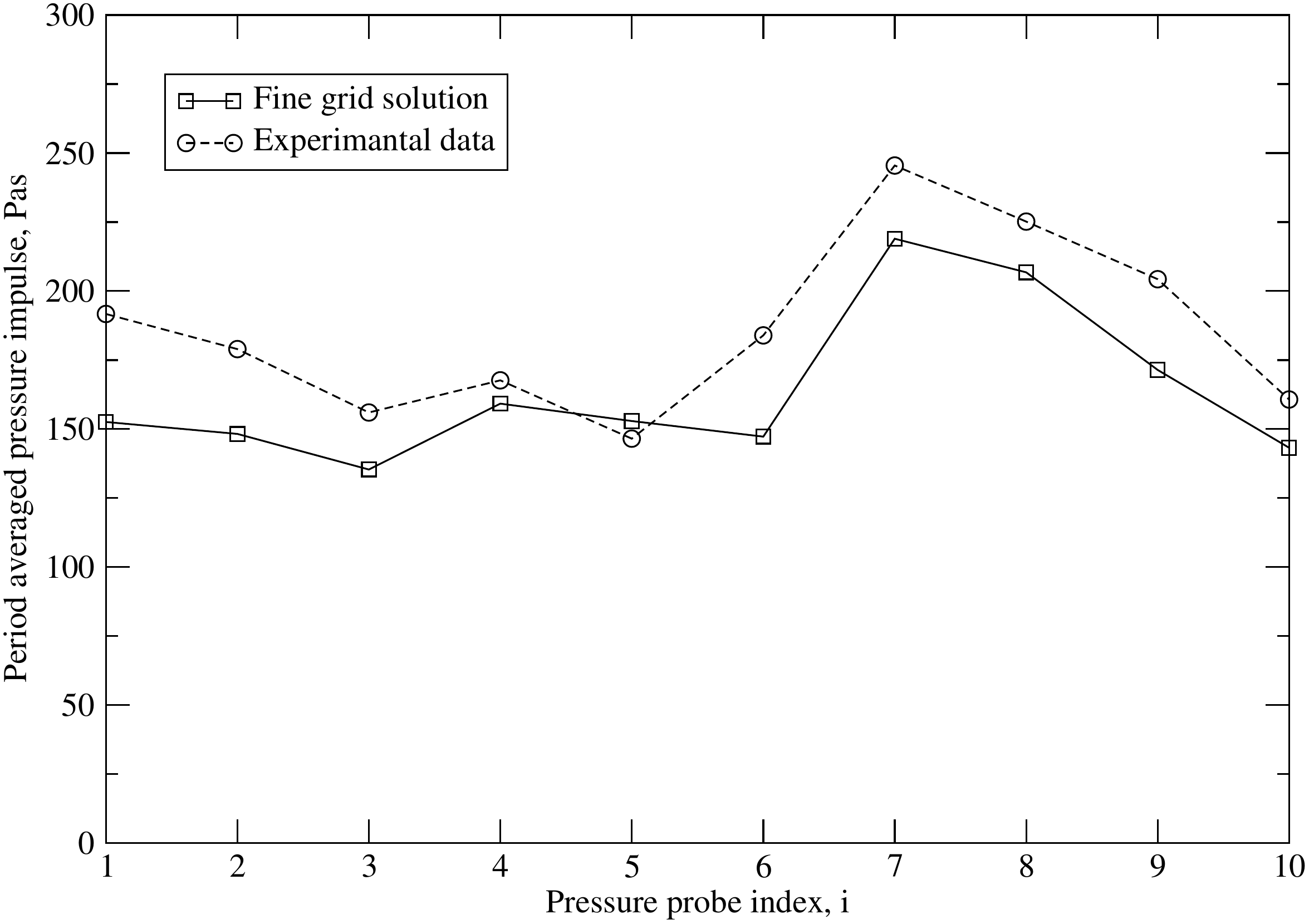}
\caption{Comparison of pressure impulses with experimental data.}
\label{fig:pressureImpulsesValidation}
\end{figure}

\noindent Having assessed numerical uncertainties, the CFD results are compared
with experimental measurements by Lee~\etal~\cite{leeEtAl2012}. Raw
experimental data has been provided by Lee~\etal~\cite{leeEtAl2012}, 
allowing us to perform identical post processing for 
period--averaged pressure impulses defined by~\eqref{eq:pImpulse} for
experimental and numerical data. The results are compared
in~\autoref{fig:pressureImpulsesValidation} for all wave gauges. The results
obtained with the fine grid are used as reference computational results
and are denoted with solid line and squares, while the experimental data is
denoted with dashed line and circles. The results obtained with the present
approach under--predict the experimental data for nine out of ten pressure
probes. The minimum relative error compared to experiments is approximately +4\%
for probe 5, while the maximum error is approximately -26\% for probe 1. On
average, the relative error between present results and experiments is
-14\%, while the estimated numerical uncertainty is approximately 18\% on
average (see~\autoref{tab:uncertaintyPressureImpulses}). The reader is referred
to Lee~\etal~\cite{leeEtAl2012} for detailed analysis and discussion on
experimental uncertainty, which is of the same order as the numerical
uncertainty.
%

\section{Conclusion and future work}
\label{sec:summary}

\noindent This paper presented a numerical model with sharp free surface
treatment within the Finite Volume framework. The model is based on a
combination of the Ghost Fluid Method for free surface jump conditions and the
\texttt{isoAdvector} geometric scheme for reconstruction and advection of the
volume fraction field. Although the method has been derived and implemented
following the arbitrary polyhedral Finite Volume approach in the open source
software OpenFOAM, all considered test cases have used structured hexahedral
grids in order to more reliably estimate the achieved orders of accuracy.\\
\indent Two test cases have been considered: moderately steep ($kH/2 = 0.174$)
wave propagation and wave breaking case on a simplified ship model. The wave
propagation case was simulated with six grids with uniform refinement ratio in
space and time, allowing us to estimate numerical uncertainty following the
latest guidelines by E\c{c}a and Hoekstra~\cite{ecaHoekstra2014}. Although the
implemented method is formally second order accurate, the observed order of
accuracy is between first and second order based on 144 measured items (12 wave
gauges and velocity probes $\times$ 3 signals (elevation, horizontal and
vertical velocity) $\times$ 4 items (first and second order amplitudes and
phases)). Averaged over all items, the achieved orders of spatial and temporal
accuracy are 1.59 and 1.60, respectively. The first order amplitudes of wave
elevation and velocity have numerical uncertainty which is 3.1\% on average,
while second order effects have higher numerical uncertainty of 18.3\% on
average, which is expected since the second order effects are an order of
magnitude smaller than the first order effects. Similarly, the average
numerical uncertainty for first order phases is $8.2^{\circ}$, while it is
$15.2^{\circ}$ for second order phases. Since the achieved order of accuracy
is between first and second order, we expect to have certain amount of
dissipation errors with the present model. Considering a domain that is 13
wave lengths long allowed us to quantify dissipation and dispersion errors
along the numerical wave tank. The smallest differences (often within 2\%
for all items) are obtained within the first four wave lengths down the
numerical wave tank. The differences tend to grow until $x = 8\lambda$
where they stabilise. The maximum error for first order amplitudes of
wave elevation and velocity is $3.8\%$ on average, while the second
order amplitudes show a discrepancy of approximately $8\%$. The maximum
dissipation error for first order effects is approximately $9.8^{\circ}$
on average for wave elevation and two velocity components, while it is
larger for second order effects: $22.7^{\circ}$.\\
\indent This extensive study revealed that the dissipation and dispersion errors
stabilise at $x \in [7\lambda, 9\lambda]$, while the errors are smaller closer
to the incoming boundary ($x = 0$). Therefore, smaller errors will be obtained
if one considers a domain with only \eg five wave lengths.\\
\indent The wave breaking case on a simplified ship model simulated with four
systematically refined grids revealed that the numerical uncertainties for
pressure impulses measured at the deck are significantly higher: ranging from
approximately 5 to 40\%. There are two reasons for higher numerical
uncertainties that are associated with irregular behaviour of pressure peaks
during successive wave periods, which shall be investigated in future:
\begin{itemize}
    \item The inability of the present model to take into account
    the air compressibility. It is possible that the compressibility of air
    is important for such a violent free surface flow, which should in turn
    smooth out the pressure peaks upon impact.
    \item Under--resolved physical scales. It is also possible that
    for such a violent flow, the nonlinearity of governing equations causes
    significant non--periodic behaviour even with periodic boundary
    conditions. Note that the finest grid considered had approximately two
    million cells, which is orders of magnitude smaller than scale--resolving
    simulations~\cite{deikeEtAl2016}.
\end{itemize}
\noindent The averaged pressure impulses measured at ten wave gauges showed good
comparison with experimental data.\\
\indent With these two test cases, the model has been extensively verified and
validated regarding wave propagation problems, although certain observations
need to be investigated in future:
\begin{itemize}
    \item Achieved orders of accuracy generally show irregular behaviour for
    different horizontal locations of wave gauges and probes. It is possible
    that the interplay of dissipation and dispersion errors needs to be
    addressed in a better way.
    \item The dispersion error for velocity shows slight oscillatory behaviour
    for different horizontal locations, while this is not observed for wave
    elevation.
    \item Quantifying the two--phase, vorticity and viscosity effects for wave
    propagation might reveal additional insights into dissipation mechanisms for
    steep waves.
\end{itemize}


\section*{Acknowledgements}
\noindent The authors are grateful to Prof. Lu\'{i}s E\c{c}a, Dr. Martin Hoekstra
and Dr. Guilherme Vaz for sharing their numerical uncertainty analysis toolbox
with the public~\cite{uncertaintyReFrescoWebsite}. The toolbox has been
extensively used in this study due to immense quantity of data, saving us quite
a lot of time and effort and enabling us to adhere to recent verification and
validation guidelines.


\bibliographystyle{packagesAndStyles/elsarticle-num}
\bibliography{vukcevikEtAl}

\begin{thebibliography}{10}
\expandafter\ifx\csname url\endcsname\relax
  \def\url#1{\texttt{#1}}\fi
\expandafter\ifx\csname urlprefix\endcsname\relax\def\urlprefix{URL }\fi
\expandafter\ifx\csname href\endcsname\relax
  \def\href#1#2{#2} \def\path#1{#1}\fi

\bibitem{yuEtAl2016}
X.~Yu, K.~Hendrickson, D.~K.-P. Yue, {Air Entrainment in Free Surface
  Turbulence}, in: {Proceedings of the 31st Symposium on Naval Hydrodynamics},
  2016.

\bibitem{deikeEtAl2016}
L.~Deike, W.~Melville, S.~Popinet, Air entrainment and bubble statistics in
  breaking waves, J. Fluid Mech. 801 (2016) 91--129.

\bibitem{larssonEtAl2013}
L.~Larsson, F.~Stern, M.~Visonneau, Numerical Ship Hydrodynamics: An assessment
  of the Gothenburg 2010 workshop, Springer, 2013.
\newblock \href {http://dx.doi.org/10.1007/978-94-007-7189-5}
  {\path{doi:10.1007/978-94-007-7189-5}}.

\bibitem{sternEtAl2012}
F.~Stern, J.~Yang, Z.~Wang, H.~Sadat-Hosseini, M.~Mousaviraad, B.~S., T.~Xing,
  {Computational Ship Hydrodynamics: Nowadays and Way Forward}, in: Proceedings
  of the 29$^{th}$ ONR Symposium on Naval Hydrodynamics, 2012, pp. 1--73.

\bibitem{higueraEtAl2015}
P.~Higuera, I.~Losada, J.~L. Lara, Three-dimensional numerical wave generation
  with moving boundaries, Coast. Eng. 101 (2015) 35--47.

\bibitem{paulsenEtAl2014b}
B.~T. Paulsen, H.~Bredmose, H.~B. Bingham, N.~G. Jacobsen, Forcing of a
  bottom-mounted circular cylinder by steep regular water waves at finite
  depth, J. Fluid Mech. 755 (2014) 1--3.
\newblock \href {http://dx.doi.org/10.1017/jfm.2014.386}
  {\path{doi:10.1017/jfm.2014.386}}.

\bibitem{vukcevicEtAl31SNH2016}
V.~Vuk\v{c}evi\'{c}, H.~Jasak, I.~Gatin, S.~Malenica, {Seakeeping Sensitivity
  Studies Using the Decomposition CFD Model Based on the Ghost Fluid Method},
  in: {Proceedings of the 31st Symposium on Naval Hydrodynamics}, 2016.

\bibitem{vukcevicJasakObliqueTokyo2015}
V.~Vuk\v{c}evi\'{c}, H.~Jasak, {Validation and Verification of Decomposition
  Model Based on Embedded Free Surface Method for Oblique Wave Seakeeping
  Simulations}, in: Proceedings of the Tokyo 2015: A Workshop on CFD in Ship
  Hydrodynamics, Vol.~3, 2015.

\bibitem{batchelor1967}
F.~R. Batchelor, An Introduction to Fluid Dynamics, Cambridge University Press,
  1967.

\bibitem{dopazo1977}
C.~Dopazo, On conditional averages for intermittent turbulent flows, J. Fluid
  Mech. 81 (1977) 433--438.

\bibitem{ubbinkIssa1999}
O.~Ubbink, R.~I. Issa, A method for capturing sharp fluid interfaces on
  arbitrary meshes, J. Comput. Phys. 153 (1999) 26--50.

\bibitem{jacobsenEtAl2012}
N.~G. Jacobsen, D.~R. Fuhrman, J.~Freds\o{}e, A wave generation toolbox for the
  open-source {CFD} library: {OpenFoam}\textregistered{}, Int. J. Numer. Meth.
  Fluids 70~(9) (2012) 1073--1088.
\newblock \href {http://dx.doi.org/10.1002/fld.2726}
  {\path{doi:10.1002/fld.2726}}.

\bibitem{higueraEtAl2013a}
P.~Higuera, J.~Lara, I.~J. Losada, Realistic wave generation and active wave
  absorption for {Navier}-{Stokes} models: {Application} to
  {OpenFoam}\textregistered{}, Coast. Eng. 71 (2013) 102--118.
\newblock \href {http://dx.doi.org/10.1016/j.coastaleng.2012.07.002}
  {\path{doi:10.1016/j.coastaleng.2012.07.002}}.

\bibitem{paulsenEtAl2014a}
B.~T. Paulsen, H.~Bredmose, H.~B. Bingham, An efficient domain decomposition
  strategy for wave loads on surface piercing circular cylinders, Coast. Eng.
  86 (2014) 57--76.
\newblock \href {http://dx.doi.org/10.1016/j.coastaleng.2014.01.006}
  {\path{doi:10.1016/j.coastaleng.2014.01.006}}.

\bibitem{lupieriEtAl2014}
G.~Lupieri, T.~Puzzer, G.~Contento, {Numerical study of the wave-wave
  interaction by viscous flow simulations with OpenFOAM}, in: {XXI. Symposium
  Sorta}, 2014.

\bibitem{fedkiwEtAl1999b}
R.~P. Fedkiw, T.~Aslam, B.~Merriman, S.~Osher, A non-oscillatory eulerian
  approach to interfaces in multimaterial flows (the ghost fluid method), J.
  Comput. Phys. 152 (1999) 457--492.

\bibitem{fedkiwEtAl1999}
R.~P. Fedkiw, T.~Aslam, S.~Xu, The ghost fluid method for deflagration and
  detonation discontinuities, J. Comput. Phys. 154~(2) (1999) 393--427.

\bibitem{wangEtAl2013}
S.~{Wang}, J.~{Glimm}, R.~{Samulyak}, X.~{Jiao}, C.~{Diao}, {{An Embedded
  Boundary Method for Two Phase Incompressible Flow}}, ArXiv e-prints.\href
  {http://arxiv.org/abs/1304.5514} {\path{arXiv:1304.5514}}.

\bibitem{olssonKreiss2005}
E.~Olsson, G.~Kreiss, A conservative level set method for two phase flow, J.
  Comput. Phys. 210~(1) (2005) 225--246.

\bibitem{desjardinsEtAl2008}
O.~Desjardins, V.~Moureau, H.~Pitsch, An accurate conservative level set/ghost
  fluid method for simulating turbulent atomization, J. Comput. Phys. 227~(18)
  (2008) 8395--8416.

\bibitem{boGrove2014}
W.~Bo, J.~W. Grove, A volume of fluid method based ghost fluid method for
  compressible multi-fluid flows, Comput. Fluids 90 (2014) 113--122.

\bibitem{lalanneEtAl2015}
B.~Lalanne, L.~R. Villegas, S.~Tanguy, F.~Risso, On the computation of viscous
  terms for incompressible two-phase flows with level set/ghost fluid method,
  J. Comput. Phys. 301 (2015) 289--307.

\bibitem{queuteyVisonneau2007}
P.~Queutey, M.~Visonneau, An interface capturing method for free--surface
  hydrodynamic flows, Comput. Fluids 36 (2007) 1481--1510.
\newblock \href {http://dx.doi.org/10.1002/j.compfluid.2006.11.007}
  {\path{doi:10.1002/j.compfluid.2006.11.007}}.

\bibitem{huangEtAl2007}
J.~Huang, P.~M. Carrica, F.~Stern, Coupled ghost fluid/two--phase level set
  method for curvilinear body--fitted grids, Int. J. Numer. Meth. Fluids 44
  (2007) 867--897.
\newblock \href {http://dx.doi.org/10.1002/fld.1499}
  {\path{doi:10.1002/fld.1499}}.

\bibitem{vukcevicEtAl2017}
V.~Vuk\v{c}evi\'{c}, H.~Jasak, I.~Gatin, {Implementation of the Ghost Fluid
  Method for Free Surface Flows in Polyhedral Finite Volume Framework}, Comput.
  Fluids 153 (2017) 1--19.
\newblock \href {http://dx.doi.org/10.1016/j.compfluid.2017.05.003}
  {\path{doi:10.1016/j.compfluid.2017.05.003}}.

\bibitem{ruschePhD2002}
H.~Rusche, Computational fluid dynamics of dispersed two - phase flows at high
  phase fractions, Ph.D. thesis, Imperial College of Science, Technology \&
  Medicine, London (2002).

\bibitem{tryggvasonEtAl2011}
G.~Tryggvason, R.~Scardovelli, S.~Zaleski, Direct Numerical Simulations of
  Gas-Liquid Multiphase Flows, Cambridge University Press, 2011.

\bibitem{aulisaEtAl2013}
E.~Aulisa, S.~Manservisi, R.~Scardovelli, S.~Zaleski, {A geometrical
  area-preserving Volume-of-Fluid advection method}, J. Comput. Phys 192~(1).
\newblock \href {http://dx.doi.org/10.1016/j.jcp.2003.07.003}
  {\path{doi:10.1016/j.jcp.2003.07.003}}.

\bibitem{roenbyEtAl2016}
J.~Roenby, H.~Bredmose, H.~Jasak, A computational method for sharp interface
  advection, Royal Society Open Science 3~(11).
\newblock \href {http://dx.doi.org/10.1098/rsos.160405}
  {\path{doi:10.1098/rsos.160405}}.

\bibitem{sethian1996}
J.~A. Sethian, Level Set Methods: Evolving Interfaces in Geometry, Fluid
  Mechanics, Computer Vision and Materials Science, Cambridge University Press,
  1996.

\bibitem{osherFedkiw2003}
S.~Osher, R.~Fedkiw, Level Set Methods and Dynamic Implicit Surfaces, Springer,
  2003.

\bibitem{olssonEtAl2007}
E.~Olsson, G.~Kreiss, S.~Zahedi, A conservative level set method for two phase
  flow ii, J. Comput. Phys. 225~(1) (2007) 785--807.

\bibitem{gomezEtAl2005}
P.~G\'{o}mez, J.~Hern\'{a}ndez, J.~L\'{o}pez, On the reinitialization procedure
  in a narrow-band locally refined level set method for interfacial flows, Int.
  J. Numer. Methods Eng. 63 (2005) 1478--1512.

\bibitem{hartmannEtAl2008}
D.~Hartmann, M.~Meinke, W.~Schr\"{o}der, Differential equation based
  constrained reinitialization for level set methods, J. Comput. Phys. 227
  (2008) 6821--6845.

\bibitem{sunBeckermann2007}
Y.~Sun, C.~Beckermann, {Sharp interface tracking using the phase--field
  equation}, {J. Comput. Phys.} 220 (2007) 626--653.
\newblock \href {http://dx.doi.org/10.1016/j.jcp.2007.05.025}
  {\path{doi:10.1016/j.jcp.2007.05.025}}.

\bibitem{sunBeckermann2008}
Y.~Sun, C.~Beckermann, {A two--phase diffusive--interface model for Hele--Shaw
  flows with large property contrasts}, {Physica D} 237 (2008) 3089--3098.
\newblock \href {http://dx.doi.org/10.1016/j.physd.2008.06.010}
  {\path{doi:10.1016/j.physd.2008.06.010}}.

\bibitem{vukcevicEtAl2016a}
V.~Vuk\v{c}evi\'{c}, H.~Jasak, S.~Malenica, {Decomposition model for naval
  hydrodynamic applications, Part I: Computational method}, Ocean Eng. 121
  (2016) 37--46.
\newblock \href {http://dx.doi.org/10.1016/j.oceaneng.2016.05.022}
  {\path{doi:10.1016/j.oceaneng.2016.05.022}}.

\bibitem{vukcevicEtAl2016b}
V.~Vuk\v{c}evi\'{c}, H.~Jasak, S.~Malenica, {Decomposition model for naval
  hydrodynamic applications, Part II: Verification and validation}, Ocean Eng.
  121 (2016) 76--88.
\newblock \href {http://dx.doi.org/10.1016/j.oceaneng.2016.05.021}
  {\path{doi:10.1016/j.oceaneng.2016.05.021}}.

\bibitem{Lopez2008}
J.~L{\'{o}}pez, C.~Zanzi, P.~G{\'{o}}mez, F.~Faura, J.~Hern{\'{a}}ndez, {A new
  volume of fluid method in three dimensions - Part II: Piecewise-planar
  interface reconstruction with cubic-B{\'{e}}zier fit}, International Journal
  for Numerical Methods in Fluids 58~(8) (2008) 923--944.
\newblock \href {http://arxiv.org/abs/fld.1} {\path{arXiv:fld.1}}, \href
  {http://dx.doi.org/10.1002/fld.1775} {\path{doi:10.1002/fld.1775}}.

\bibitem{Hernandez2008}
J.~Hern{\'{a}}ndez, J.~L{\'{o}}pez, P.~G{\'{o}}mez, C.~Zanzi, F.~Faura, {A new
  volume of fluid method in three dimensions - Part I: Multidimensional
  advection method with face-matched flux polyhedra}, International Journal for
  Numerical Methods in Fluids 58~(8) (2008) 897--921.
\newblock \href {http://arxiv.org/abs/fld.1} {\path{arXiv:fld.1}}, \href
  {http://dx.doi.org/10.1002/fld.1776} {\path{doi:10.1002/fld.1776}}.

\bibitem{Ahn2007}
H.~T. Ahn, M.~Shashkov, {Multi-material interface reconstruction on generalized
  polyhedral meshes}, Journal of Computational Physics 226~(2) (2007)
  2096--2132.
\newblock \href {http://dx.doi.org/10.1016/j.jcp.2007.06.033}
  {\path{doi:10.1016/j.jcp.2007.06.033}}.

\bibitem{Xie2014}
B.~Xie, S.~Ii, F.~Xiao, {An efficient and accurate algebraic interface
  capturing method for unstructured grids in 2 and 3 dimensions: The THINC
  method with quadratic surface representation}, International Journal for
  Numerical Methods in Fluids 76~(12) (2014) 1025--1042.
\newblock \href {http://arxiv.org/abs/fld.3968} {\path{arXiv:fld.3968}}, \href
  {http://dx.doi.org/10.1002/fld.3968} {\path{doi:10.1002/fld.3968}}.

\bibitem{Roenby2017}
J.~Roenby, B.~E. Larsen, H.~Bredmose, H.~Jasak, {A New Volume-of-Fluid Method
  in Openfoam}, in: VII International Conference on Computational Methods in
  Marine Engineering, MARINE 2017, 2017, pp. 1--12.

\bibitem{higueraEtAl2013b}
P.~Higuera, J.~Lara, I.~J. Losada, Simulating coastal engineering processes
  with {OpenFoam} \textregistered{}, Coast. Eng. 71 (2013) 119--134.
\newblock \href {http://dx.doi.org/10.1016/j.coastaleng.2012.06.002}
  {\path{doi:10.1016/j.coastaleng.2012.06.002}}.

\bibitem{roache1997}
P.~Roache, Quantification of uncertainty in computational fluid dynamics, Ann.
  Rev. Fluid. Mech. 29 (1997) 123--160.

\bibitem{sternEtAl2001}
F.~Stern, R.~V. Wilson, H.~W. Coleman, E.~G. Paterson, {Comprehensive Approach
  to Verification and Validation of CFD Simulations--Part 1: Methodology and
  Procedures}, J. Fluids. Eng 123(4) (2001) 793--802.
\newblock \href {http://dx.doi.org/10.1115/1.1412235}
  {\path{doi:10.1115/1.1412235}}.

\bibitem{ecaHoekstra2014}
{E\c{c}a, L. and Hoekstra, M.}, A procedure for the estimation of the numerical
  uncertainty of cfd calculations based on grid refinement studies, J. Comput.
  Phys. 262 (2014) 104--130.
\newblock \href {http://dx.doi.org/10.1016/j.jcp.2014.01.006}
  {\path{doi:10.1016/j.jcp.2014.01.006}}.

\bibitem{leeEtAl2012}
H.-H. Lee, H.-J. Lim, S.~H. Rhee, {Experimental investigation of green water on
  deck for a CFD validation database}, Ocean Engineering 42 (2012) 47--60.
\newblock \href {http://dx.doi.org/10.1016/j.oceaneng.2011.12.026}
  {\path{doi:10.1016/j.oceaneng.2011.12.026}}.

\bibitem{vukcevicPhD2016}
V.~Vuk\v{c}evi\'{c}, Numerical modelling of coupled potential and viscous flow
  for marine applications - in preparation, Ph.D. thesis, Faculty of Mechanical
  Engineering and Naval Architecture, University of Zagreb, {PhD Thesis}
  (2016).
\newblock \href {http://dx.doi.org/10.13140/RG.2.2.23080.57605}
  {\path{doi:10.13140/RG.2.2.23080.57605}}.

\bibitem{jasakPhD1996}
H.~Jasak, Error analysis and estimation for the finite volume method with
  applications to fluid flows, Ph.D. thesis, Imperial College of Science,
  Technology \& Medicine, London (1996).

\bibitem{patankarSpalding1972}
S.~V. Patankar, D.~B. Spalding, A calculation procedure for heat, mass and
  momentum transfer in three-dimensional parabolic flows, Int. J. Heat Mass
  Transf. 15 (1972) 1787--1806.

\bibitem{issa1986}
R.~I. Issa, Solution of the implicitly discretised fluid flow equations by
  operator-splitting, J. Comput. Phys. 62 (1986) 40--65.

\bibitem{rhieChow1983}
C.~M. Rhie, W.~L. Chow, A numerical study of the turbulent flow past an
  isolated airfoil with trailing edge separation, AIAA J. 21 (1983) 1525--1532.

\bibitem{tukovicJasak2012}
Z.~Tukovi\'{c}, H.~Jasak, A moving mesh finite volume interface tracking method
  for surface tension dominated interfacial fluid flow, Comput. Fluids 55
  (2012) 70--84.

\bibitem{demirdzic2015}
I.~Demird\v{z}i\'{c}, {On the Discretization of the Diffusion Term in
  Finite--Volume Continuum Mechanics}, Numer. Heat Transfer, Part B 68 (2015)
  1--10.
\newblock \href {http://dx.doi.org/10.1080/10407790.2014.985992}
  {\path{doi:10.1080/10407790.2014.985992}}.

\bibitem{ferzigerPeric1996}
J.~H. Ferziger, M.~Peric, Computational Methods for Fluid Dynamics, Springer,
  1996.

\bibitem{jasakEtAlCFDWTOT2015}
H.~Jasak, V.~Vuk\v{c}evi\'{c}, I.~Gatin, {Numerical Simulation of Wave Loads on
  Static Offshore Structures}, in: {CFD for Wind and Tidal Offshore Turbines},
  Springer Tracts in Mechanical Engineering, 2015, pp. 95--105.

\bibitem{Denner2014}
F.~Denner, B.~G. {Van Wachem}, {Fully-coupled balanced-force VOF framework for
  arbitrary meshes with least-squares curvature evaluation from volume
  fractions}, Numerical Heat Transfer, Part B: Fundamentals 65~(3) (2014)
  218--255.
\newblock \href {http://arxiv.org/abs/1405.0829} {\path{arXiv:1405.0829}},
  \href {http://dx.doi.org/10.1080/10407790.2013.849996}
  {\path{doi:10.1080/10407790.2013.849996}}.

\bibitem{rieneckerFenton1981}
M.~M. Rienecker, J.~D. Fenton, {A Fourier approximation method for steady water
  waves}, J. Fluid Mech. 104 (1981) 119--137.

\bibitem{deanDalrymple2010}
R.~G. Dean, R.~A. Dalrymple, Water Wave Mechanics for Engineers and Scientists,
  Vol. 2: Advanced Series on Ocean Engineering, World Scientific, 2010.

\bibitem{uncertaintyReFrescoWebsite}
{ReFRESCO V\&V Tools},
  \url{http://www.refresco.org/verification-validation/utilitiesvv-tools/},
  [Online; accessed 10 October 2017] (2017).

\bibitem{ecaEtAlNUTTS2017}
{E\c{c}a, L. and Vaz, G. and Hoekstra, M.}, {Iterative Errors in Unsteady Flow
  Simulations: Are they Really Negligible?}, in: Proceedings of the 20th
  Numerical Towing Tank Symposium (NUTTS2017), 2017.

\bibitem{gatinEtAl2017a}
{Gatin, I. and Vuk\v{c}evi\'{c}, V. and Jasak, H. and Seo, J. and Rhee, S.-H.},
  {CFD Verification and Validation of Green Sea Loads}, Ocean Eng.Accepted for
  publication.

\end{thebibliography}

\end{document}